\documentclass[a4paper,fleqn,usenatbib]{mnras}

\usepackage{graphicx}

\usepackage{times}
\usepackage{aas_macros}
\usepackage{pdfpages}
\usepackage{grffile}
\usepackage{amsmath,amssymb}
\usepackage{amsfonts}
\usepackage{bm}
\usepackage{bbm}
\usepackage[ruled,linesnumbered]{algorithm2e}
\usepackage{bbold}
 \usepackage{url}
 \usepackage{arydshln}
\usepackage{booktabs}

\usepackage{xcolor,colortbl}

\usepackage{tikz}
\usetikzlibrary{positioning}
\usepackage{multirow}
\usepackage{pifont}% http://ctan.org/pkg/pifont
\newcommand{\cmark}{\ding{51}}%
\newcommand{\xmark}{\ding{55}}%

\newtheorem{theorem}{Theorem}[section]

\newtheorem{proposition}[theorem]{Proposition}
\newtheorem{lemma}[theorem]{Lemma}
\newtheorem{remark}[theorem]{Remark}
\newtheorem{example}[theorem]{Example}
\newtheorem{corollary}[theorem]{Corollary}

\def\argmin{\mathop{\rm argmin}}
\def\argmax{\mathop{\rm argmax}}

\newcommand{\vect}[1]{\boldsymbol{#1}}

\newcommand{\tabL}{1.5mm}

\definecolor{Gray}{gray}{0.8}
\definecolor{Gray1}{gray}{0.9}
\newcolumntype{H}{>{\columncolor{Gray}}c}
\newcolumntype{I}{>{\columncolor{Gray1}}c}

\newcommand{\blue}[1]{\textcolor{blue}{#1}}

\newcommand{\yellow}[1]{\textcolor{yellow}{#1}}

\SetKwRepeat{Do}{do}{while}

\title[Uncertainty quantification for RI imaging I]{Uncertainty quantification for radio interferometric imaging:\\ I.~proximal MCMC methods}
\author[Cai, Pereyra and McEwen]{Xiaohao Cai$^{1}$\thanks{E-mail:~x.cai@ucl.ac.uk~(XC);~m.pereyra@hw.ac.uk~(MP); \newline jason.mcewen@ucl.ac.uk (JDM)},
Marcelo Pereyra$^{2}$\blue{\footnotemark[1]} 
and Jason D. McEwen$^{1}$\blue{\footnotemark[1]} \\
$^{1}$Mullard Space Science Laboratory,  University College London (UCL), Surrey RH5 6NT, United Kingdom  \\
$^{2}$Maxwell Institute for Mathematical Sciences, Heriot-Watt University, Edinburgh EH14 4AS, United Kingdom\\
} 

\begin{document}

%\date{\today}
\date{Accepted ---. Received ---; in original form ---}
\pagerange{\pageref{sec:intro}--\pageref{lastpage}}
\pubyear{2017}

\maketitle

\begin{abstract}  
Uncertainty quantification is a critical missing component in radio
interferometric imaging that will only become increasingly important as the
big-data era of radio interferometry emerges.  Since radio interferometric
imaging requires solving a high-dimensional, ill-posed inverse problem,
uncertainty quantification is difficult but also critical to the accurate
scientific interpretation of radio observations.  Statistical sampling approaches to perform Bayesian inference, like Markov Chain Monte Carlo (MCMC)
sampling, can in principle recover the full posterior distribution of the image,
from which uncertainties can then be quantified.  However, traditional
high-dimensional sampling methods are generally limited to smooth (\textit{e.g.}
Gaussian) priors and cannot be used with sparsity-promoting priors.  Sparse
priors, motivated by the theory of compressive sensing, have been shown
to be highly effective for radio interferometric imaging.   In this article
\mbox{proximal} MCMC methods are developed for radio interferometric imaging,
leveraging proximal calculus to support non-differential priors, such as sparse
priors, in a Bayesian framework.  Furthermore, three strategies to quantify
uncertainties using the recovered posterior distribution are developed: (i)
local (pixel-wise) credible intervals to provide error bars for each individual
pixel; (ii) highest posterior density credible regions;  and (iii) hypothesis
testing of image structure.  These forms of uncertainty quantification provide
rich information for analysing radio interferometric  observations in a
statistically robust manner.
\end{abstract}


%-------------------------------------------------------------------
\section{Introduction}\label{sec:intro}
%-------------------------------------------------------------------
Radio interferometric (RI) telescopes provide a wealth of valuable information for astrophysics and cosmology \citep{RV46,ryl60,tho08} since they allow observation of the radio emission of the sky with high angular resolution and sensitivity. 
The measured visibilities acquired by the telescope relate to Fourier measurements of the sky image of interest
(the Fourier model may be modified to account for, {\it e.g.}, wide fields of view, co-planer baselines, and other directional 
dependent effects). Imaging observations made by radio telescopes requires solving an ill-posed linear inverse problem \citep{tho08}, 
which is an important first step in many subsequent scientific analyses. 
Since the inverse problem is ill-posed (sometimes seriously), uncertainty information regarding reconstructed images 
(\textit{e.g.} error estimates) is critical.  Nevertheless, uncertainty information is currently lacking in all RI imaging techniques used in practice.

Classical imaging techniques were developed in the field to solve the RI reconstruction problem, 
such as CLEAN and its multi-scale variants \citep{hog74,BC04,cor08,SFM11}.
In particular, CLEAN builds a model image by iteratively removing point source components from the residuals of the acquired data 
(at each iteration). CLEAN-based algorithms, however, are {typically slow (generally requiring computationally demanding major cycles; \textit{cf.} Clark CLEAN)}, requiring fine-tuning and supervision, while
providing suboptimal imaging quality \citep[see, \textit{e.g.},][]{li11a,CMW12}.
Another classical technique is the maximum entropy method (MEM) \citep{A74,GD78}, extended to RI imaging by \citet{CE85}.  
The MEM approach of \citet{CE85} developed for RI imaging considers a regularisation problem consisting of a relative entropic prior, 
a (Gaussian) likelihood term and an additional flux constraint.  In principle, MEM requires less fine-tuning and supervision compared 
to CLEAN and can therefore alleviate part of the shortcomings of CLEAN-based algorithms. However, an optimal metric -- 
expressed as an entropy functional -- is not known in advance and therefore needs to be chosen individually \citep{SMQB01,MHL04}.  
Indeed, it is widely known that MEM fails to reconstruct sharp and smooth image features simultaneously.  
Recently, the theory of compressed sensing (CS) has suggested the use of sparse representation and regularisation approaches 
for the recovery of sparse signals from 
incomplete linear measurements \citep{D06,can08,can10}, which has shown great success.
CS techniques based on sparse regularisation were ushered into RI imaging for image reconstruction
{\citep{S09,wia09a,wia09b,WMPBR10,mce11,li11a,li11b,CMW12,car14,wol13,dab15,DWPMW17,gar15,OCRMTPW16,ODW17,PMdCOW16,KCTW17}} 
and have shown promising results and improvements compared to traditional approaches such as CLEAN-based methods and MEM.  
In general, such approaches can recover sharp and smooth image features simultaneously \citep[\textit{e.g.}][]{CMW12}.  {While sparse approaches have been shown to be highly effective, the best approach to image different sources remains an open question.}
Algorithms have been developed to scale sparse approaches to big-data \citep{car14,OCRMTPW16,ODW17,KCTW17,CPraM17}, 
such as that anticipated from the Square Kilometre Array (SKA\footnote{\url{http://www.skatelescope.org/}}).
However, CLEAN-based methods, MEM, and CS-based methods, unfortunately, do not provide any uncertainty quantification 
about the accuracy of recovered images. 

Statistical sampling methods to perform Bayesian inference, like Markov chain Monte Carlo (MCMC) methods, 
which sample the full posterior distribution, have the ability to provide
uncertainty information.  However, this comes at a considerable computational cost.
A proof of concept application of MCMC sampling to RI imaging was performed by \citet{SWMBKKTTZ14}, 
using Gibbs sampling with Gaussian process priors.  Uncertainty information in the form of the posterior image variance was considered.  
However, an idealised telescope model was adopted and the technique has yet to be applied to real observational data.
In general MCMC sampling techniques that scale to high-dimensional settings (like RI imaging), place restrictions on the priors that can be considered.  
Gibbs sampling, for example, requires the ability to draw from conditional distributions.  Two of the most effective classes of MCMC methods for 
high-dimensional settings include Hamiltonian Monte Carlo (HMC) \citep{N12} and the unadjusted Langevin algorithm (ULA) \citep{RT96}.  When a Metropolis-Hasting (MH) accept-reject step is added to ULA, one obtains the Metropolis-adjusted Langevin algorithm (MALA) \citep{RC04}.  HMC, ULA and MALA exploit gradients to capture local properties of the target density in order to explore high-dimensional parameter spaces efficiently.  However, a significant limitation of HMC, MALA and ULA is that the priors considered must be smooth, 
which prohibits their use for priors that promote sparseness.
An alternative Bayesian approach to RI imaging using Information Field Theory \citep{EFK09} has been presented in the form of 
the \mbox{RESOLVE} algorithm \citep{JBSE16,GVJE17}.  This approach assumes a log-normal prior and recovers a 
\textit{maximum a posteriori} (MAP) estimate, proving uncertainty information in the form of an approximate posterior covariance.  
However, the method remains computationally demanding.

Uncertainty quantification is an important missing component in RI imaging for quantitative imaging, scientific inquiry, and decision-making.  
Moreover, since the RI imaging problem is often (severely) ill-posed, uncertainty quantification becomes increasingly important.  
No existing RI imaging techniques that are used in practice provide uncertainty quantification.  
Also, those approaches that do provide some form of uncertainty quantification {in RI imaging} cannot scale to big-data.  Moreover, such approaches 
only support restrictive classes of priors (typically Gaussian or log-normal, which lead to poor reconstruction results relative to sparse priors).  
In summary, no existing approach can support the sparse priors that have been shown in practice to be highly effective for 
RI imaging \citep[\textit{e.g.}][]{PMdCOW16}, while also providing 
uncertainty quantification, in a manner that can scale to big-data.  We present new techniques that fulfil precisely these criteria.

In two companion articles, we present novel RI imaging techniques that support the sparsity-promoting priors that have been shown to 
be highly effective in practice, provide various forms of uncertainty quantification, and that scale to big-data.  In the current article 
we show how to support uncertainty quantification for sparse priors via proximal MCMC methods.  
In the companion article \citep{CPMp217}, we show how to scale uncertainty quantification with sparse priors to big-data.

In this article, two proximal MCMC methods, Moreau-Yosida ULA (MYULA) \citep{DMP16} and proximal MALA (Px-MALA) \citep{M15}, 
are introduced for RI imaging.  These algorithms are direct extensions of ULA and MALA that exploit proximity mappings
Moreau-Yosida envelopes, and Moreau approximations.  
Most importantly, due to the versatility of proximity mappings, these two algorithms are able to sample 
high-dimensional distributions with a variety of different types of priors, including the non-differentiable sparse priors that have been widely 
used in RI imaging but yet cannot be tackled by standard MCMC methods.
Specifically, Px-MALA can sample the posterior distribution with high accuracy (formally, it is guaranteed to converge to the target distribution), 
but the MH accept-reject step embedded in it induces a high computation overhead.  
MYULA, on the other hand, eliminates the MH accept-reject step by introducing well-controlled approximations 
(formally, the bias introduced by such approximations can be made arbitrarily small), and thus has a lower computational overhead.

The uncertainty quantification strategy considered in this article proceeds as follows.
Firstly, using Bayesian inference, two unconstrained inverse models -- analysis and synthesis forms --
with sparse priors are presented to address the RI imaging problem. 
Then, full posterior distributed samples corresponding to these two unconstrained models 
are generated by the sampling methods \mbox{Px-MALA} and MYULA.
After that, three ways of quantifying uncertainty information for RI imaging are constructed, including: 
(i) local (pixel-wise) credible intervals (\textit{cf.} error bars) computed from the generated posterior samples; 
(ii) highest posterior density (HPD) credible regions computed using the generated posterior samples; 
and (iii) hypothesis testing of image structure using the HPD credible regions.
Moreover, comparisons between the performance of Px-MALA and MYULA, and between the analysis and synthesis
models are presented.

The remainder of this article is organised as follows. In Section \ref{sec:ri} we introduce the RI imaging problem, 
the Bayesian inference approach to imaging, and the regularisation approach to imaging, elaborating the relationship 
between various approaches and various algorithms (\textit{e.g.} CLEAN and MEM).
In Section~\ref{sec:mcmc} we discuss Bayesian inference for sparse priors by proximal MCMC methods
and in Section~\ref{sec:imp-mcmc} derive the detailed implementation of the proximal MCMC methods for RI imaging problems.  Uncertainty quantification for RI imaging is formulated in Section~\ref{sec:alg}.
Numerical results evaluating the performance of our uncertainty quantification methods are reported in Section \ref{sec:exp}.
Finally, we conclude in Section \ref{sec:con} with a brief description of the main contributions, a discussion of 
planned extensions of this work, and elucidate connections with the companion article \citep{CPMp217}. 

%-------------------------------------------------------------------
\section{Radio interferometric imaging}\label{sec:ri}
%-------------------------------------------------------------------
To start, we first recall the RI imaging problem and then review sparse representations, which are often exploited in modern approaches
to solve this problem. We model the RI imaging problem from the perspective of Bayesian inference and, finally, 
elaborate the relationship between Bayesian inference and regularisation on which CLEAN, MEM, and CS approaches are based.

%-------------------------------
\subsection{Radio interferometry}
%-------------------------------
The sky intensity can be imaged by radio interferometric telescopes that measure the radio emission of the sky using an array of spatially separated antennas.
When the baselines in an array are co-planar and the field of view is narrow,
the visibility $\vect y$ can be measured by correlating the signals from pairs of antennas,
separated by the baseline components $\vect u = (u, v)$. The general RI equation for obtaining $\vect y$ reads as  \citep{tho08}
\begin{equation} \label{eqn:ri}
\vect y (\vect u) = \int A(\vect l) \vect x(\vect l) {\rm e}^{-2\pi i \vect u \cdot \vect l} {\rm d}^2 \vect l,
\end{equation}
where $\vect x$ represents the sky brightness distribution, described in coordinates $\vect l = (l, m)$ 
(the coordinates of the plane of the sky, centred on the pointing direction of the telescope), 
and $A(\vect l)$ represents the primary beam of the telescope. While not considered further in this article, wide fields and other direction dependent effects can be incorporated 
(see {\it e.g.} \citealt{MS08,CGB08,BCGU08,wia09b,mce11,wol13,OMHB14,DWPMW17}).

In RI imaging, the goal is to recover the sky intensity signal $\vect x$ from the measured visibilities $\vect y$ acquired according to \eqref{eqn:ri}. Precisely, we consider the estimation of a vector $\vect x \in \mathbb{R}^N$ representing a sampled image on a discrete grid of $N$ points in real space, from a measurement vector $\vect y \in \mathbb{C}^M$ gathering the $M$ visibilities observed in a complex vector space, related to $\vect x$ by the linear observation model
\begin{equation}\label{eqn:y}
{\vect y}=\bm{\mathsf{\bm{\mathsf{\Phi}}}} {\vect x} + {\vect n},
\end{equation}
where $\bm{\mathsf{\Phi}} \in \mathbb{C}^{M\times N}$ is a linear measurement operator modelling the realistic acquisition of 
the sky brightness components and ${\vect n} \in \mathbb{C}^{M}$ is the instrumental noise. Without loss of generality, 
we assume independent and identically distributed (i.i.d.) Gaussian noise. The estimation of $\vect x$ is therefore a linear inverse problem, 
which is challenging because the operator $\bm{\mathsf{\Phi}}$ is ill-posed and ill-conditioned, and because of the high dimensionality involved \citep{RBVC09}. 

%-------------------------------
\subsection{Sparse representation}
%-------------------------------
RI imaging methods typically use prior knowledge about $\vect x$ to regularise the estimation problem and deliver more accurate estimation results. In particular, many new methods use the fact that natural signals and images in general, and RI images in particular, often exhibit a sparse representation in some bases
({\it e.g.} a point source basis or a multi-scale basis such as wavelets). Let 
\begin{equation}\label{eqn:x}
{\vect x} = \bm{\mathsf{\Psi}} {\vect a} = \sum_{i} \bm{\mathsf{\Psi}}_i a_i,
\end{equation}
where $\bm{\mathsf{\Psi}} \in \mathbb{C}^{N\times L}$ is a dictionary ({\it e.g.} 
a wavelet basis or an overcomplete frame) and ${\vect a} = (a_1, \cdots, a_L)^\top$ is the vector of the synthesis coefficients
of ${\vect x}$ under $\bm{\mathsf{\Psi}}$. Then ${\vect x}$ is said to be sparse if ${\vect a}$ contains only $K$ non-zero coefficients, 
{\it i.e.}, $\|\vect a\|_0 = K$ (recall $\|\vect a\|_0$ gives the number of non-zero components of $\vect a$), where $K\ll N$. 
Similarly, $\vect x$ is called compressible under $\bm{\mathsf{\Psi}}$ if many coefficients of $\vect a$ are nearly zero, {\it i.e.}, 
its sorted coefficients $a_i$ satisfy a power law decay.
In practice, it is ubiquitous that natural signals and images $\vect x$ are sparse or compressible. 

%-------------------------------
\subsection{Bayesian inference}
\label{sec:ri:bayesian}
%-------------------------------
The inverse problem presented in \eqref{eqn:y} can be addressed elegantly in the Bayesian statistical inference framework, which in addition to allowing one to derive estimates of $\vect x$ also provides tools to analyse and quantify the uncertainty in the solutions obtained. Let $p(\vect y | \vect x)$ be the likelihood function of the statistical model associated with \eqref{eqn:y}. In the case of i.i.d.\ Gaussian noise the likelihood function reads
\begin{equation} \label{eqn:baye-like}
p(\vect y|\vect x) \propto {\rm exp}(-\|\vect y - \bm{\mathsf{\Phi}} \vect x \|_2^2 /2\sigma^2),
\end{equation}
where $\sigma$ represents the standard deviation of the noise level.

As mentioned previously, recovering $\vect x$ solely from $\vect y$ is not possible because the problem is not well posed. Bayesian methods address this difficulty by exploiting prior knowledge -- represented by a prior distribution $p(\vect x)$ -- to regularise the problem, reduce uncertainty, and improve estimation results.  Typically priors of the form $p(\vect x) \propto {\rm exp}\left(-\phi({\cal B} \vect x) \right)$ are considered,
for some linear operator ${\cal B}$ and potential function $\phi$. Various forms for $\phi$ can be considered, for example:
Tikhonov regularisation \citep{GHO99,CCZ13}, used to promote smoothness, 
corresponds to the Gaussian prior of $p(\vect x) \propto {\rm exp}(-\mu \|\vect x\|_2^2)$; 
the entropic prior of $p(\vect x) \propto {\rm exp}(-\mu \vect x^{\dagger} {\rm log} \vect x)$ \citep{A74,GD78,CE85};
and the $\ell_p$ norm with $0\le p \le 1$ used as a regulariser to promote sparseness 
\citep{CSS16,CFNSS15,wia09a,wia09b,mce11,D06,can08}.  Here $\mu > 0$ is a regularisation parameter.  We refer to such priors as \emph{analysis} priors because they operate on the canonical coordinate system of ${\vect x}$. Alternatively, it is also possible to adopt a so-called \emph{synthesis} approach and use \eqref{eqn:x} to express the prior knowledge for $\vect x$ via a prior distribution $p(\vect a)$ on the synthesis coefficients $\vect a$. 

In this article we consider both analysis and synthesis formulations because they are both widely used in RI imaging. For analysis models we consider Laplace-type priors of the form
\begin{equation} \label{eqn:baye-like-x}
p(\vect x) \propto {\rm exp}(-\mu \|\mathsf{\Psi}^\dagger \vect x\|_1),
\end{equation}
where $\mathsf{\Psi}^\dagger$ denotes the adjoint of $\mathsf{\Psi}$, $\mu > 0$ is a regularisation parameter, and $\|\cdot\|_1$ is the $\ell_1$ norm; 
while for synthesis models we consider the Laplace prior
\begin{equation} \label{eqn:baye-like-a}
p(\vect a) \propto {\rm exp}(-\mu \|\vect a\|_1).
\end{equation}
Observe that both formulations are equivalent when $\bm{\mathsf{\Psi}}$ is an orthogonal basis. However, for redundant dictionaries the approaches have very different
properties. Further discussions about the analysis and synthesis forms can be found, for example, in \citet{MHL04}, \citet{EMR07} and \citet{CJP12}.

Prior and observed information can then be combined by using Bayes' theorem to obtain the posterior distribution. For analysis formulations the posterior is given by
\begin{equation}\label{eqn:baye-x}
p(\vect x | \vect y) = \frac{p(\vect y | \vect x) p(\vect x)}{p_{\rm a}({\vect y})},
\end{equation}
which models our knowledge about $\vect x$ after observing $\vect y$, where $p_{\rm a}({\vect y}) =  \int_{\mathbb{R}^N}p(\vect y | \vect x) p(\vect x) {\rm d} \vect x$ is the marginal likelihood (or Bayesian evidence) of the analysis model. Similarly, for synthesis models the posterior reads
\begin{equation}\label{eqn:baye-a}
p(\vect a | \vect y) = \frac{p(\vect y | \vect a) p(\vect a)}{p_{\rm s}({\vect y})},
\end{equation}
with $p(\vect y | \vect a) = p(\vect y | \vect x)$ for $\vect x = \bm{\mathsf{\Psi}} {\vect a}$, where $p_{\rm s}({\vect y}) =  \int_{\mathbb{R}^N}p(\vect y | \vect a) p(\vect a) {\rm d} \vect a$ is the model's marginal likelihood.

Note that the denominators $p_{\rm a}({\vect y})$ in \eqref{eqn:baye-x} and $p_{\rm s}({\vect y})$ in \eqref{eqn:baye-a}, {\it i.e.} the marginal likelihoods, 
are unrelated to $\vect x$ and $\vect a$, respectively, and therefore constants with respect to (w.r.t.) parameter inference. 
It follows that the unnormalised posterior distributions for the analysis and synthesis formulations read
\begin{equation}\label{eqn:baye-post-x}
p(\vect x | \vect y) \propto {\rm exp} \Big\{ - \big(\mu \|\bm{\mathsf{\Psi}}^\dagger {\vect x}\|_1 + \|{\vect y}-\bm{\mathsf{\Phi}} {\vect x}\|_2^2/2\sigma^2 \big) \Big\}
\end{equation}
and 
\begin{equation}\label{eqn:baye-post-a}
p(\vect a | \vect y) \propto {\rm exp} \Big\{ - \big(\mu \|{\vect a}\|_1 + \|{\vect y}-\bm{\mathsf{\Phi}}\bm{\mathsf{\Psi}} {\vect a}\|_2^2/2\sigma^2 \big) \Big\},
\end{equation}
respectively, where the first terms ({\it i.e.} the $\ell_1$ norm terms) in the exponentials of each equation correspond to the prior and the second ({\it i.e.} the $\ell_2$ norm terms) correspond to the likelihood.

Drawing conclusions directly from $p(\vect x | \vect y)$ or $p(\vect a | \vect y)$ can be difficult because of the high dimensionality involved. Instead, Bayesian methods often derive solutions by computing estimators that summarise $p(\vect x | \vect y)$ or $p(\vect a | \vect y)$. In particular, it is often common practice to compute maximum-a-posteriori (MAP) estimators given by
\begin{equation}\label{eqn:ir-un-af}
\begin{split}
\hat{\vect x}_{\rm map} &= \argmax_{\vect x} p(\vect x|\vect y) \\
&=\argmin_{\vect x} \Big\{\mu \|\bm{\mathsf{\Psi}}^\dagger {\vect x}\|_1 + \|{\vect y}-\bm{\mathsf{\Phi}} {\vect x}\|_2^2/2\sigma^2 \Big\},
\end{split}
\end{equation}
for the analysis model, and 
\begin{equation}\label{eqn:ir-un-sf}
\begin{split}
  \hat{\vect a}_{\rm map} &= \argmax_{\vect a} p(\vect a|\vect y) \\
&= \argmin_{{\vect a}} \Big\{ \mu \|{\vect a}\|_1 + \|{\vect y}-\bm{\mathsf{\Phi}}\bm{\mathsf{\Psi}} {\vect a}\|_2^2/2\sigma^2 \Big\},\\
 \end{split}
\end{equation}
which is then mapped to canonical coordinates by using \eqref{eqn:x}, for the synthesis model. 
A main computational advantage of the MAP estimators \eqref{eqn:ir-un-af} and \eqref{eqn:ir-un-sf} is that they can 
be formulated as a convex optimisation problem that can be solved very efficiently, even in high dimensions, 
by using modern convex optimisation techniques \citep{Green2015}. Also, there is abundant empirical evidence 
that these estimators deliver accurate reconstruction results, and that they promote solutions that 
are sparse under $\bm{\mathsf{\Psi}}$ in agreement with our prior knowledge about $\vect x$. 
See \cite{Pereyra:2016b} for a theoretical analysis of MAP estimation.

{The regularisation parameter $\mu$ appearing in the analysis and synthesis formulations  controls the balance between the likelihood and the prior information, and plays an important role in terms of image reconstruction quality.  Typically, setting $\mu$ is performed by visual cross-validation.} 
{However, there exist more advanced Bayesian strategies to address the problem of unknown $\mu$.  For example, hierarchical Bayesian strategies allow estimating $\mu$ jointly with $\vect{x}$ (or $\vect{\alpha}$) from $\vect{y}$, or removing $\mu$ from the model by marginalisation followed by inference with the marginal model (see \citealt{MBF15} for details). Alternatively, empirical Bayesian approaches set regularisation parameters by marginal maximum likelihood estimation \citep{JBSE16, AF2018} or by MCMC sampling \citep{SWMBKKTTZ14}. The selection of a regularisation parameter was also studied by \cite{SG91} in the context of maximum entropy methods, where the marginal distribution of the regularisation parameter is again maximised.}

To compute other Bayesian estimators or quantifies of interest {beyond MAP estimators} it is typically necessary to use more advanced 
Bayesian computation tools, such as MCMC sampling methods. These methods compute probabilities 
and expectations w.r.t.\ $p(\vect x | \vect y)$ or $p(\vect a | \vect y)$ and can be used to 
calculate moments and Bayesian confidence regions useful for uncertainty quantification. 
This is the main purpose of this article and thus will be detailed subsequently.

\subsection{Connections with alternative approaches}

It is worth noticing that many RI imaging techniques can be seen as regularisation techniques
and many of them can be viewed as MAP estimation for appropriate priors. 
While this interpretation is not always precise, the resulting approximate unifying Bayesian framework is useful to aid intuition.

%-------------------------------
\subsubsection{Compressive sensing and $\ell_1$-regularised regression}
%-------------------------------
The theory of CS (compressive sensing) led to an important breakthrough in the recovery of sparse signals from 
incomplete linear measurements \citep{D06,can08,can10}. CS goes beyond the traditional Nyquist sampling paradigm, 
where its acquisition approaches can save a huge amount of time and memory thanks to the fact that natural signals 
often exhibit a sparse representation in multi-scale bases.
CS can be implemented for signal reconstruction by regularising 
the resulting ill-posed inverse problem through a sparsity-promoting prior, resulting in a convex optimisation problem that can be solved by leveraging techniques from the field of convex optimisation.
Briefly speaking, the theoretical framework of CS motivates sparse regularisation approaches such as the ones used 
in \eqref{eqn:ir-un-af} and \eqref{eqn:ir-un-sf}. In fact, the MAP estimators \eqref{eqn:ir-un-af} and \eqref{eqn:ir-un-sf} 
are equivalent to the $\ell_1$ regularised least-squares estimators used extensively in CS. 
In the literature and henceforth, the discussion of CS-based methods for RI imaging typically refers to sparse regularisation approaches, 
even though RI imaging models such as \eqref{eqn:ir-un-af} and \eqref{eqn:ir-un-sf} may not satisfy the idealised CS setting.

%-------------------------------
\subsubsection{CLEAN}
%-------------------------------
CLEAN, the most well-known and standard RI image reconstruction algorithm, is a non-linear deconvolution method based on local iterative beam removal. 
In general, it can be operated iteratively in two steps, {\it i.e.} major and minor cycles.
Let $\chi^2 = \|\vect y - \bm{\mathsf{\Phi}} \vect x\|^2_2$ and denote the gradient of $\chi^2$ at iteration $t$
by $\vect r^{(t)} = \bm{\mathsf{\Phi}}^{\dagger}(\vect y - \bm{\mathsf{\Phi}} \vect x^{(t)})$.
The major cycle of CLEAN computes the residual image $\vect r^{(t)}$, 
followed by the minor cycle of deconvolving the brightest sources in $\vect r^{(t)}$, 
represented by ${\cal T} (\vect r^{(t)})$, yielding the iterative form 
 \begin{equation}
 \vect x^{(t+1)} =  \vect x^{(t)} + {\cal T} (\vect r^{(t)})
 \end{equation}
to reconstruct an image $\vect x$. 
  
Extensions of CLEAN have also been considered to achieve better reconstruction. 
For example, multi-scale versions of CLEAN: MS-CLEAN (\citealt{cor08});
and ASP-CLEAN (\citealt{BC04}).
For further variants of CLEAN, please refer to \cite{RBVC09} and references therein.

CLEAN implicitly involves a sparse prior on the original signal in real space.  Moreover, a close connection has been shown between CLEAN and the well-known 
Matching Pursuit algorithm in the CS literature \citep{cor88,wia09a,RBVC09}; in other words, CLEAN is essentially $\ell_0$ regularisation with a point source basis. 
The performance of CLEAN, however, is empirically found to be similar to $\ell_1$ regularisation with a point source basis (\citealt{wia09a}). As a proxy for CLEAN, $\ell_1$ regularisation with a point source basis is equivalent to MAP estimation involving a Laplace prior.

%-------------------------------
\subsubsection{Maximum entropy method (MEM)}
%-------------------------------
Another important method for RI imaging is MEM, which is, 
mildly speaking, a special case of the MAP method.
The MEM approach for RI imaging \citep{CE85} differs to the original MEM formulation \citep{A74,GD78}, in that not only does the regularisation problem considered consist of a relative entropic prior
and a (Gaussian) likelihood, but an an additional flux constraint is also incorporated.
In particular, an entropic prior, ${\rm exp}(-\mu \vect x^{\dagger} {\rm log} \vect x)$, on the image is adopted.

%-------------------------------
\subsubsection{Constrained regularisation}
%-------------------------------

In addition to the unconstrained optimisation problems of \eqref{eqn:ir-un-af} and \eqref{eqn:ir-un-sf}, many CS-based approaches consider constrained forms of the analysis and synthesis models, which are, respectively, given by
\begin{equation} \label{eqn:ir-af}
\min_{\vect x} \|\bm{\mathsf{\Psi}}^\dagger {\vect x}\|_1, \quad {\rm s.t.} \ \ \|{\vect y}-\bm{\mathsf{\Phi}} {\vect x}\|_2^2 \le \epsilon
\end{equation}
and 
\begin{equation}\label{eqn:ir-sf}
\min_{\vect a} \|{\vect a}\|_1, \quad {\rm s.t.} \ \ \|{\vect y}-\bm{\mathsf{\Phi}}\bm{\mathsf{\Psi}} {\vect a}\|_2^2 \le \epsilon,
\end{equation}
where $\epsilon$ is an upper-bound related to the noise level present in $\vect y$.
CS approaches based on constrained optimisation problems, solved via convex optimisation techniques, have been applied broadly in RI imaging 
\citep{wia09a,wia09b,mce11,li11a,li11b,CMW12,car14,OCRMTPW16,PMdCOW16}.
These techniques have shown promising results, with improvements in terms of image fidelity and flexibility compared to traditional approaches 
such as CLEAN-based methods and MEM.  For these constrained regularisation approaches, parallel implementation structures have also been explored \citep{car14,OCRMTPW16}.
Compared with the unconstrained analysis and synthesis models, constrained approaches are 
parameterised by $\epsilon$ (related to noise level) which controls the error of the reconstruction explicitly; in contrast, unconstrained models 
use regularisation parameter $\mu$ to impose a tradeoff between the prior and data fidelity.  {The constrained approach therefore avoids the problem of unknown regularisation parameter $\mu$, replacing it with the problem of estimating the noise bound $\epsilon$.  The latter can be performed in a principled manner by noting that for Gaussian noise the $\ell_2$ norm data fidelity term follows a $\chi^2$ distribution with $2M$ degrees of freedom (see, \textit{e.g.}, \citealt{CMW12}).} While constrained problems do not afford a straightforward Bayesian interpretation, the constrained and 
unconstrained models are closely related \citep{N16}.

%-------------------------------------------------------------------
\section{Bayesian inference with sparse priors by proximal MCMC sampling}\label{sec:mcmc}
%-------------------------------------------------------------------
Sparse regularisation, motivated by CS, has been shown to be a powerful framework for solving inverse problems
and has been used to deal with the recovery of sparse signals from incomplete linear measurements ({\it e.g.}, \citealt{D06}). 
It has been demonstrated that sparse signals can be recovered accurately from incomplete data under some conditions.
Sparse priors have also been ushered into RI imaging for image reconstruction ({\it e.g.} \citealt{wia09a,mce11}), 
and have shown promising results on real RI data \citep{PMdCOW16}.
Unfortunately, CS-based techniques do not provide any uncertainty information regarding their point estimates. This is also a limitation of CLEAN-based methods and MEM. 

From an inferential viewpoint, the lack of uncertainty quantification is problematic, particularly because RI problems are ill-posed and hence solutions have significant intrinsic uncertainty. As explained previously, in this article we apply recent developments in Bayesian methodologies to analyse uncertainty in RI imaging. Precisely, we use new MCMC Bayesian computation algorithms to compute probabilities and expectations  w.r.t.\ the posterior distribution of interest, \emph{i.e.}, $p(\vect x| \vect y)$ or $p(\vect a | \vect y)$ given by \eqref{eqn:baye-x} and \eqref{eqn:baye-a}, depending on whether an analysis or a synthesis formulation is used. This involves constructing a Markov chain that generates samples from the distribution of interest, and then using the samples to approximate probabilities and expectations by Monte Carlo integration \citep{RC04}. Computing such Markov chains in large-scale settings is computationally challenging, and we address this difficulty by using state-of-the-art MCMC methods tailored for these types of problems \citep{M15, DMP16}. In this section we introduce these MCMC algorithms. To ease presentation, all symbols and dimensions specified here corresponds to the analysis model \eqref{eqn:ir-un-af}, however these can be straightforwardly adapted  to the synthesis model \eqref{eqn:ir-un-sf}.

%-------------------------------
\subsection{Preliminaries} \label{sec:mcmc-pre}
%-------------------------------
A function $g : \mathbb{C}^{N} \rightarrow (-\infty, \infty]$ is said to be lower semicontinuous (l.s.c.) if for all $M \in \mathbb{R}$, $\{g < M\}$ is a closed subset of $\mathbb{C}^{N}$. Let $\mathcal{C}^1(\mathbb{C}^{N})$ be the class of continuously differentiable functions on $\mathbb{C}^{N}$. If $g \in \mathcal{C}^1 (\mathbb{C}^{N})$, denote by $\nabla g$ the gradient of $g$. Also, $\nabla g$ is said to be Lipchitz continuous with constant $\beta_{\rm Lip} \in (0, \infty)$ if 
\begin{equation}
\|\nabla g(\hat{\vect z}) - \nabla g (\bar{\vect z})\| \le \beta_{\rm Lip} \|\hat{\vect z} - \bar{\vect z} \|, 
	\quad \forall (\hat{\vect z}, \bar{\vect z}) \in \mathbb{C}^{N} \times \mathbb{C}^{N}.
\end{equation}

Moreover, let $h : \mathbb{C}^{N} \rightarrow (-\infty, \infty]$ be a convex l.s.c. function and $\lambda > 0$. The $\lambda$-Moreau-Yosida envelope of $h$ is a carefully regularised approximation of $h$ given by
\begin{equation} 
h^{\lambda} ({\vect z})  \equiv \min_{{\vect u}\in \mathbb{R}^N}  \left \{ h({\vect u}) + \|{\vect u} - {\vect z}\|^2/2\lambda \right \}.
\end{equation}
The approximation $h^{\lambda}$ can be made arbitrarily close to $h$ by adjusting $\lambda$, \emph{i.e.}, $\underset{\lambda\rightarrow 0}{\lim} h^{\lambda} ({\vect z}) = h({\vect z})$ (see \citealt{PB14}). Also, by construction $h^{\lambda}  \in \mathcal{C}^1$, with $\lambda$-Lipchitz gradient given by
\begin{equation} 
\nabla h^{\lambda} ({\vect z}) = \left({\vect z} - {\rm prox}_h^{\lambda} ({\vect z}) \right)/\lambda,
\end{equation}
where ${\rm prox}_h^{\lambda} ({\vect z})$ is the {\it proximity operator} of $h$ at $\vect z$ defined as
\begin{equation} \label{eqn:prox-ope}
{\rm prox}_h^{\lambda} ({\vect z})  \equiv \argmin_{{\vect u}\in \mathbb{R}^N}  \left \{ h({\vect u}) + \|{\vect u} - {\vect z}\|^2/2\lambda \right \}.
\end{equation}
It can be verified easily that 
${\rm prox}_h^{\lambda} ({\vect z})  = {\rm prox}_{\lambda h} ({\vect z}).$
For simplicity, we represent ${\rm prox}_h^{1} ({\vect z})$ by ${\rm prox}_h({\vect z})$.
This operator generalises the projection operator defined as
\begin{equation} 
{\cal P}_{C} ({\vect z})  \equiv \argmin_{{\vect u}\in \mathbb{R}^N}  \left \{ \iota_{C}({\vect u}) + \|{\vect u} - {\vect z}\|^2/2 \right \},
\end{equation}
where $ \iota_{C}$ is the characteristic function for the convex set ${C}$ defined by $\iota_{C}({\vect u}) = \infty$ if $\vect u \notin {C}$ and 0 otherwise. 

%-------------------------------
\subsection{Langevin MCMC}
%-------------------------------
Let $\pi$ be a probability density (or a user-specified target density), such as the posteriors $p(\vect x | \vect y)$ or $p(\vect a | \vect y)$.
When $\pi$ is defined on $\mathbb{C}^n$ and assume $\pi \in \mathcal{C}^1$ with Lipchitz gradient,
the Langevin diffusion on $\mathbb{C}^n$ associated with $\pi$ is a stochastic process defined as
\begin{equation} \label{eqn:ldp}
{\rm d}{\cal L}(t) = \frac{1}{2} \nabla \log \pi[{\cal L}(t)] {\rm d}t + {\rm d} {\cal W}(t)\, , 
\end{equation}
where ${\cal W}$ is the Brownian motion on $\mathbb{C}^n$. This process converges to $\pi$ as $t$ increases, and is therefore useful for generating samples from $\pi$. Unfortunately, simulating ${\cal L}(t)$ in continuous time is generally not possible, so instead we use discrete-time approximations. In particular, ULA (unadjusted Langevin algorithm) is based on a forward Euler-Maruyama approximation with step-size $\delta > 0$, resulting in the Markov chain
\begin{equation} \label{eqn:ldp-d}
{\vect l}^{(m+1)} = {\vect l}^{(m)} + \frac{\delta}{2} \nabla \log  {\pi}[{\vect l}^{(m)}] + \sqrt{\delta} {\vect w}^{(m+1)},   
\end{equation}
where ${\vect w}^{(m+1)} \sim {\cal N} (0,\mathbb{1}_N)$ (an $N$-sequence of standard Gaussian random variables).
Under appropriate regularity conditions, the chain generated by ULA converges to an ergodic measure which is close to ${\pi}$.  In MALA (Metropolis-adjusted Langevin Algorithm), this approximation error is corrected by complementing ULA with an MH (Metropolis-Hasting) accept-reject step targeting ${\pi}$, which removes the asymptotic bias due to the discretisation at the expense of some additional estimation variance \citep{RT96}. Theoretical and empirical results show that ULA and MALA scale very efficiently to high dimensions. 

However, a main limitation of ULA and MALA (and generally MCMC methods based on gradients) is the requirement that $\log{\pi}$ is continuously differentiable with Lipchitz gradient, otherwise the Markov chain \eqref{eqn:ldp-d} fails to converge. As explained previously, this prohibits their application to image processing models with non-smooth densities, {\it e.g}, involving the term $\phi(\cdot) = \|\cdot\|_1$. In \cite{M15}, this limitation of ULA and MALA is addressed by using the Moreau-Yosida envelope of $\log{\pi}$ to regularise the diffusion process to handle non-smoothness, {\it e.g.} sparse priors.

%-------------------------------
\subsection{Moreau-Yosida regularised ULA (MYULA)}
%-------------------------------
We consider models of the form $\pi(\vect x) \propto \exp{\{-f(\vect x) -g(\vect x)\}}$, where $f \notin \mathcal{C}^1$ is l.s.c. convex 
with operator ${\rm prox}_f^{\lambda}(\vect z)$ tractable $\forall \vect z \in \mathbb{C}^N$, and $g \in \mathcal{C}^1$ is l.s.c. convex 
with $\nabla g$ and $\beta_{\rm Lip}$-Lipchitz continuous.  Typically $f$ corresponds to the log-prior and $g$ to the log-likelihood.

We wish to use the Langevin diffusion \eqref{eqn:ldp} to generate samples from $\pi$ but this is not directly possible since $f$ is not smooth, {\it i.e.} $f \notin \mathcal{C}^1$. The key idea underpinning proximal ULA and MALA is to carefully regularise $f$ to guarantee that \eqref{eqn:ldp} and its discrete-time approximation \eqref{eqn:ldp-d} have good convergence properties \citep{M15}. This is achieved by defining an approximation
\begin{equation}
\pi_\lambda (\vect x) = \frac{\exp{\{- f^\lambda(\vect x) -g(\vect x)\}}}{\int  \exp{\{-f^\lambda(\vect x) -g(\vect x) \}} \textrm{d} \vect x}\, ,
\end{equation}
where the non-smooth term $f$ is replaced by its Moreau-Yosida envelope $f^\lambda$. Since $\nabla \log \pi_\lambda = - \nabla f^\lambda -\nabla g $ is Lipchitz continuous, the Langevin diffusion associated with $\pi_\lambda$ is well posed and leads to a Markov chain \eqref{eqn:ldp-d} with good convergence properties. Precisely, the MYULA chain is defined by
\begin{equation}\label{eqn:myula-ite}
\begin{split}
{\vect l}^{(m+1)} = & \ \left (1 - \frac{\delta}{\lambda}\right ) {\vect l}^{(m)} + \frac{\delta}{\lambda}  {\rm prox}_{f}^{\lambda} ({\vect l}^{(m)})   - \delta \nabla g({\vect l}^{(m)})\\
			    &  + \sqrt{2\delta} {\vect w}^{(m)},
\end{split}	 
\end{equation}
where we have noted that $\nabla f^{\lambda} ({\vect z}) = \left({\vect z} - {\rm prox}_f^{\lambda} ({\vect z}) \right)/\lambda$.

The MYULA chain \eqref{eqn:myula-ite} scales well in high dimensions and efficiently delivers samples that are approximately distributed according to $\pi$. The approximation error involved can be made arbitrarily small by reducing the value of $\lambda$ and by increasing the number of iterations \citep{DMP16}.

Finally, in our experiments we implement \eqref{eqn:myula-ite} with $f(\vect x) = \mu \|\bm{\mathsf{\Psi}}^\dagger {\vect x}\|_1$, 
$g(\vect x) = \|{\vect y}-\bm{\mathsf{\Phi}} {\vect x}\|_2^2/2\sigma^2$ for the analysis model \eqref{eqn:ir-un-af} 
(the setting for the synthesis model \eqref{eqn:ir-un-sf} is analogous), 
and by setting $\lambda = 2/\beta_{\rm Lip}$ and $\delta \in [1/5\beta_{\rm Lip}, 1/2\beta_{\rm Lip}]$, as suggested by \cite{DMP16}. 

%-------------------------------
\subsection{Proximal MALA (Px-MALA)}
%-------------------------------
In a manner akin to MALA, the Px-MALA combines MYULA with an MH step targeting the desired density $\pi$ which is not differentiable \citep{M15}. 
At each iteration of the algorithm a new candidate ${\vect l}^*$ is generated by using one MYULA iteration as proposal mechanism. The candidate is then accepted with probability
\begin{equation} \label{eqn:mh}
\rho = \min \left\{ 1, \frac{q({\vect l}^{(m)} | {\vect l}^*) \pi({\vect l}^*)} {q({\vect l}^* | {\vect l}^{(m)}) \pi({\vect l}^{(m)})}  \right\},
\end{equation}
where $q(\cdot | \cdot)$ is the MYULA transition kernel defined by \citep{PSC16} 
\begin{equation} \label{eqn:mh-ker}
q({\vect l}^* | {\vect l}^{(m)}) \sim {\rm exp} \Bigg (- \frac{\Big({\vect l}^* - {\vect l}^{(m)} - \frac{\delta}{2} \nabla \log \pi({\vect l}^{(m)}) \Big)^2} {2\delta}  \Bigg ).
\end{equation}
Regarding computational efficiency, for the models considered here Px-MALA inherits the good convergence properties of MYULA and scales efficiently in high dimensions. However, note that the MH correction removes the asymptotic estimation bias at the expense of increasing the correlation of the Markov chain and hence the estimation variance (this is observed clearly in the experiments reported in Section \ref{sec:exp}). Also note that Px-MALA iterations are more expensive than MYULA iterations because of the computational overhead associated with the MH step.

Finally, in our experiments, following the setting in \cite{M15}, we implement Px-MALA with 
\mbox{$f(\vect x) = \|{\vect y}-\bm{\mathsf{\Phi}} {\vect x}\|_2^2/2\sigma^2 + \mu \|\bm{\mathsf{\Psi}}^\dagger {\vect x}\|_1, g(\vect x) = 0$ }
for the analysis model \eqref{eqn:ir-un-af} (the setting for the synthesis model \eqref{eqn:ir-un-sf} is analogous), 
and by setting $\lambda = 2/\beta_{\rm Lip}$ and adjusting $\delta$ for an acceptance probability of approximately $0.5$. Other settings w.r.t.\ the definitions of $f$ and $g$, {\it e.g.} as used in MYULA, could also be considered. Also note that the efficient computation of ${\rm prox}_{f}^{\lambda}$ often involves some approximations, which we also correct with the MH step. We discuss such approximations for the analysis and synthesis models in Section \ref{sec:imp-mcmc}.

%-------------------------------------------------------------------
\section{Proximal MCMC methods for RI imaging}\label{sec:imp-mcmc}
%-------------------------------------------------------------------

This section presents the implementation details of MYULA and Px-MALA for the analysis model \eqref{eqn:ir-un-af} and the synthesis 
model \eqref{eqn:ir-un-sf}. We first consider the computation of the  proximity operator of $f$, for different forms of $f$. Computing the proximity operator of $f$ requires solving an optimisation problem, which must be performed efficiently since it needs to be computed to generate each sample by \eqref{eqn:myula-ite}.  We then summarise the sampling procedures for the two proximal MCMC methods. 
Note that computing the gradient of $g$ in \eqref{eqn:myula-ite} is straightforward since it is differentiable. 
For clarity, we henceforth use the label \ $\bar{}$ \ for symbols related to the analysis model, and \ $\hat{}$  \
for symbols related to the synthesis model. Although not essential, we also assume  
$\bm{\mathsf{\Psi}}^\dagger \bm{\mathsf{\Psi}} = \bm{\mathsf{ I}}$ (where $\bm{\mathsf{ I}}$ 
is the identity matrix), unless otherwise stated.

%-------------------------------
\subsection{Computing proximity operators}
%-------------------------------
%%%%%%%
\begin{algorithm} 
\caption{Sample generation by MYULA}
\label{alg:ee-myula-sam}
  \textbf{Input:} visibility ${\vect y} \in \mathbb{C}^M$, ${\vect x}^{(0)} \in \mathbb{R}^N$, ${\vect a}^{(0)} \in \mathbb{C}^L$, $K$, 
 	$K_{\rm gap}$, $K_{\rm burn}$, $P_{\rm type} \in \{{\tt analysis}, {\tt synthesis}\}$, and $m=0, j=1$ \\
 \textbf{Output:} $K$ samples $\{{\vect x}^{(j)}\}_{j=1}^K$ or $\{\bm{\mathsf{\Psi}}{\vect a}^{(j)}\}_{j=1}^K$  \vspace{0.05in} \\ 
\Do{$j \le K$}{
	{\bf if} $P_{\rm type} == {\tt analysis}$ \\
		\quad compute  ${\vect x}^{(m+1)} $ \\
		\hspace{0.4in} $= {\vect x}^{(m)} - \delta \bm{\mathsf{\Phi}}^\dagger (\bm{\mathsf{\Phi}} {\vect x}^{(m)} 
						- {\vect y})/2\sigma^2 + \sqrt{\delta} {\bar{\vect w}}^{(m)} $ \\
		\hspace{0.4in} $ + \frac{\delta}{\lambda} \bm{\mathsf{\Psi}} \left ( {\rm soft}_{\lambda \mu/2}(\bm{\mathsf{\Psi}}^\dagger 
						{\vect x}^{(m)}) - \bm{\mathsf{\Psi}}^\dagger {\vect x}^{(m)}) \right )$ \\
		\quad set ${\vect z} = {\vect x}^{(m+1)}$ \\
	{\bf elseif} $P_{\rm type} == {\tt synthesis}$  \\
		\quad compute ${\vect a}^{(m+1)} $ \\
		\hspace{0.4in} $ = (1 - \frac{\delta}{\lambda}) {\vect a}^{(m)} + \frac{\delta}{\lambda}  {\rm soft}_{\lambda\mu/2} ({\vect a}^{(m)})$  \\
		\hspace{0.4in} $- \delta \bm{\mathsf{\Psi}}^\dagger\bm{\mathsf{\Phi}}^\dagger (\bm{\mathsf{\Phi}}\bm{\mathsf{\Psi}} {\vect a} - {\vect y})/2\sigma^2 
			    + \sqrt{\delta} {\hat{\vect w}}^{(m)}$ \\
			    
		\quad set ${\vect z} = {\vect a}^{(m+1)}$ \\
	{\bf endif}  \\
	{\bf if} $m$ satisfies \eqref{eqn:mh-con} \\
		\quad {\bf if} $P_{\rm type} == {\tt analysis}$ \\
			\quad \quad set ${\vect x}^{(j)} = {\vect z}$ \\
		\quad {\bf elseif} $P_{\rm type} == {\tt synthesis}$  \\
			\quad \quad set ${\vect a}^{(j)} = {\vect z}$ \\
		\quad {\bf endif}  \\
		\quad $j=j+1$  \\
	{\bf endif}  \\
        $m=m+1$ 
}
\end{algorithm}

\begin{algorithm} 
\caption{Sample generation by Px-MALA}
\label{alg:ee-pmala-sam}
 \textbf{Input:} visibility ${\vect y} \in \mathbb{C}^M$, ${\vect x}^{(0)} \in \mathbb{R}^N$, ${\vect a}^{(0)} \in \mathbb{C}^L$, $K$, 
 	$K_{\rm gap}$, $K_{\rm burn}$, $P_{\rm type} \in \{{\tt analysis}, {\tt synthesis}\}$, and $m=0, j=1$ \\
 \textbf{Output:} $K$ samples $\{{\vect x}^{(j)}\}_{j=1}^K$ or $\{\bm{\mathsf{\Psi}}{\vect a}^{(j)}\}_{j=1}^K$  \vspace{0.05in} \\ 
\Do{$j \le K$}{
	{\bf if} $P_{\rm type} == {\tt analysis}$ \\
		\quad compute ${\vect x}^{(m+1)} = {\rm prox}_{\bar f}^{\delta/2} ({\vect x}^{(m)}) + \sqrt{\delta} \bar{\vect w}^{(m)}$ \\
		\quad set ${\vect z} = {\vect x}^{(m+1)}$, ${\vect z}^\prime = {\vect x}^{(j-1)}$ \\
	{\bf elseif} $P_{\rm type} == {\tt synthesis}$  \\
		\quad compute ${\vect a}^{(m+1)} = {\rm prox}_{\hat f}^{\delta/2} ({\vect a}^{(m)}) + \sqrt{\delta} \hat{\vect w}^{(m)}$ \\
		\quad set ${\vect z} = {\vect a}^{(m+1)}$, ${\vect z}^\prime = {\vect a}^{(j-1)}$ \\
	{\bf endif}  \\
	{\bf if} $m$ satisfies \eqref{eqn:mh-con} \\
		\quad {\bf if} ${\tt MH}\big({\vect z}, {\vect z}^\prime \big) == 1$   \quad // Metropolis-Hasting step   \\
			\quad \quad {\bf if} $P_{\rm type} == {\tt analysis}$ \\
				\quad \quad \quad set ${\vect x}^{(j)} = {\vect z}$ \\
			\quad \quad {\bf elseif} $P_{\rm type} == {\tt synthesis}$  \\
				\quad \quad \quad set ${\vect a}^{(j)} = {\vect z}$ \\
			\quad \quad {\bf endif}  \\
		\quad\quad $j=j+1$  \\
		\quad {\bf endif} \\
	{\bf endif}  \\
        $m=m+1$ 
}  \vspace{0.1in}

{\bf function} ${\tt MH}\big({\vect l}^*, {\vect l} \big)$  \\
\quad Compute the acceptance probability \\
\qquad $\rho = \min \left\{ 1, \frac{q({\vect l} | {\vect l}^*) \pi({\vect l}^*)} {q({\vect l}^* | {\vect l}) \pi({\vect l})}  \right\}$ \\
\quad Generate a threshold  $u \sim {\cal U}(0, 1)$ \\
\quad {\bf if} $u \le \rho$ \\
\qquad {\bf return} 1 \quad // Accept the candidate \\
\quad {\bf elseif} \\
\qquad  {\bf return} 0  \quad // Reject the candidate \\
\quad {\bf endif} \\
{\bf end function} 
\end{algorithm}
%%%%%%%

Before considering the computation of various proximity operators for the analysis and synthesis forms,
define, $\forall \vect z \in \mathbb{R}^L$, the soft-thresholding operator with threshold $\beta_{\rm th}$ as
\begin{equation}
{\rm soft}_{\beta_{\rm th}}(\vect z) = \left ({\rm soft}_{\beta_{\rm th}}(z_1), \cdots, {\rm soft}_{\beta_{\rm th}}(z_L) \right), 
\end{equation}
where for $i = 1, \ldots, L$,
\begin{equation}
{\rm soft}_{\beta_{\rm th}}(z_i) = 
\begin{cases}
0, & {\rm if} \ |z_i| \le \beta_{\rm th}, \\
z_i (|z_i| - \beta_{\rm th})/ |z_i|, & {\rm otherwise}.
\end{cases}
\end{equation}

%--
\subsubsection{Analysis form: MYULA} 
To implement MYULA for the analysis model \eqref{eqn:ir-un-af}, we set $\bar{f}(\vect x) = \mu \|\bm{\mathsf{\Psi}}^\dagger {\vect x}\|_1$
and $\bar g(\vect x ) = \|{\vect y}-\bm{\mathsf{\Phi}} {\vect x}\|_2^2/2\sigma^2$. Then, to compute the iteration \eqref{eqn:myula-ite} it is 
necessary to evaluate ${\rm prox}^\lambda_{{\bar f}}(\vect x)$ and $\nabla \bar g(\vect x)$.

To evaluate ${\rm prox}^\lambda_{{\bar f}}(\vect x)$ we use the closed-form representation \cite[see Table 1]{CP10},
\begin{equation} \label{eqn:prox-a}
\begin{split}
{\rm prox}^\lambda_{{\bar f}} (\vect x) 
	&=  \argmin_{{\vect u} \in \mathbb{R}^N}  \lambda \mu \|\bm{\mathsf{\Psi}}^\dagger {\vect u}\|_1 + \|{\vect u} - \vect x\|^2/2   \\
	&=  \vect x + \bm{\mathsf{\Psi}} \left ( {\rm prox}^\lambda_{ \mu \| \cdot \|_1}(\bm{\mathsf{\Psi}}^\dagger \vect x) - \bm{\mathsf{\Psi}}^\dagger {\vect 	x} \right )  \\
	&= \vect x + \bm{\mathsf{\Psi}} \left ( {\rm soft}_{\lambda \mu}(\bm{\mathsf{\Psi}}^\dagger \vect x) - \bm{\mathsf{\Psi}}^\dagger \vect x \right ).
\end{split}
\end{equation}
Moreover,
\begin{equation} \label{eqn:diff-a}
\nabla{{\bar g}} (\vect x) = \nabla(\|{\vect y}-\bm{\mathsf{\Phi}} \vect x\|_2^2/2\sigma^2)
	= \bm{\mathsf{\Phi}}^\dagger (\bm{\mathsf{\Phi}} \vect x - {\vect y})/\sigma^2.
\end{equation}

%---
\begin{remark} \label{rmk-op}
If $\bm{\mathsf{\Psi}}^\dagger \bm{\mathsf{\Psi}} \neq \bm{\mathsf{ I}}$, the case where $\bm{\mathsf{\Psi}}$ is overcomplete, 
${\rm prox}^\lambda_{ {\bar f}} (\vect x)$ can be computed in an iterative manner:
\begin{align} 
{\vect u}^{(t+\frac{1}{2})} &= \lambda_{\rm ite}^{(t)} ({\vect 1} - {\rm prox}^\lambda_{\|\cdot \|_1/ \lambda_{\rm ite}^{(t)}}) 
	\left(\frac{{\vect u}^{(t-\frac{1}{2})}}{ \lambda_{\rm ite}^{(t)}} + \bm{\mathsf{\Psi}}^{\dagger} {\vect u}^{(t)}  \right), \\
{\vect u}^{(t+1)} &= \vect x - \bm{\mathsf{\Psi}} {\vect u}^{(t+\frac{1}{2})},
\end{align}
where $ \lambda_{\rm ite}^{(t)} \in (0, 2/\beta_{\rm Par})$ ($\beta_{\rm Par}$ is a constant satisfying 
$\|\bm{\mathsf{\Psi}} \vect z \|^2 \le \beta_{\rm Par} \|\vect z \|^2, \forall \vect z \in \mathbb{R}^L$)  
is a predefined step size and ${\vect u}^{(t)} \rightarrow {\rm prox}^\lambda_{ {\bar f}} (\vect x)$;
refer to \citet{FS09} and \citet{JHF11} for details.
\end{remark}
%---

%--
\subsubsection{Analysis form: Px-MALA} 
%--
To implement Px-MALA for the analysis model \eqref{eqn:ir-un-af}, we set 
$\bar f(\vect x) = \|{\vect y}-\bm{\mathsf{\Phi}} {\vect x}\|_2^2/2\sigma^2 + \mu \|\bm{\mathsf{\Psi}}^\dagger {\vect x}\|_1$
and $\bar g(\vect x) = 0$. Therefore, at each iteration of the algorithm it is necessary to evaluate
\begin{equation}
	  {\rm prox}_{\bar f}^{\lambda} (\vect x) = \argmin_{\vect u \in \mathbb{R}^N} \left \{\mu \|\bm{\mathsf{\Psi}}^\dagger \vect u\|_1 
	 	\!+\! \frac{\|{\vect y} - \bm{\mathsf{\Phi}} \vect u\|_2^2}{2\sigma^2} \!+\! \frac{\|\vect u - \vect x\|_2^2}{2 \lambda} \right \}. 
\end{equation}
By the Taylor expansion of $\|{\vect y} - \bm{\mathsf{\Phi}} {\vect u}\|_2^2$ at point $\vect x$,
\begin{equation}
\begin{split}
\|{\vect y} - \bm{\mathsf{\Phi}} {\vect u}\|_2^2
	\approx & \|{\vect y} - \bm{\mathsf{\Phi}} {\vect x}\|_2^2 + (\vect u - {\vect x})^\top \nabla \! \left ( \|{\vect y} - \bm{\mathsf{\Phi}} {\vect x}\|_2^2 \right ) \\
 	= & \|{\vect y} - \bm{\mathsf{\Phi}} {\vect x}\|_2^2 + 2 (\vect u - {\vect x})^\top \bm{\mathsf{\Phi}}^\dagger (\bm{\mathsf{\Phi}} {\vect x} - {\vect y}),
\end{split}	
\end{equation}
and we obtain the following approximation of ${\rm prox}_{\bar f}^{\lambda} (\vect x)$,
\begin{align}
	 &   \argmin_{\vect u \in \mathbb{R}^N} \! \Big \{ \mu \|\bm{\mathsf{\Psi}}^\dagger \vect u\|_1 + 
	 		\frac{\|\vect u - {\vect x}\|_2^2}{2\lambda} +  \frac{\|{\vect y} - \bm{\mathsf{\Phi}} {\vect x}\|_2^2}{2\sigma^2} \nonumber \\
	   &             \hspace{0.62in} \hspace{0.4in}  + (\vect u - {\vect x})^\top \bm{\mathsf{\Phi}}^\dagger (\bm{\mathsf{\Phi}} {\vect x} - {\vect y})/\sigma^2 \Big \}  
	   \nonumber \\
	& \approx   \argmin_{\vect u \in \mathbb{R}^N}  \left \{\mu \|\bm{\mathsf{\Psi}}^\dagger \vect u\|_1 \!+\! 
	 		\frac{\|\vect u - {\vect x} + \delta \bm{\mathsf{\Phi}}^\dagger (\bm{\mathsf{\Phi}} {\vect x} - {\vect y})/2\sigma^2 \|_2^2 }{2\lambda} \right \}
			\nonumber \\
	& =   {\rm prox}_{\mu \|\bm{\mathsf{\Psi}}^\dagger \cdot \|_1}^{\lambda} \left({\vect x} - \lambda \bm{\mathsf{\Phi}}^\dagger (\bm{\mathsf{\Phi}} {\vect x} - {\vect y})/\sigma^2 \right).
\end{align}
Let $\bar{\vect v} = {\vect x} - \lambda \bm{\mathsf{\Phi}}^\dagger (\bm{\mathsf{\Phi}} {\vect x} - {\vect y})/\sigma^2$, using \eqref{eqn:prox-a}, we have    
\begin{equation} \label{eqn:ir-un-af-prox}
{\rm prox}_{\bar f}^{\lambda} ({\vect x}) \approx \bar{\vect v} + \bm{\mathsf{\Psi}} \left ( {\rm soft}_{\mu\lambda}(\bm{\mathsf{\Psi}}^\dagger \bar{\vect v}) - \bm{\mathsf{\Psi}}^\dagger \bar{\vect v}) \right ).
\end{equation}
Note that ${\rm prox}_{\bar f}^{\lambda} ({\vect x})$ here can be computed in the same manner as the one mentioned in 
remark \ref{rmk-op} if $\bm{\mathsf{\Psi}}^\dagger \bm{\mathsf{\Psi}} \neq \bm{\mathsf{ I}}$.

\begin{remark} \label{rmk-prox-a}
The approximation shown in \eqref{eqn:ir-un-af-prox} can be regarded as one iteration of the forward-backward algorithm \citep{CP10} minimising 
objective function $\bar{f} + \bar{g}$. The Taylor approximation performed above makes the assumptions in performing 
a single forward-backward iteration explicit.
\end{remark}

%--
\subsubsection{Synthesis form: MYULA}
To implement MYULA for the synthesis model \eqref{eqn:ir-un-sf}, we set $\hat{f}(\vect a) = \mu \| {\vect a}\|_1$ and 
$\hat g(\vect a ) = \|{\vect y}-\bm{\mathsf{\Phi}} \bm{\mathsf{\Psi}} {\vect a}\|_2^2/2\sigma^2$. 
Then, to compute the iteration \eqref{eqn:myula-ite} it is necessary to evaluate
\begin{equation} \label{eqn:prox}
\begin{split}
	{\rm prox}^\lambda_{\mu \|{\vect \cdot}\|_1} (\vect a)  
	& =  \argmin_{{\vect u} \in \mathbb{R}^L} \Big\{\mu \|{\vect u}\|_1 + \|{\vect u} - {\vect a}\|^2/2\lambda  \Big\}\, ,\\
	& = {\rm soft}_{\lambda \mu}({\vect a})\, ,
\end{split}	
\end{equation}
and
\begin{equation} \label{eqn:diff-s}
\nabla{{\hat g}} ({\vect a}) = \nabla(\|{\vect y}-\bm{\mathsf{\Phi}}\bm{\mathsf{\Psi}} {\vect a}\|_2^2/2\sigma^2)
	= \bm{\mathsf{\Psi}}^\dagger\bm{\mathsf{\Phi}}^\dagger (\bm{\mathsf{\Phi}}\bm{\mathsf{\Psi}} {\vect a} - {\vect y})/\sigma^2.
\end{equation}

\subsubsection{Synthesis form: Px-MALA}
To implement Px-MALA for the synthesis model \eqref{eqn:ir-un-sf}, we set 
$\hat f(\vect a) = \|{\vect y}-\bm{\mathsf{\Phi}}\bm{\mathsf{\Psi}} {\vect a}\|_2^2/2\sigma^2 + \mu \|{\vect a}\|_1$ 
and $\hat g(\vect a) = 0$. Therefore, at each iteration of the algorithm it is necessary to evaluate
\begin{equation}
	 {\rm prox}_{\hat f}^{\lambda} (\vect a) = \argmin_{\vect u \in \mathbb{R}^L}  \left \{ \mu \|{\vect u}\|_1 \!+\! 
	   \frac{\|{\vect y}  -  \bm{\mathsf{\Phi}}\bm{\mathsf{\Psi}} {\vect u}\|_2^2}{2\sigma^2}  \!+\!  \frac{\|\vect u - \vect a\|_2^2}{2\lambda} \right \}.
\end{equation}
By proceeding similarly to \eqref{eqn:ir-un-af-prox} we obtain
\begin{align} 
{\rm prox}_{\hat f}^{\lambda} (\vect a)  
	  & \approx {\rm prox}_{ \mu \| {\vect \cdot}\|_1}^{\lambda} \left(\vect a - \lambda  \bm{\mathsf{\Psi}}^\dagger\bm{\mathsf{\Phi}}^\dagger (\bm{\mathsf{\Phi}}\bm{\mathsf{\Psi}} {\vect a} - {\vect y})/\sigma^2 \right)   
	 	 \nonumber \\
	  & \approx {\rm soft}_{\mu\lambda}  \left(\vect a - \lambda  \bm{\mathsf{\Psi}}^\dagger\bm{\mathsf{\Phi}}^\dagger (\bm{\mathsf{\Phi}}\bm{\mathsf{\Psi}} {\vect a} - {\vect y})/\sigma^2  \right),\label{eqn:ir-un-sf-prox}
\end{align}
where the first line of \eqref{eqn:ir-un-sf-prox} follows by \eqref{eqn:prox}.

\begin{remark}
Similar to Remark \ref{rmk-prox-a}, the approximation shown in \eqref{eqn:ir-un-sf-prox} can be regarded 
as one iteration of the forward-backward algorithm \citep{CP10} minimising $\hat{f} + \hat{g}$.
Again, the above derivations make the corresponding assumptions explicit.
\end{remark}

%-----
\subsection{Sampling by proximal MCMC methods} 
%-----
Using formulas \eqref{eqn:diff-a} and \eqref{eqn:diff-s} which compute gradient operators,
formulas \eqref{eqn:prox-a} and \eqref{eqn:prox} which compute proximity operators according to sparse regularisations, 
and the MYULA iterative formula \eqref{eqn:myula-ite}, a set of full posterior samples for the analysis model  \eqref{eqn:ir-un-af} 
and synthesis model  \eqref{eqn:ir-un-sf} can be generated by
\begin{equation} \label{eqn:ir-un-af-myula}
\begin{split}
{\vect x}^{(m+1)} = & \ {\vect x}^{(m)} + \frac{\delta}{\lambda} \bm{\mathsf{\Psi}} \left ( {\rm soft}_{\lambda \mu/2}(\bm{\mathsf{\Psi}}^\dagger {\vect x}^{(m)}) 
			    - \bm{\mathsf{\Psi}}^\dagger {\vect x}^{(m)}) \right )   \\
			   &  - \delta \bm{\mathsf{\Phi}}^\dagger (\bm{\mathsf{\Phi}} {\vect x}^{(m)} - {\vect y})/2\sigma^2 + \sqrt{\delta} {\bar{\vect w}}^{(m)}
\end{split}	 
\end{equation}
and
\begin{equation} \label{eqn:ir-un-sf-myula}
\begin{split}
{\vect a}^{(m+1)} = & \ (1 - \frac{\delta}{\lambda}) {\vect a}^{(m)} + \frac{\delta}{\lambda}  {\rm soft}_{\lambda\mu/2} ({\vect a}^{(m)})   \\
			    & - \delta \bm{\mathsf{\Psi}}^\dagger\bm{\mathsf{\Phi}}^\dagger (\bm{\mathsf{\Phi}}\bm{\mathsf{\Psi}} {\vect a} - {\vect y})/2\sigma^2 + \sqrt{\delta} {\hat{\vect w}}^{(m)},
\end{split}	 
\end{equation}
respectively, where  $\bar{\vect w}^{(m)} \in \mathbb{R}^N  \sim {\cal N} (0,\mathbb{1}_N)$ and 
$\hat{\vect w}^{(m)} \in \mathbb{R}^L  \sim {\cal N} (0,\mathbb{1}_L)$.

Analogously, using formulas \eqref{eqn:ir-un-af-prox} and \eqref{eqn:ir-un-sf-prox}, the Px-MALA 
iterative forms generating samples as to the analysis and synthesis models can be written as
\begin{equation}\label{eqn:pula-ite-x}
{\vect x}^{(m+1)} =  {\rm prox}_{\bar f}^{\delta/2} ({\vect x}^{(m)}) + \sqrt{\delta} \bar{\vect w}^{(m)},
\end{equation}
and
\begin{equation}\label{eqn:pula-ite-a}
{\vect a}^{(m+1)} =  {\rm prox}_{\hat f}^{\delta/2} ({\vect a}^{(m)}) + \sqrt{\delta} \hat{\vect w}^{(m)},
\end{equation}
respectively.
After a proper candidate generated by \eqref{eqn:pula-ite-x} or \eqref{eqn:pula-ite-a}, 
Px-MALA includes an MH accept-reject step with an acceptance probability $\rho$, specified by \eqref{eqn:mh},
to ensure the sequence converges to the target distribution.

To generate $K$ samples using the proximal MCMC methods proposed, two parameters controlling sample candidates should
be assigned: (i) the number of initial or \emph{burn-in} iterations, $K_{\rm burn} \in {\mathbb Z}$ 
(denotes the previous number of iterations that are discarded); 
and (ii) the chain's \emph{thinning} factor or number of intermediate iterations between samples, $K_{\rm gap} \in {\mathbb Z}$
(denotes the intermediate number of iterations that are discarded; used to reduce correlations between samples and the algorithm's memory footprint).
Because of memory limitations we do not store all samples
(generated by \eqref{eqn:ir-un-af-myula},  \eqref{eqn:ir-un-sf-myula}, \eqref{eqn:pula-ite-x} or \eqref{eqn:pula-ite-a}), 
and only store 1-in-$K_{\rm gap}$ samples if
\begin{equation} \label{eqn:mh-con}
m>K_{\rm burn} \quad \text{and} \quad {\tt mod}(m - K_{\rm burn}, K_{\rm gap}) = 0,
\end{equation}
where ${\tt mod}(\cdot, \cdot)$ represents the modulus after division.

We conclude this section by summarising the MYULA and Px-MALA implementations for RI imaging
in Algorithms \ref{alg:ee-myula-sam} and \ref{alg:ee-pmala-sam}, respectively.
Note that symbol $P_{\rm type} \in \{{\tt analysis}, {\tt synthesis}\}$ specifies the problem type considered.
Moreover, after obtaining the sets of samples corresponding to the analysis and synthesis models using
Algorithms \ref{alg:ee-myula-sam} and \ref{alg:ee-pmala-sam}, the posterior mean (or median) of each set of samples  
can be computed as a point estimator to represent the recovered sky image of interest and thus address the original ill-posed reconstruction problem.

%-------------------------------------------------------------------
\section{Bayesian uncertainty quantification: proximal MCMC methods}\label{sec:alg}
%-------------------------------------------------------------------
In this section we describe a range of uncertainty quantification analyses that are of interest for RI imaging. The analyses require calculating summary statistics w.r.t.\ the posterior $p(\vect x|\vect y)$, which we compute using the samples $\{{\vect x}^{(j)}\}_{j=1}^K$ generated by MYULA or Px-MALA (in the case of synthesis we generate samples $\{{\vect a}^{(j)}\}_{j=1}^K$ from $p(\vect a | \vect y)$ and map them to the image space by using $\bm{\mathsf{\Psi}}$). 

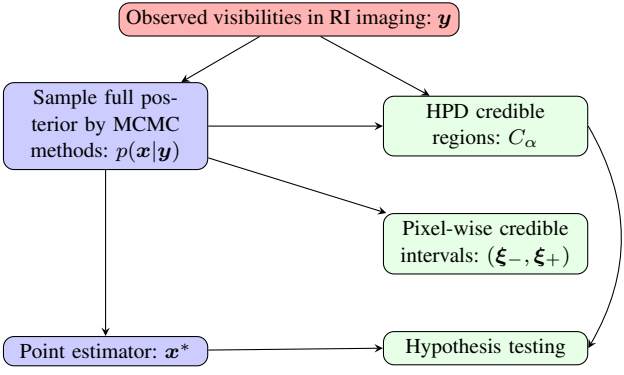
\begin{figure}
  \begin{center}
    \begin{tabular}{c}
  	 % figure of our uncertainty quantification procedure

\begin{tikzpicture}[>=stealth,every node/.style={shape=rectangle,draw,rounded corners},]
    % create the nodes
     \node (c0){};
    \node (c1) [fill=red!30, right of = c0, node distance=1cm]{Observed visibilities in RI imaging: $\vect y$};
    \node (c2) [fill=blue!20, below left of = c0, node distance=2cm, text width=2.5cm,align=center]{Sample full posterior by MCMC methods: $p(\vect x | \vect y)$ };
    \node (c3) [fill=green!10, right of = c2, node distance=5cm, text width=2.5cm,align=center]{HPD credible regions: $C_{\alpha}$ };
    \node (c4) [fill=blue!20, below = 2.2cm of c2, text width=2.5cm,align=center]{Point estimator: ${\vect x}^*$};
    \node (c5) [fill=green!10, below = 0.75cm of c3,text width=2.5cm,align=center]{Pixel-wise credible intervals: $({\vect \xi}_-, {\vect \xi}_+)$ };
    \node (c6) [fill=green!10, below = 0.75cm of c5,text width=2.5cm,align=center]{Hypothesis testing };

    % connect the nodes
    \draw[->] (c1) -- (c2);
    \draw[->] (c1) -- (c3);
    \draw[->] (c2) -- (c3);
    \draw[->] (c2) -- (c4);
    \draw[->] (c3.east) to [out=-60,in=60] (c6.east);
    \draw[->] (c2) -- (c5);
    \draw[->] (c4) -- (c6.west);
\end{tikzpicture}
    \end{tabular}
  \end{center}
  \caption{Our proposed uncertainty quantification procedure for RI imaging based on proximal MCMC sampling. 
  The light green areas on the right show the types of uncertainty quantification developed.
  Firstly, the full posterior distribution of the image is sampled by MCMC methods, such as MYULA and Px-MALA. 
  Then, various forms of uncertainty quantification are performed. 
  Pixel-wise credible intervals (\textit{cf.} error bars) are computed using the posterior samples. Global Bayesian credible regions are computed, again using the posterior samples, and are then used to perform hypothesis testing of image structure   
  to test whether a structure of interest is either physical or an artefact. }
  \label{fig:uq_diag}
\end{figure}

The diagram in Figure \ref{fig:uq_diag} shows the main components of our proposed uncertainty quantification 
methodology based on (proximal) MCMC methods.
As is shown, firstly, the full posterior distribution of the image is sampled by MCMC methods, such as MYULA and Px-MALA as adopted in this article. 
Then, various forms of uncertainty quantification are performed. 
Firstly, pixel-wise credible intervals are computed using the posterior samples. After that, global Bayesian credible regions are computed, and are then used to perform hypothesis testing of image structure   
to test whether a structure of interest is either physical or an artefact. 

%--------
\subsection{Pixel-wise credible intervals}
%--------
The first analyse we consider is the set of marginal credible intervals of each image pixel, denoted by 
$ [\xi_{i-}, \xi_{i+}]$ for pixel $x_i$. 
These intervals specify the range of values that the image pixels take with probability $(1-\alpha)$, \emph{i.e.},
\begin{equation}
p(x_i \in [\xi_{i-}, \xi_{i+}] | \vect y) = 1- \alpha\, , \quad i = 1,\ldots,N\, .
\end{equation}
Pixel-wise intervals are useful for analysing local information relevant to small image structures and for identifying regions of the image with high uncertainty. For example, these can be conveniently visualised by constructing an image with the quantities $\{\xi_{i+} - \xi_{i-}\}_{i = 1}^N$ related to the length of the intervals. 

To compute the marginal credible interval we simply calculate:
\begin{align} \label{eqn:cr-local}
       (\bar{\xi}_{i-}, \bar{\xi}_{i+})  &= {\tt quantile} \left (\big\{{x_i}^{(j)}\big\}_{j=1}^K, \left \{\frac{\alpha}{2}, 1 - \frac{\alpha}{2} \right\}  \right),\\
	(\hat{\xi}_{i-}, \hat{\xi}_{i+})  &=  {\tt quantile} \left (\big\{({\bm{\mathsf{\Psi}} \vect a }^{(j)})_i \big\}_{j=1}^K, \left \{\frac{\alpha}{2}, 1 - \frac{\alpha}{2} \right\}  \right),
\end{align}
depending on whether an analysis or a synthesis formulation is used, respectively; we have used the fact that samples can be marginalised implicitly by projection. 

\begin{remark}
Function ${\tt quantile}(\cdot, \cdot)$ is a standard function built into many programming languages, which, {\it e.g.}, in \eqref{eqn:cr-local} computes the quantile thresholds
$\bar{\xi}_{i-}$ and $\bar{\xi}_{i+}$ at probabilities $\alpha/2$ and $(1- \alpha/2)$, respectively.
In detail, $\bar{\xi}_{i-}$ and $\bar{\xi}_{i+}$ can be computed respectively from the following definitions:
\begin{equation} \label{quantil-}
\begin{split}
\bar{\xi}_{i-} & = {\inf} \left \{{\xi}_{i-} : p(z_i \le {\xi}_{i-} | \vect y ) \ge {\alpha}/{2} \right \}, \\
\bar{\xi}_{i+} & = {\inf} \left \{{\xi}_{i+}: p(z_i \le {\xi}_{i+} | \vect y ) \ge 1- {\alpha}/{2} \right \},
\end{split}
\end{equation}
where $z_i $ denotes $i$-th image pixel in the canonical coordinate system.
Refer to, {\it e.g.}, \cite{KJ78} for more details about computing quantile thresholds.

\end{remark}

%-------------------------------
\subsection{Highest posterior density (HPD) credibility regions}
%-------------------------------
Pixel-wise intervals are useful for analysing local image structures. To perform more sophisticated analyses it is more convenient to compute credible regions that operate at an image level. Precisely, in Bayesian decision theory, a set $C_{\alpha} \subset \mathbb{R}^N$ with $\alpha \in (0, 1)$ is a posterior credible region with confidence level 
$100(1-\alpha)\%$ if
\begin{equation} \label{eqn:cr-b}
p(\vect x \in C_{\alpha} | \vect y ) = \int_{\mathbb{R}^N} p(\vect x | \vect y) \mathbb{1}_{C_{\alpha}} (\vect x) {\rm d} {\vect x}  
	= 1-\alpha,
\end{equation}
where $\mathbb{1}_{C}$ is the indicator function for the set ${C}$ defined by $\mathbb{1}_{C}({\vect u}) = 1$ if $\vect u \in {C}$ and 0 otherwise. 

There are infinitely many regions $C_{\alpha}$ that satisfy the above property. The optimal region, in the sense of compactness, is the so-called highest posterior density (HPD) region 
\begin{equation} \label{eqn:cr-hpd}
C_{\alpha} = \{\vect x: f(\vect x) + g(\vect x) \le  \gamma_{\alpha} \},
\end{equation}
where the threshold $\gamma_{\alpha}$ is set such that \eqref{eqn:cr-b} holds, and we recall that $p(\vect x | \vect y) \propto \exp\{-f(\vect x) - g(\vect x)\}$.  The threshold $\gamma_{\alpha}$ defines an isocontour or level-set of the log-posterior. This region is decision-theoretically optimal in the sense of minimum volume \citep{R01}.

The value of $\gamma_{\alpha}$ such that \eqref{eqn:cr-b} and \eqref{eqn:cr-hpd} holds is easily estimated from the MCMC samples. 
Precisely, let $\bar{C}_{\alpha}$ and $\hat{C}_{\alpha}$ represent the HPD regions associated with the set of 
samples $\{{\vect x}^{(j)}\}_{j=1}^K$ and $\{{\vect a}^{(j)}\}_{j=1}^K$ generated with 
MYULA or Px-MALA for the analysis and synthesis models, respectively. To calculate the thresholds 
$\bar{\gamma}_{\alpha}$ and $\hat{\gamma}_{\alpha}$ we use the estimators:
\begin{equation} \label{eqn:hpd-r}
\begin{split}
\bar{\gamma}_{\alpha} &=  {\tt quantile}\left(\big\{({\bar f}+{\bar g})({\vect x}^{(j)})\big\}_{j=1}^K, 1- \alpha \right),\\
\hat{\gamma}_{\alpha} &=  {\tt quantile}\left(\big\{({\hat f}+{\hat g})({\vect a}^{(j)})\big\}_{j=1}^K, 1- \alpha \right)\, .
\end{split}
\end{equation}

Notice that $C_{\alpha}$ is a joint credible region operating at the image level (as opposed to the pixel level), 
and therefore we use it to analyse larger image structures. In addition,
we use $C_{\alpha}$ for posterior checks to analyse the degree of confidence in specific structure
observed in reconstructions, as discussed in the following section.

%--------
\subsection{Hypothesis testing of image structure}\label{sss:gyp-testing}
%--------
We now describe a \emph{knock-out} posterior check to assess specific areas or structures of interest in reconstructed images. The \mbox{rationale} for this test is that if the data supports a specific feature that we observe in a reconstructed image, {\it e.g.} ${\vect x}_{\rm map}$, then removing this feature from the image is likely to lead to a point that is outside the HPD credible region. Precisely, we use a segmentation-inpainting procedure to carefully replace the feature of interest with background information (although alternative procedures can certainly be considered).  If the segmented-inpainted image lies outside of the HPD region this indicates that the likelihood strongly disagrees with the modification, and hence that the data support the feature or structure under consideration. Conversely, if the segmented-inpainted image is within the HPD region, this suggests that the likelihood is not too sensitive to the modification, and therefore that the data does not strongly support the feature or structure being scrutinised.

Algorithmically, the first step of this two-step procedure is to generate a meaningful surrogate test image ${\vect x}^{*, {\rm sgt}}$. We achieve this by taking a point estimator ${\vect x}^{*}$ ({\it e.g.},  the posterior mean $\bar{\vect x}^{*} = \sum_{j=1}^K  {\vect x}^{(j)} / K$,  or $\hat{\vect x}^{*} = \sum_{j=1}^K  \bm{\mathsf{\Psi}} {\vect a}^{(j)} / K$ if a synthesis model is used) and masking out the structure of interest. This region of the image is then filled by inpainting with background information. Here we use a classical inpaiting approach \citep{JCS08} based on a recursive wavelet filter
\begin{equation} \label{eqn:inpaint}
{\vect x}^{(m+1), {\rm sgt}} = {\vect x}^{*}  \mathbb{1}_{\Omega - \Omega_D} 
	+ \bm{\mathsf{\Lambda}}^{\dagger}{\rm soft}_{\lambda_{\rm th}} (\bm{\mathsf{\Lambda}}{\vect x}^{(m), {\rm sgt}}) \mathbb{1}_{\Omega_D},
\end{equation} 
where $\Omega$ is the image domain, $\Omega_D$ is the masked region, $\mathsf{\Lambda}$ is a wavelet filter operator, $\lambda_{\rm th}$ is a prefixed threshold, and ${\vect x}^{(m+1), {\rm sgt}}$ is the inpainted result obtained 
at iteration $m$ (generally 100 iterations suffice to achieve convergence). 
The second step of the procedure is simply to check if $\bar{\vect x}^{*, {\rm sgt}} \notin \bar{C}_{\alpha}$ by using \eqref{eqn:cr-hpd} and \eqref{eqn:hpd-r}, {\it i.e.} by evaluating ${\bar f}(\bar{\vect x}^{*, {\rm sgt}})+ {\bar g}(\bar{\vect x}^{*, {\rm sgt}})$ and comparing to $\bar{\gamma}_{\alpha}$ (or to check if $\hat{\vect x}^{*, {\rm sgt}} \notin \hat{C}_{\alpha}$ in the synthesis setting). 

Finally, note that if the test involves a large structure then the choice of the point estimator used to construct ${\vect x}^{*, {\rm sgt}}$ is usually not important. However, for small structures we recommend using the posterior median as it is closer to the boundaries of ${C}_{\alpha}$ than the posterior mean and the MAP estimates.

%-------------------------------------------------------------------
\section{Experimental results}\label{sec:exp}
%-------------------------------------------------------------------
In this section we demonstrate MYULA and Px-MALA on a range of experiments with simulated RI observations. The generated samples are then used to compute Bayesian point estimators and to perform various forms of uncertainty quantification.

%--------
\subsection{Simulations}
%--------
\begin{figure}
	\centering
	\begin{tabular}{c}
		\includegraphics[trim={{.25\linewidth} {.15\linewidth} {.25\linewidth} {.05\linewidth}}, clip, width=0.47\linewidth, height = 0.47\linewidth]
		{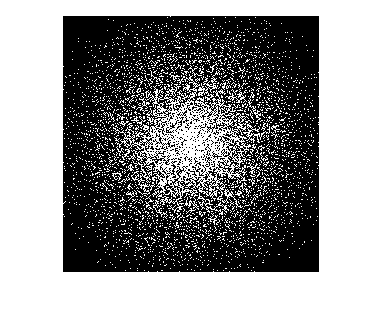}
        \end{tabular}
	\caption{A randomly generated visibility coverage (10\% of Fourier coefficients) with size of $256\times 256$.
	}
	\label{fig-mask}
\end{figure}

\begin{figure*}
	\centering
	\begin{tabular}{cccc}
		\includegraphics[trim={{.15\linewidth} {.07\linewidth} {.02\linewidth} {.07\linewidth}}, clip, width=0.24\linewidth, height = 0.21\linewidth]
		{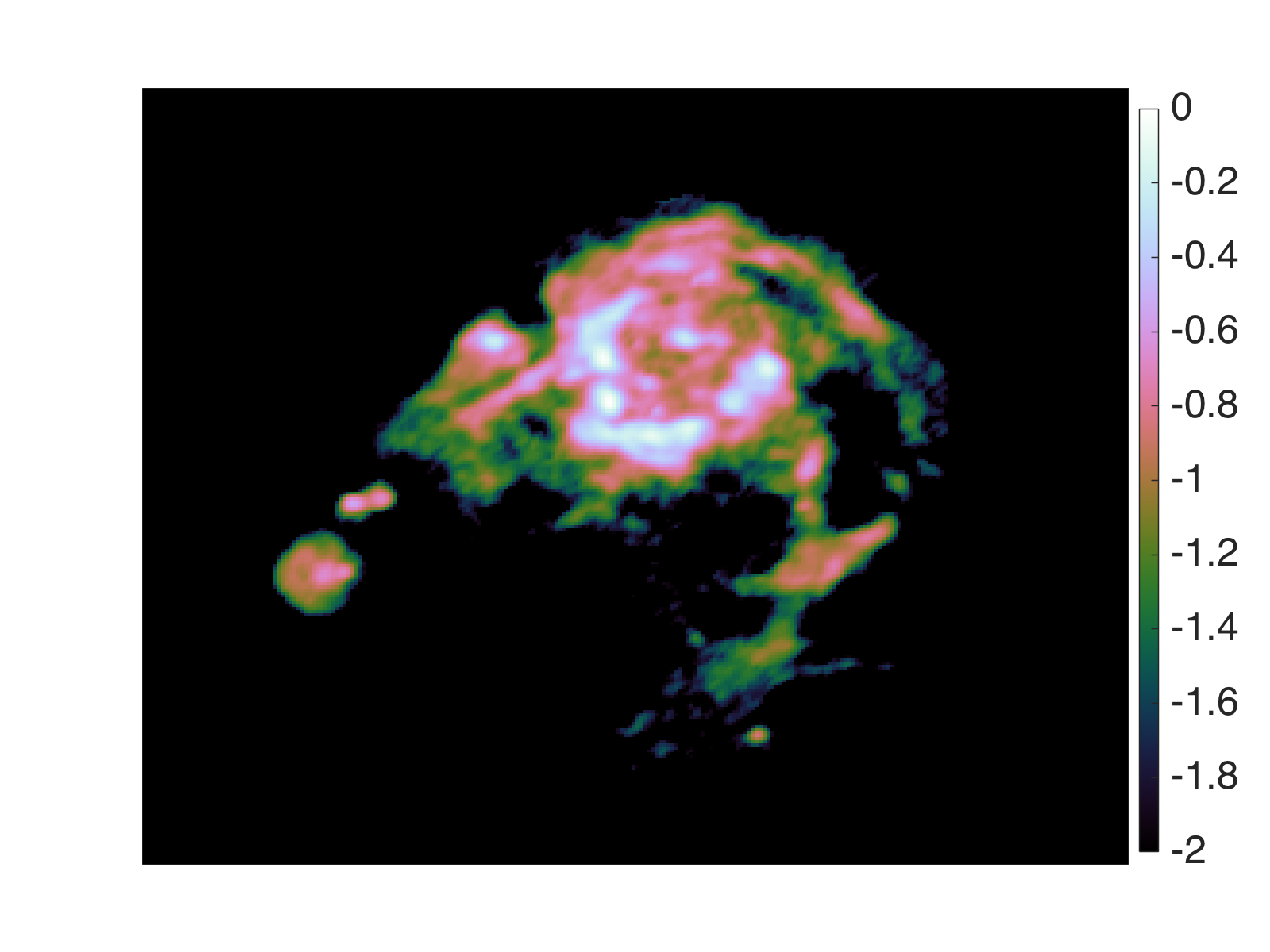} &
		\includegraphics[trim={{.15\linewidth} {.07\linewidth} {.03\linewidth} {.073\linewidth}}, clip, width=0.24\linewidth, height = 0.21\linewidth]
		{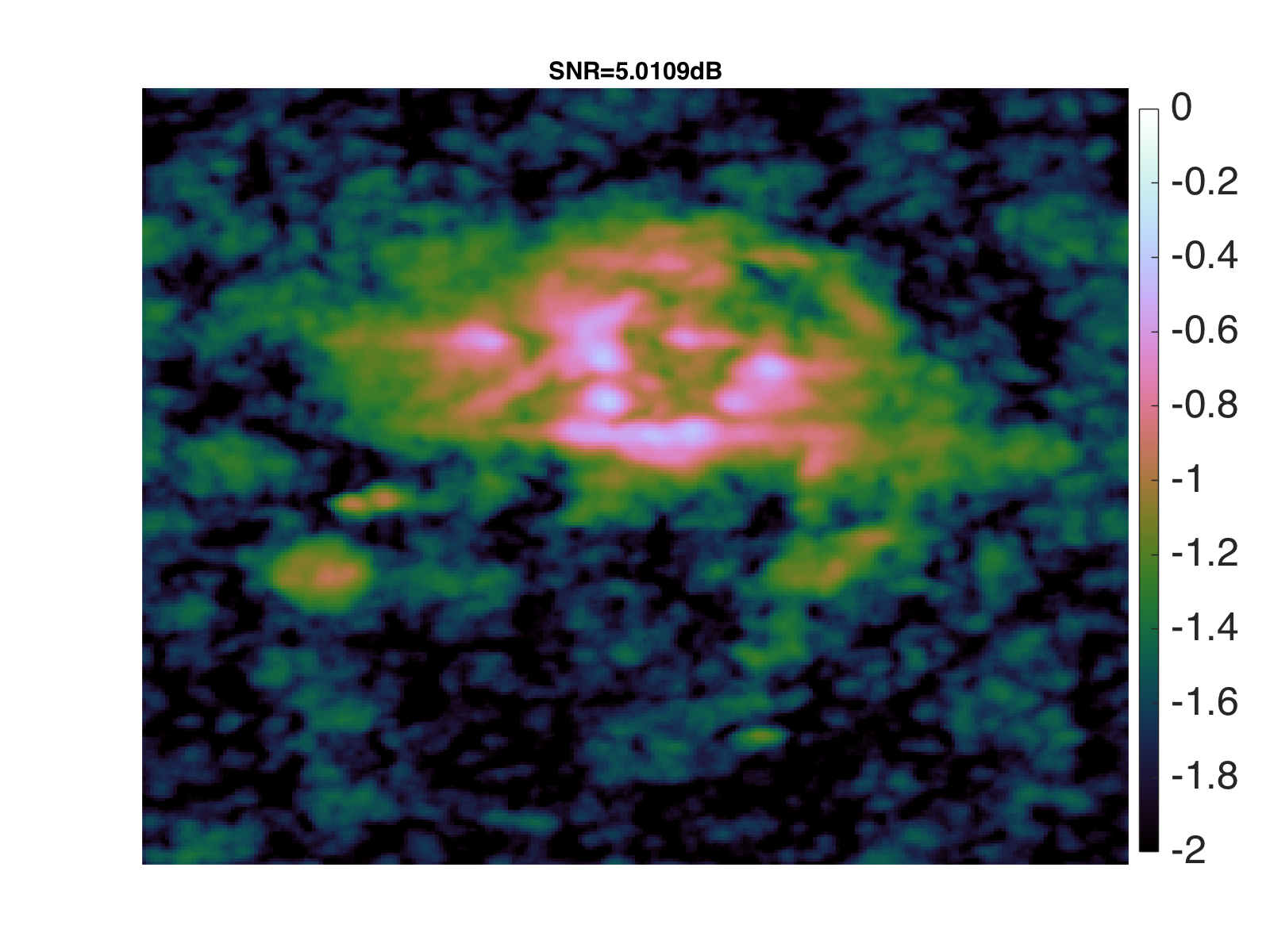} &
		\includegraphics[trim={{.15\linewidth} {.07\linewidth} {.02\linewidth} {.07\linewidth}}, clip, width=0.24\linewidth, height = 0.21\linewidth]
		{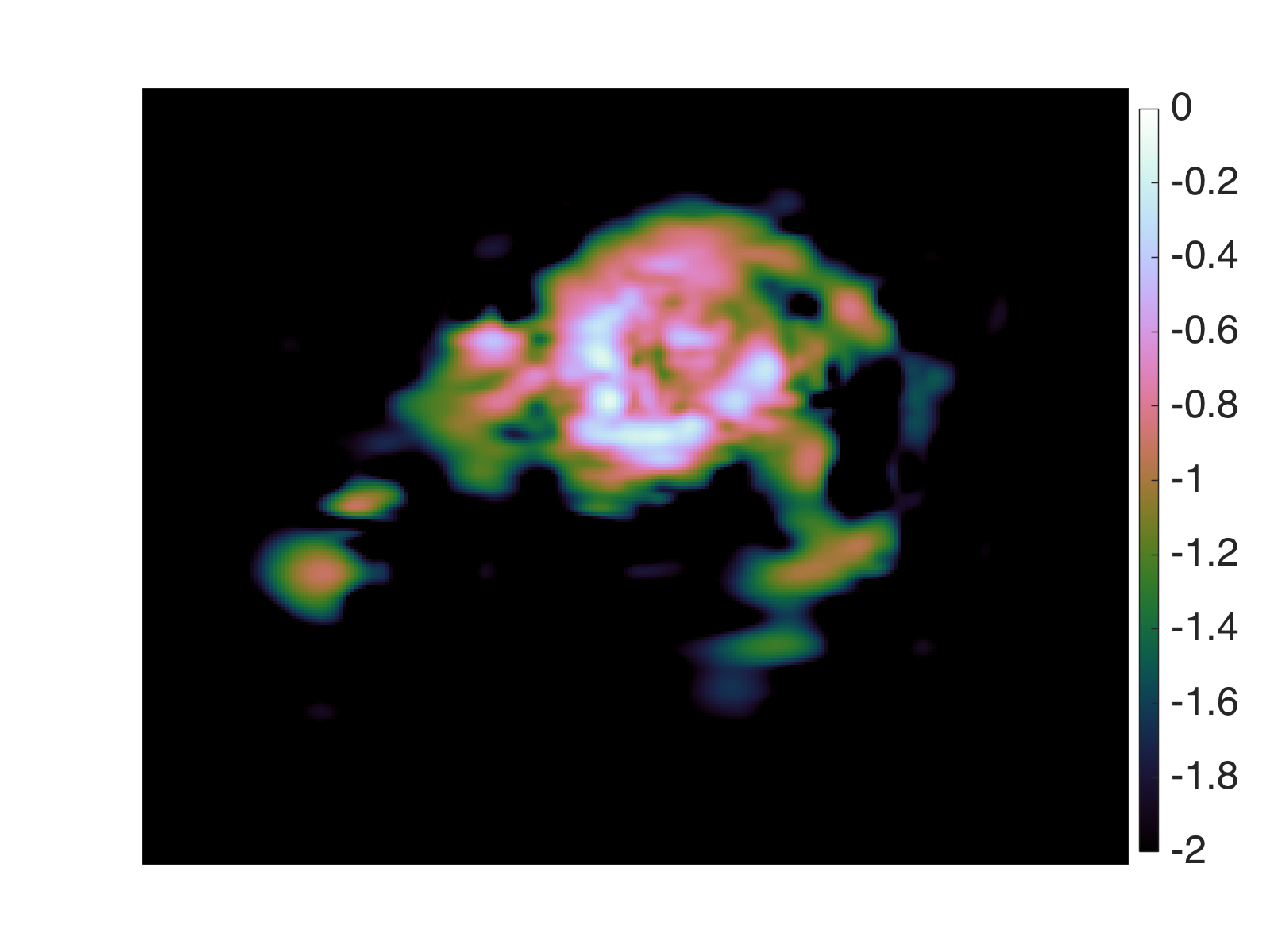} &
		\includegraphics[trim={{.15\linewidth} {.07\linewidth} {.02\linewidth} {.073\linewidth}}, clip, width=0.24\linewidth, height = 0.21\linewidth]
		{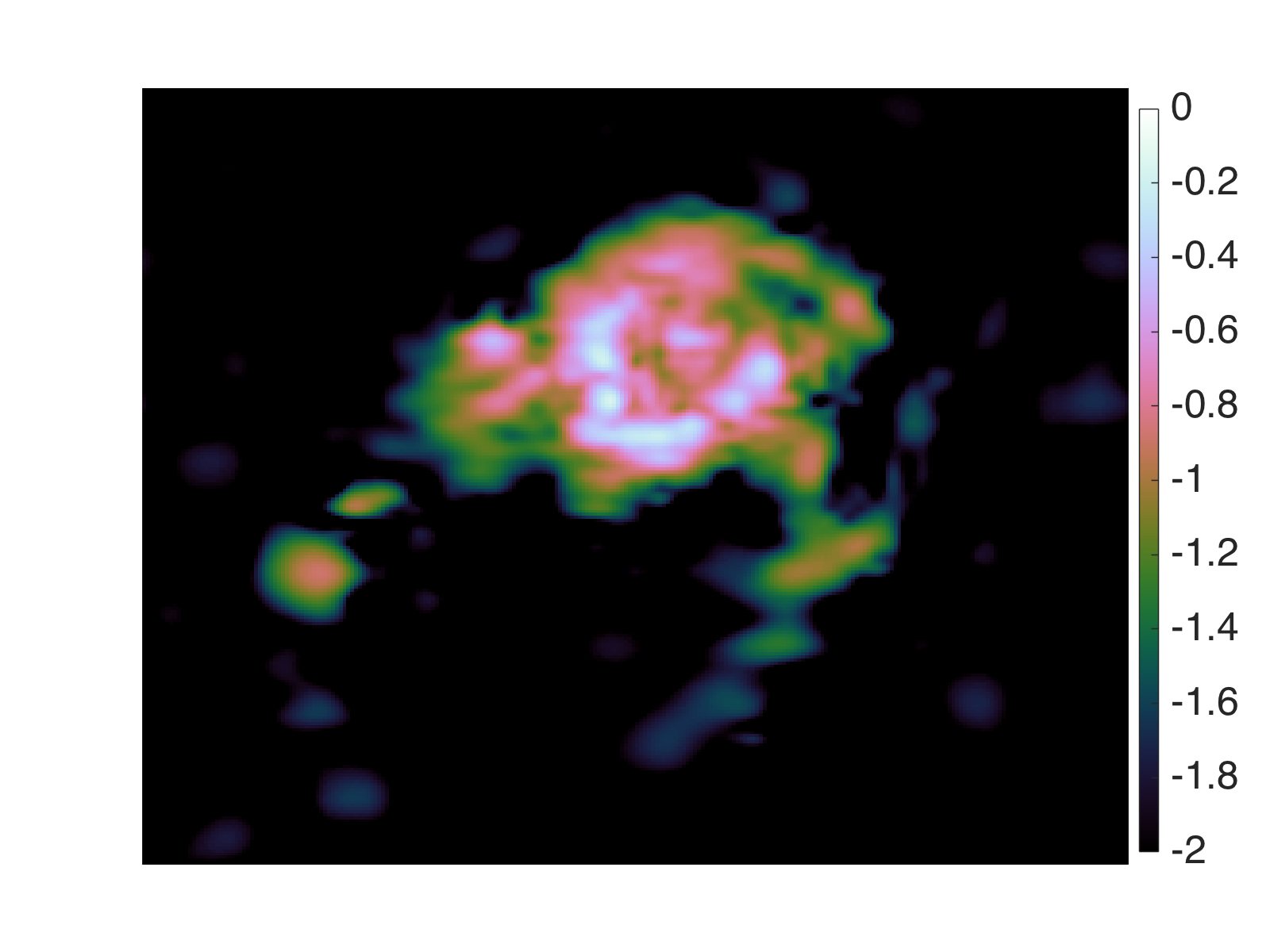} 
		\\
		{\small (a) ground truth} & {\small (b) dirty image} & {\small (c) MYULA for analysis model } & {\small (d) Px-MALA for analysis model }  
		\\
		& &
		\includegraphics[trim={{.15\linewidth} {.07\linewidth} {.02\linewidth} {.07\linewidth}}, clip, width=0.24\linewidth, height = 0.21\linewidth]
		{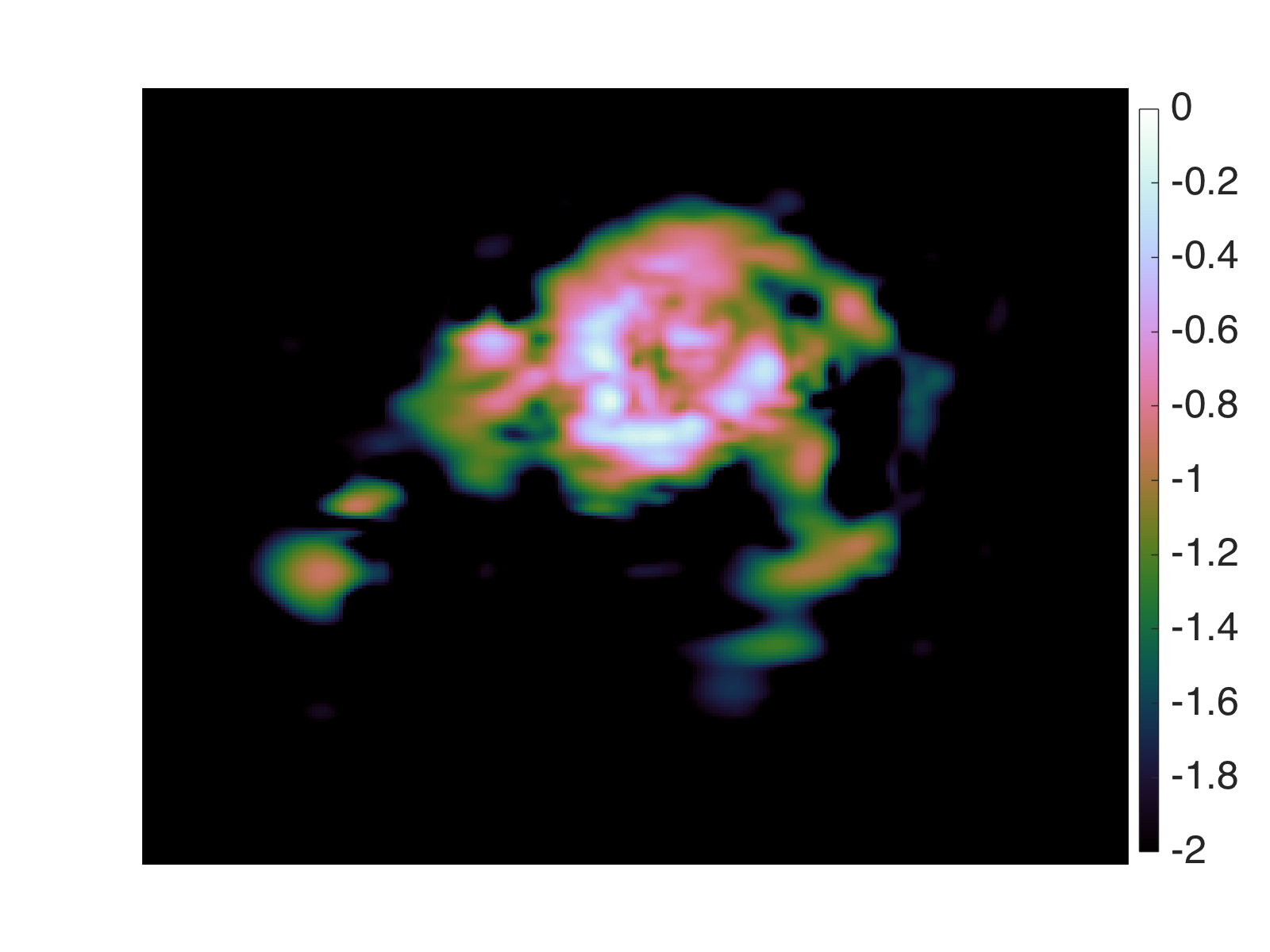} &
		\includegraphics[trim={{.15\linewidth} {.07\linewidth} {.02\linewidth} {.073\linewidth}}, clip, width=0.24\linewidth, height = 0.21\linewidth]
		{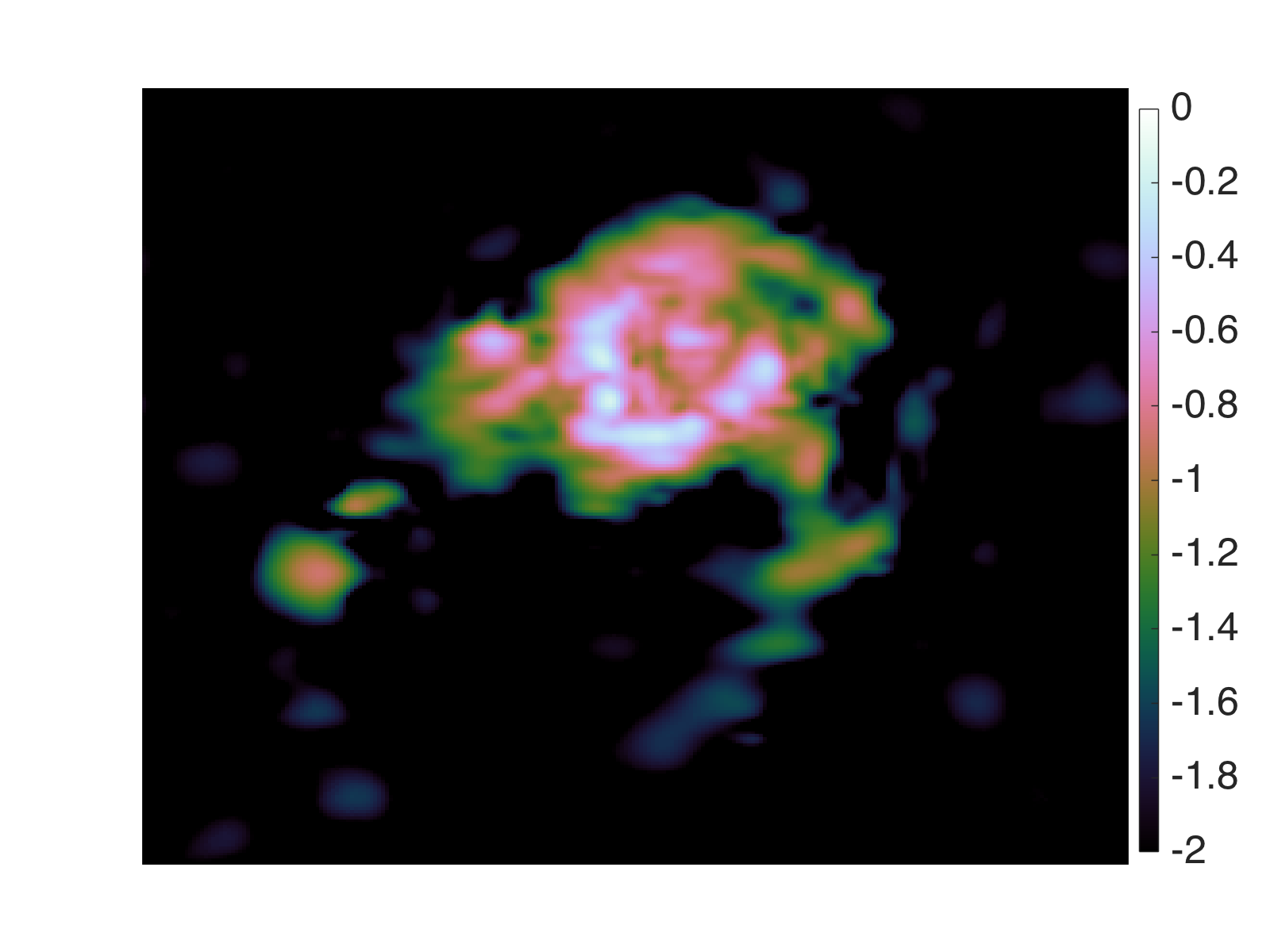} 
		\\
		& &{\small (e) MYULA for synthesis model } & {\small (f)  Px-MALA for synthesis model }  	 
        \end{tabular}
	\caption{Image reconstructions for M31 (size $256\times 256$). All images are shown in ${\tt log}_{10}$ scale {(\textit{i.e.} the numeric labels on the colour bar are the logarithms of the image intensity)}. Panel (a): ground truth; 
	(b): dirty image (reconstructed by inverse Fourier transform);
	(c) and (d): point estimators recovered from the mean of the samples generated by MYULA and Px-MALA for the analysis model \eqref{eqn:ir-un-af}, respectively; 
	(e) and (f): the same as (c) and (d) but for the synthesis model \eqref{eqn:ir-un-sf}.
	Clearly, consistent results between MYULA and Px-MALA, and between the analysis and synthesis models, are obtained.  See further discussion in main text.
	}
	\label{fig-m31}
\end{figure*}

The following four images are used in our experiments: the HI region of the M31 galaxy (size $256\times 256$ pixels) shown in Figure~\ref{fig-m31}~(a); 
the Cygnus A radio galaxy (size $256\times 512$ pixels) shown in Figure~\ref{fig-others}~(a, top); the W28 supernova remnant (size $256\times 256$ pixels) shown in Figure~\ref{fig-others}~(a, middle); and the 3C288 radio galaxy 
(size $256\times 256$ pixels) shown in Figure~\ref{fig-others}~(a, bottom). The hardware used to perform these simulations and subsequent numerical experiments is a workstation with 24 CPU cores, x86\_64 architecture, and 256 GB memory. All the codes are run on {\sc Matlab} R2015b.

To generate visibilities, a $u$-$v$ coverage is generated randomly through the variable density sampling profile \citep{PVW11} in half
the Fourier plane with 10\% of Fourier coefficients of each ground truth image; see Figure~\ref{fig-mask} for an example of the sampling profile. 
The visibilities are then corrupted by zero mean complex Gaussian noise with standard deviation $\sigma$ computed by  
$\sigma = \|f\|_{\infty} 10^{-\textrm{SNR}/20}$, where $\|\cdot \|_{\infty}$ is the infinity norm (the maximum absolute value
of components of $f$), and SNR (signal to noise ratio) is set to 30 dB for all simulations.

The dictionary $\bm{\mathsf{\Psi}}$ in the analysis and synthesis models \eqref{eqn:ir-un-af} and \eqref{eqn:ir-un-sf} 
is set to Daubechies 8 wavelets (therefore, we do not expect appreciable difference between the results of the analysis and synthesis models), 
which is implemented by using the {\sc Matlab} built-in function {\tt wavedec2}; 
complex wavelets or their hybrids, such as those with overcomplete bases, are suggested for better reconstruction.  
The $\ell_1$ regularisation parameter $\mu$ in the analysis and synthesis models is fixed to $10^{4}$ by visual cross-validation.  
Note that, in practice, parameter $\mu$ generally needs to be selected carefully either manually or automatically according 
to some appropriate criterion {(see the discussion in 
Section \ref{sec:ri:bayesian}).  This is beyond the scope of the current article but application of the hierarchical Bayesian strategies developed by \citet{MBF15} will be considered in future work.}

In all experiments MYULA and Px-MALA are implemented using the same algorithm parameters. Precisely, we use each algorithm to generate $10^3$ samples from the posterior distributions \eqref{eqn:baye-x} and \eqref{eqn:baye-a}, with $10^5$ burn-in iterations (these iterations correspond to the chains' transient period and are discarded), and a thinning factor of $10^3$ iterations between samples (with these settings each algorithm runs for $1.1\times 10^6$  iterations to produce $10^3$ samples). We have used these settings to simplify comparisons between MYULA and Px-MALA, however in all our experiments MYULA converged very quickly and could have been implemented with a significantly lower numbers of iterations. The other parameters are set as follows: the maximum iteration number used in \eqref{eqn:inpaint} for segmented-inpainting is set to 200; the range of values of $\alpha$ in \eqref{eqn:cr-b} is fixed to [0.01, 0.99]; the credible intervals \eqref{eqn:cr-local} are computed at level $95\%$ with $\alpha = 0.05$; 
and $\alpha$ is set to 0.01 (corresponding to the 99\% confidence level) in \eqref{eqn:hpd-r} for hypothesis testing.

%%%%
\begin{table}
\begin{center}
\caption{CPU time in minutes for MYULA and Px-MALA, for the M31, Cygnus A, W28 and 3C288 experiments, with respect to
the analysis and synthesis models \eqref{eqn:ir-un-af} and \eqref{eqn:ir-un-sf}. The results show that MYULA is much more economical than 
Px-MALA, requiring approximately half the computation time of Px-MALA.  However, by including an MH (Metropolis-Hastings) 
accept-reject step Px-MALA removes asymptotic bias. 
	} \label{tab:time}
 \vspace{-0.05in}
\begin{tabular}{ccrr}
\toprule  
 \multirow{2}{*}{Images}  & \multirow{2}{*}{Methods} &  \multicolumn{2}{c}{CPU time (min)} \\ 
 & & Analysis & Synthesis 
\\ \toprule
\multirow{2}{*}{M31 (Fig. \ref{fig-m31} ) } & MYULA & 618 & 581  
\\ 
&  Px-MALA &  $1307$ &  $944$   
\\ \midrule
\multirow{2}{*}{Cygnus A  (Fig. \ref{fig-others} ) } & MYULA & 1056  & 942 
\\ 
&  Px-MALA & $2274$  & $1762$
\\ \midrule
\multirow{2}{*}{W28  (Fig. \ref{fig-others} ) }  &  MYULA& 646 &  598
\\ 
& Px-MALA & $1122 $    & $879$  
\\ \midrule
\multirow{2}{*}{3C288 (Fig. \ref{fig-others} ) }  &  MYULA & 607 & 538
 \\ 
 &  Px-MALA & $1144 $ &  $881$ 
\\ \bottomrule
\end{tabular}
\end{center}
\end{table}
%%%%

%%%%
\addtolength{\tabcolsep}{-\tabL}
\addtolength{\tabcolsep}{\tabL}
%%%%

%%%%
\addtolength{\tabcolsep}{-\tabL}
{ \renewcommand{\arraystretch}{0.0}
\begin{figure*}
	\centering
	\begin{tabular}{cccc}
		\includegraphics[trim={{.15\linewidth} {.07\linewidth} {.025\linewidth} {.07\linewidth}}, clip, width=0.24\linewidth, height = 0.13\linewidth]
		{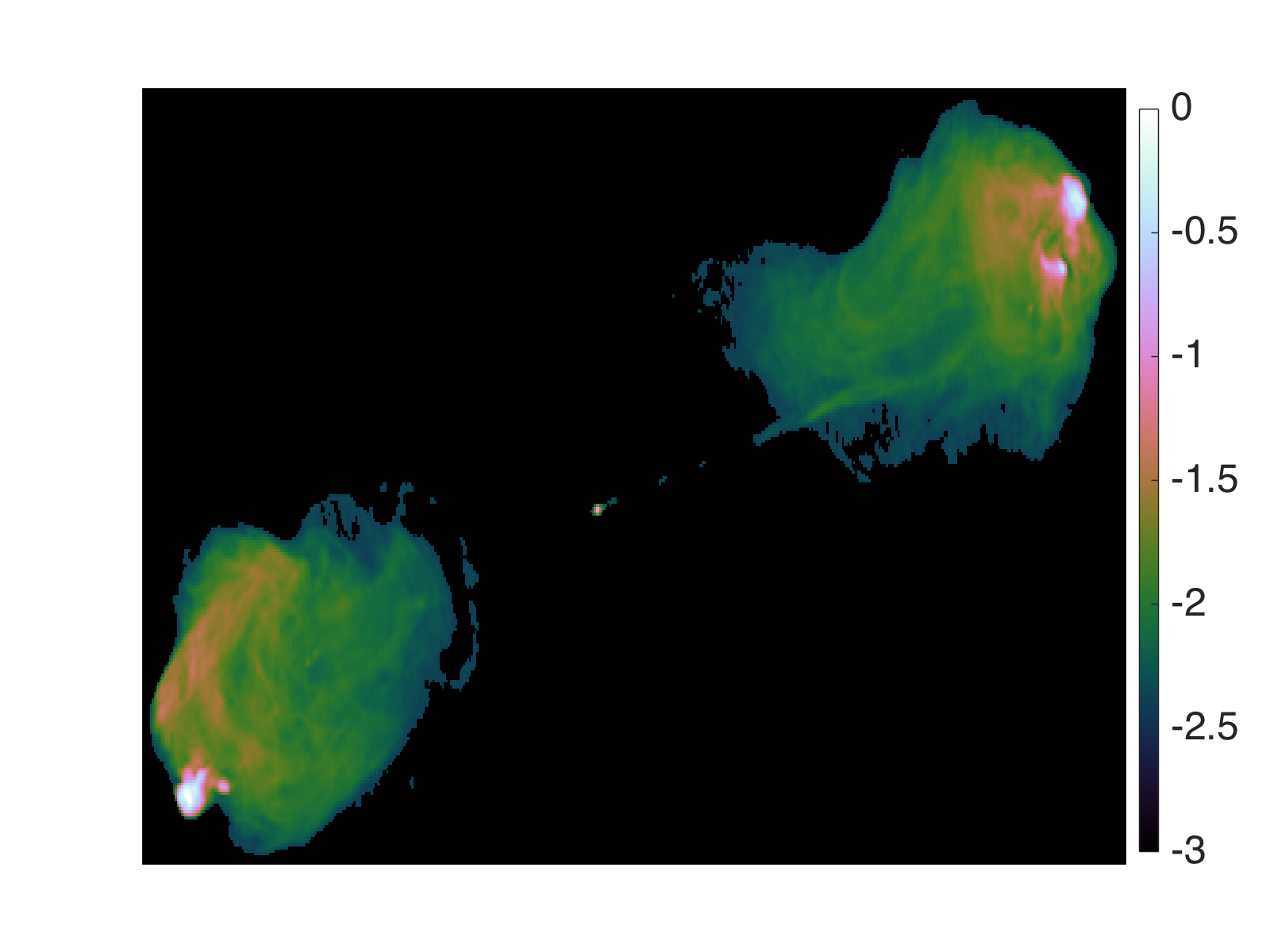}   \put(-130,11){\rotatebox{90}{ Cygnus A}} &
		\includegraphics[trim={{.15\linewidth} {.07\linewidth} {.025\linewidth} {.072\linewidth}}, clip, width=0.24\linewidth, height = 0.13\linewidth]
		{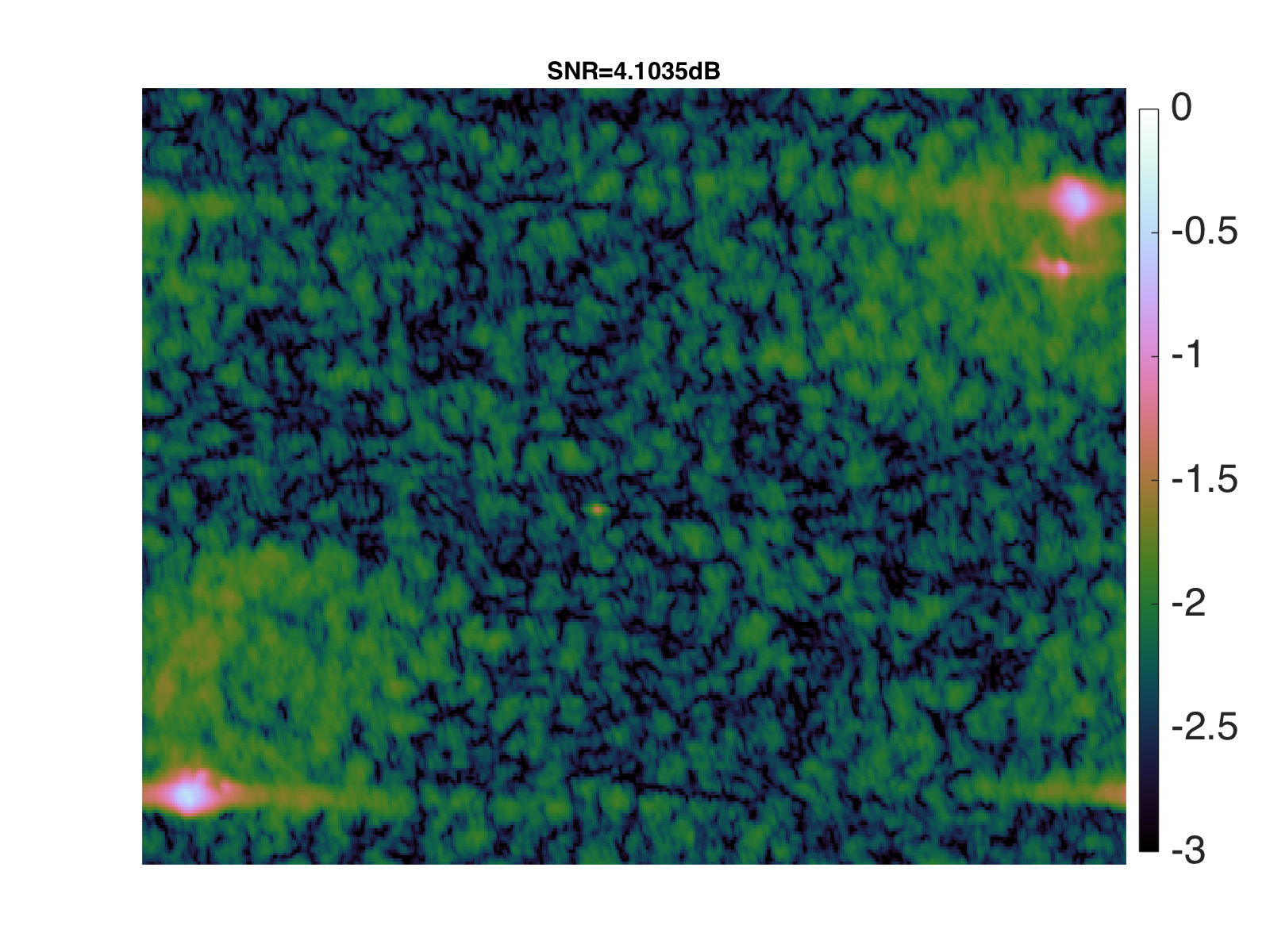} &
		\includegraphics[trim={{.15\linewidth} {.07\linewidth} {.025\linewidth} {.07\linewidth}}, clip, width=0.24\linewidth, height = 0.13\linewidth]
		{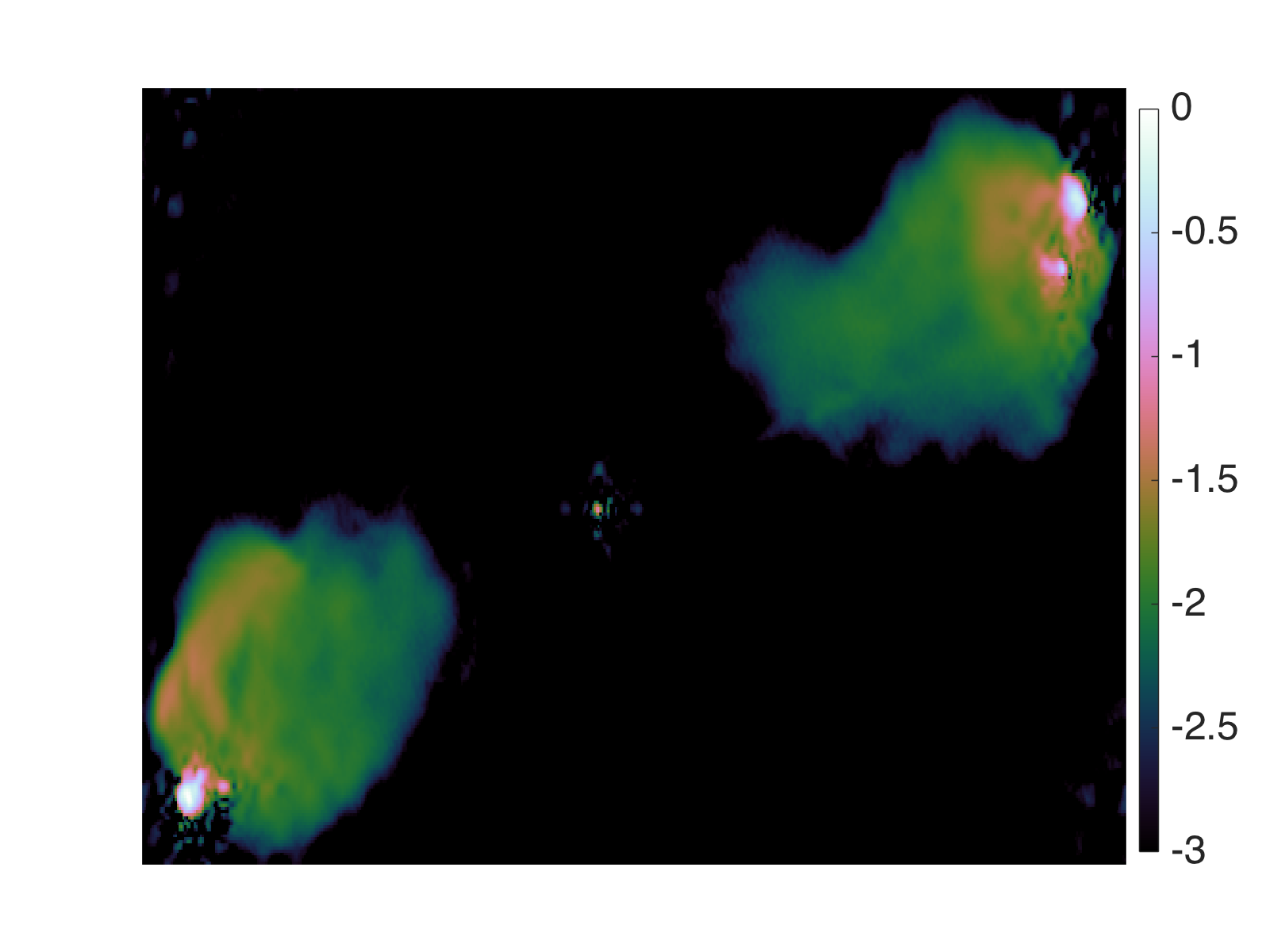}  &
		\includegraphics[trim={{.15\linewidth} {.07\linewidth} {.025\linewidth} {.072\linewidth}}, clip, width=0.24\linewidth, height = 0.13\linewidth]
		{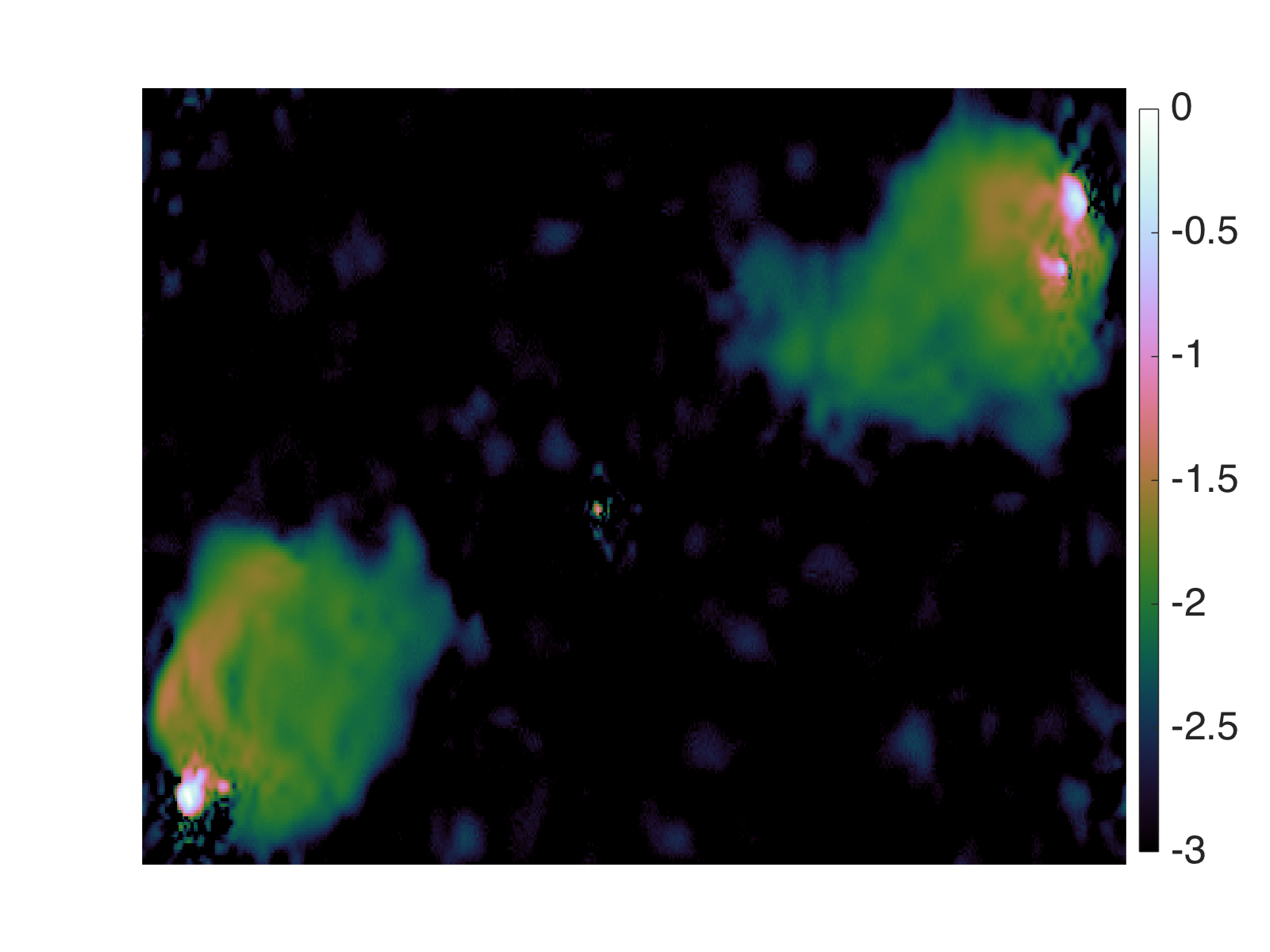} 
		\\
		\includegraphics[trim={{.15\linewidth} {.07\linewidth} {.02\linewidth} {.07\linewidth}}, clip, width=0.24\linewidth, height = 0.21\linewidth]
		{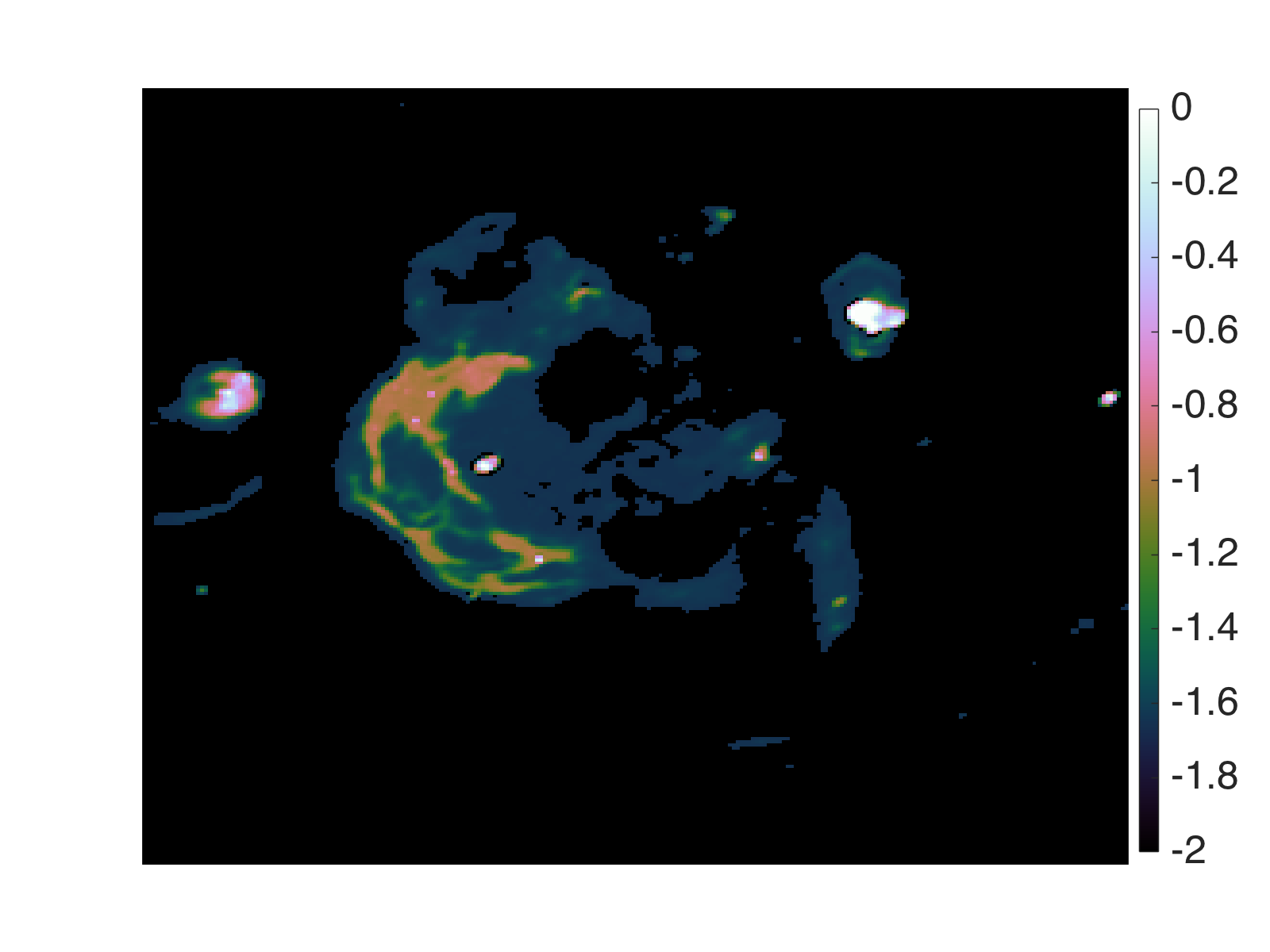}  \put(-130,40){\rotatebox{90}{ W28}} &
		\includegraphics[trim={{.15\linewidth} {.07\linewidth} {.03\linewidth} {.073\linewidth}}, clip, width=0.24\linewidth, height = 0.21\linewidth]
		{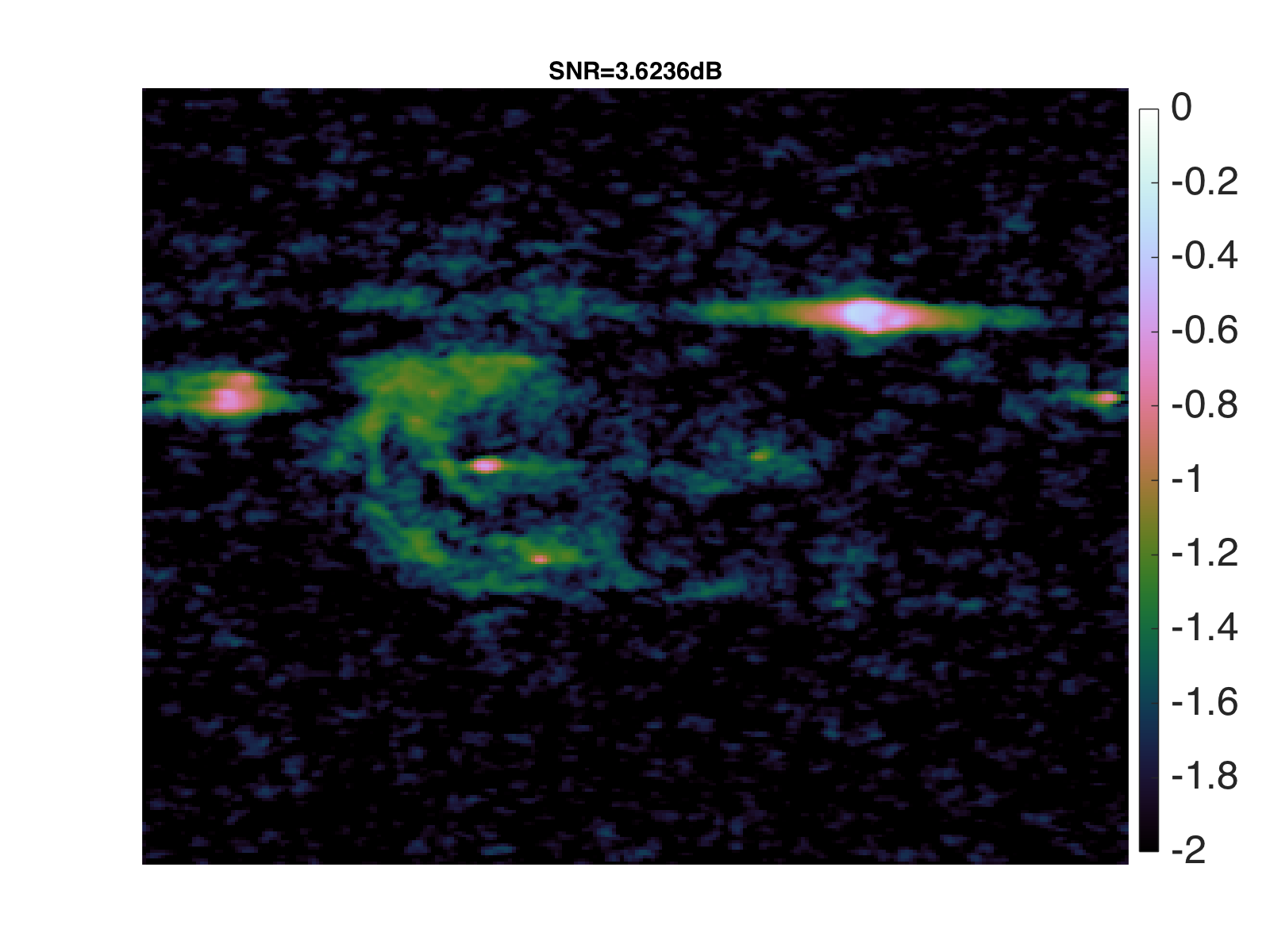} &
		\includegraphics[trim={{.15\linewidth} {.07\linewidth} {.02\linewidth} {.07\linewidth}}, clip, width=0.24\linewidth, height = 0.21\linewidth]
		{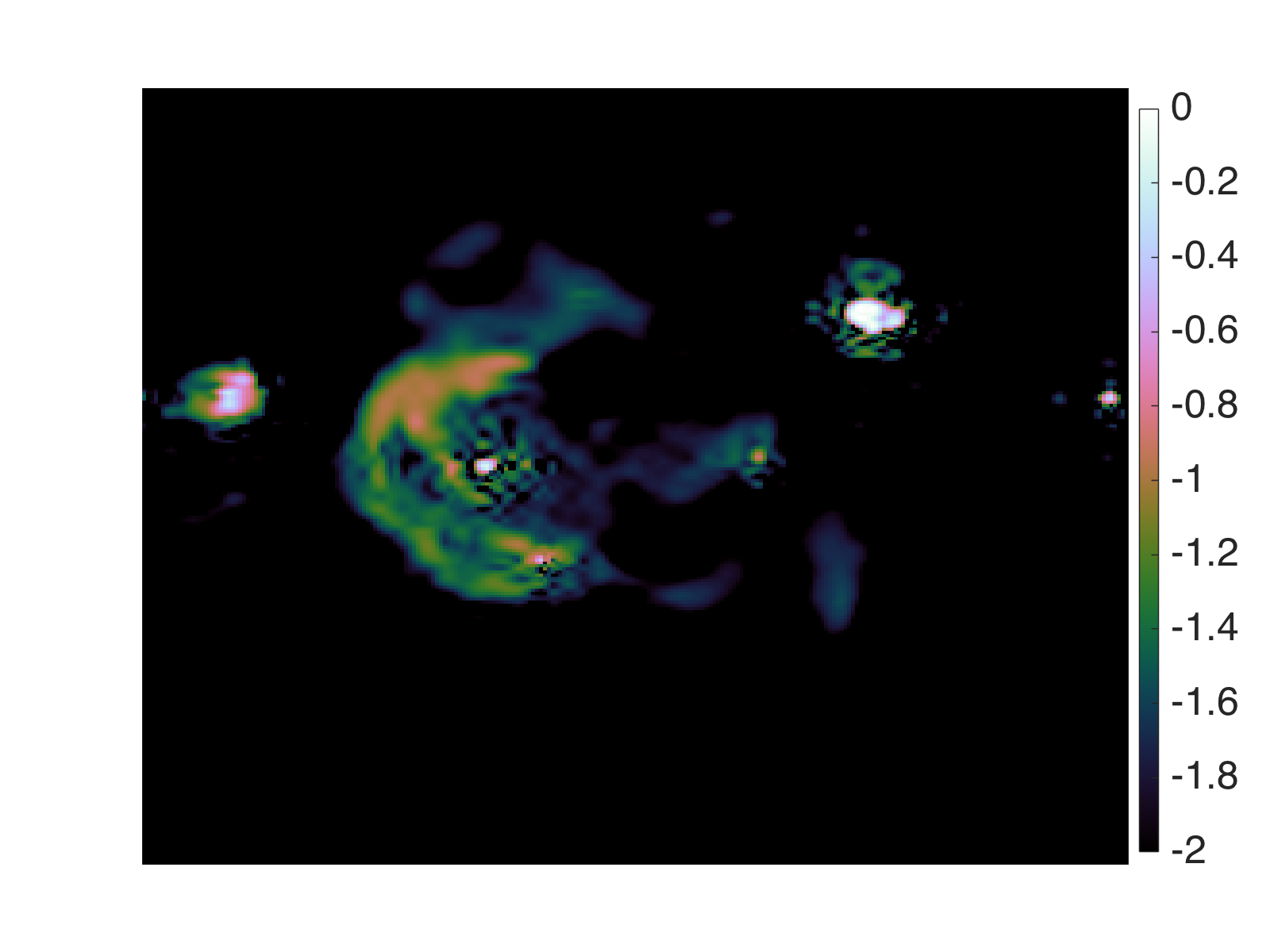} &
		\includegraphics[trim={{.15\linewidth} {.07\linewidth} {.02\linewidth} {.073\linewidth}}, clip, width=0.24\linewidth, height = 0.21\linewidth]
		{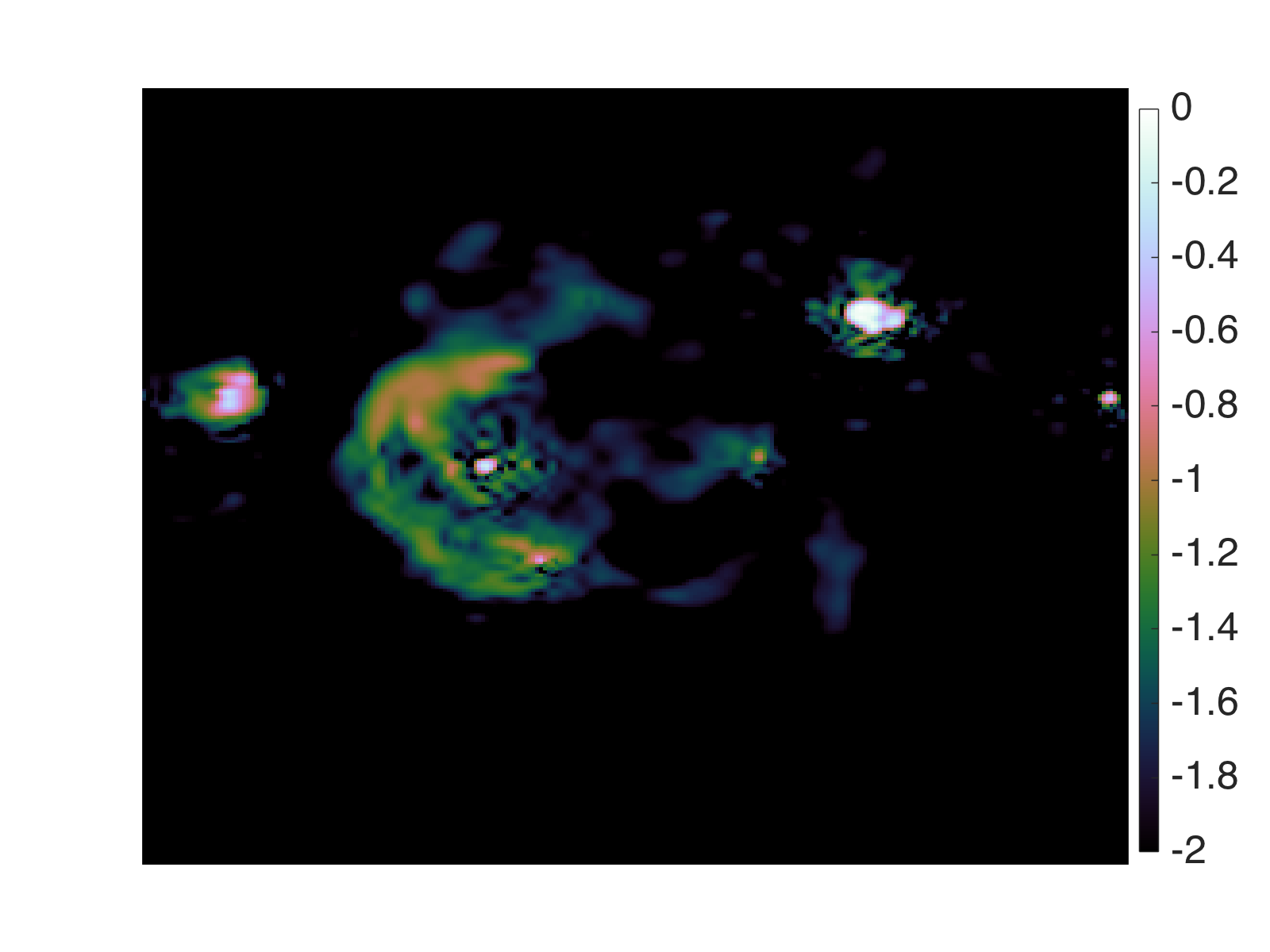} 
		\\
		\includegraphics[trim={{.15\linewidth} {.07\linewidth} {.02\linewidth} {.07\linewidth}}, clip, width=0.24\linewidth, height = 0.21\linewidth]
		{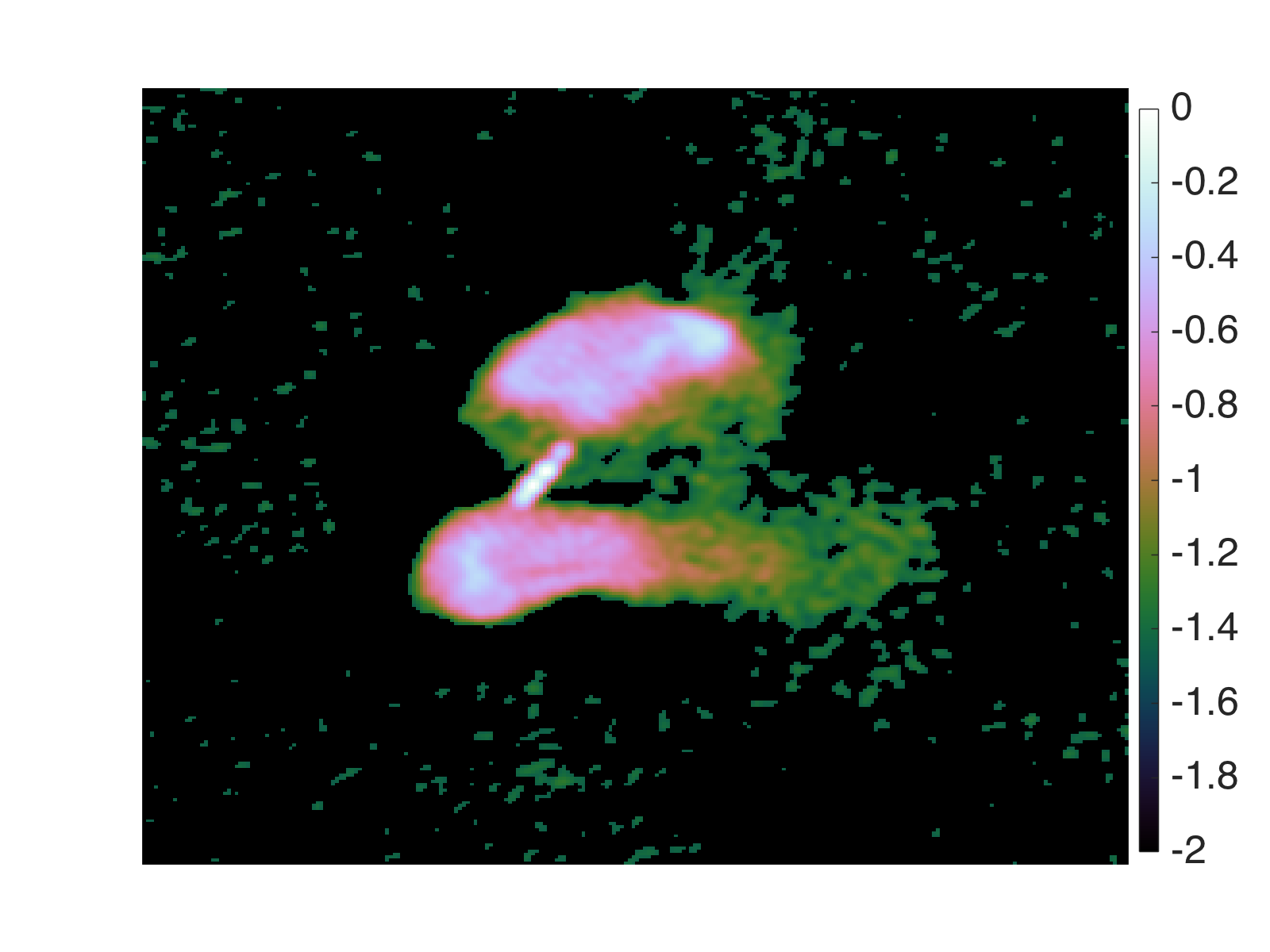} \put(-130,39){\rotatebox{90}{ 3C288}} &
		\includegraphics[trim={{.15\linewidth} {.07\linewidth} {.03\linewidth} {.073\linewidth}}, clip, width=0.24\linewidth, height = 0.21\linewidth]
		{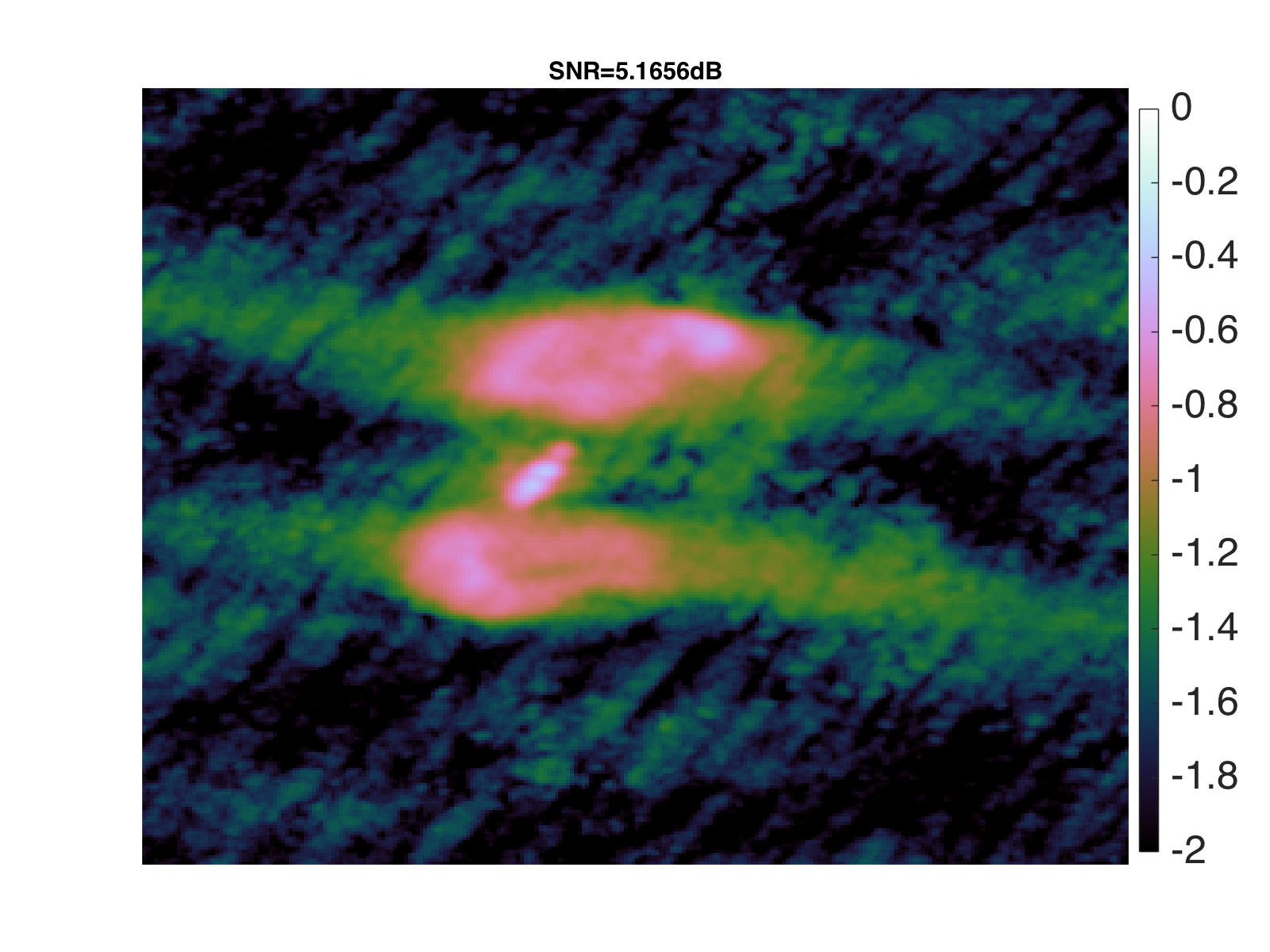} &
		\includegraphics[trim={{.15\linewidth} {.07\linewidth} {.02\linewidth} {.07\linewidth}}, clip, width=0.24\linewidth, height = 0.21\linewidth]
		{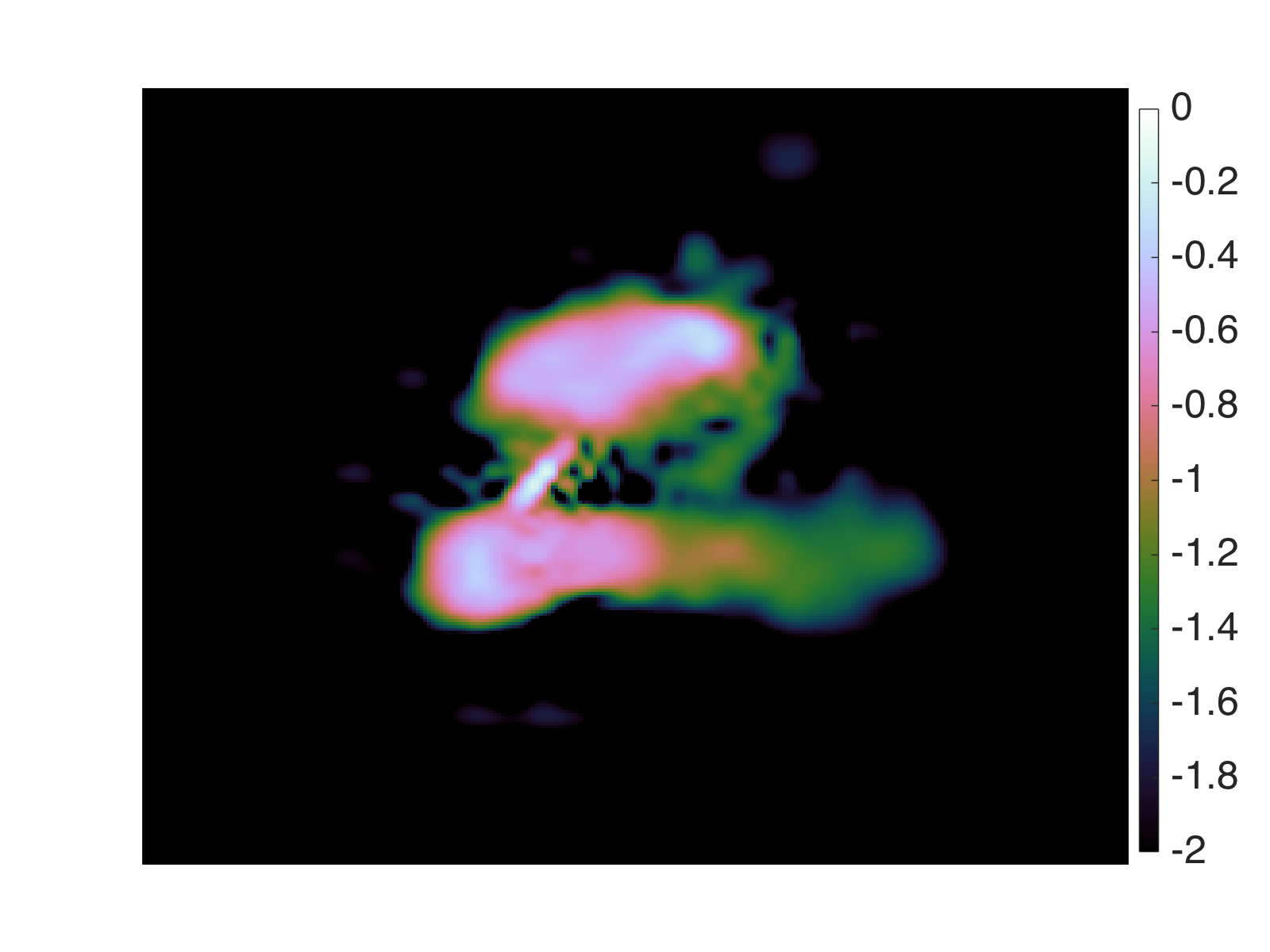} &
		\includegraphics[trim={{.15\linewidth} {.07\linewidth} {.02\linewidth} {.073\linewidth}}, clip, width=0.24\linewidth, height = 0.21\linewidth]
		{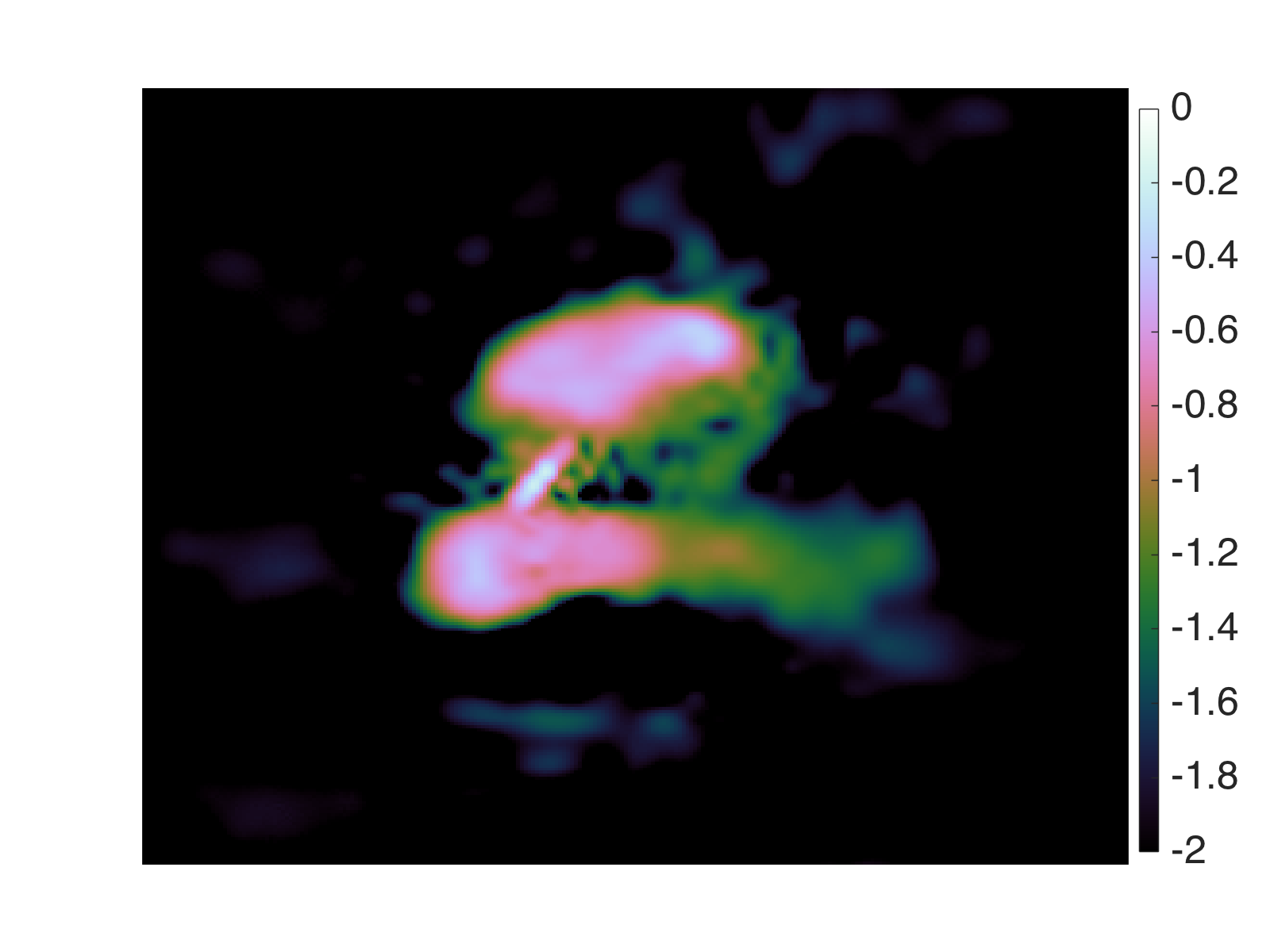} 		
		\\ \vspace{0.05in}\\
		{\small (a) ground truth} & {\small (b) dirty image} & {\small (c) MYULA for analysis model }  & {\small (d) Px-MALA for analysis model }
        \end{tabular}
	\caption{Image reconstructions for Cygnus A (size $256\times 512$), W28 (size $256\times 256$), and 3C288 (size $256\times 256$) (first to third rows). 
	All images are shown in ${\tt log}_{10}$ scale.
	First column: (a) ground truth. Second to forth columns:  (b) dirty images, (c) and (d) point estimators for
	the analysis model \eqref{eqn:ir-un-af} using samples generated by MYULA and Px-MALA, respectively.
	Clearly, consistent results between MYULA and Px-MALA are obtained.  See further discussion in main text.
	}
	\label{fig-others}
\end{figure*}
}
\addtolength{\tabcolsep}{\tabL}
%%%%

%--------
\subsection{Image reconstruction}
%--------
In our first experiment we apply MYULA and Px-MALA to the M31 data and use the samples generated to compute the posterior mean for the synthesis and the analysis models. For comparison, we also report the dirty reconstruction 
obtained directly via inverse Fourier transform of the visibilities $\vect y$.  The dirty image is shown in Figure~\ref{fig-m31}~(b) and compares poorly with the ground truth in Figure~\ref{fig-m31}~(a). The posterior means associated with the models \eqref{eqn:baye-x} and \eqref{eqn:baye-a} obtained with MYULA and Px-MALA are displayed in panels~(c)--(f). All of these results demonstrate accurate and similar reconstruction performance. In detail, MYULA provides slightly superior reconstruction quality. Moreover, as we can see from Figure \ref{fig-m31}, the difference between the results with respect 
to the analysis and synthesis models is negligible (due to an orthogonal basis $\bm{\mathsf{\Psi}}$ being used). Figure \ref{fig-others} shows the results obtained for the Cygnus A, W28, and 3C288 data with the analysis model, observing that these results support the conclusions obtained from the M31 data presented in Figure \ref{fig-m31} (results for the synthesis model are not reported here to avoid redundancy because the results are very similar to those of the analysis model).

In summary, both MYULA and Px-MALA perform well for image reconstruction and produce accurate point estimation results. MYULA provides slightly superior reconstruction performance. This is related to the fact that while Px-MALA has more accurate asymptotic properties than MYULA, the superior convergence properties of MYULA mean that it performs better in practice for a fixed number of samples. Furthermore, to generate the same number of samples, MYULA requires approximately half the computation time of Px-MALA; see Table \ref{tab:time} for the CPU time cost in detail.

%%%%
\addtolength{\tabcolsep}{-\tabL}
{ \renewcommand{\arraystretch}{0.0}
\begin{figure*}
	\centering
	\begin{tabular}{cccc}
		\includegraphics[trim={{.19\linewidth} {.07\linewidth} {.06\linewidth} {.017\linewidth}}, clip, width=0.245\linewidth, height = 0.215\linewidth]
		{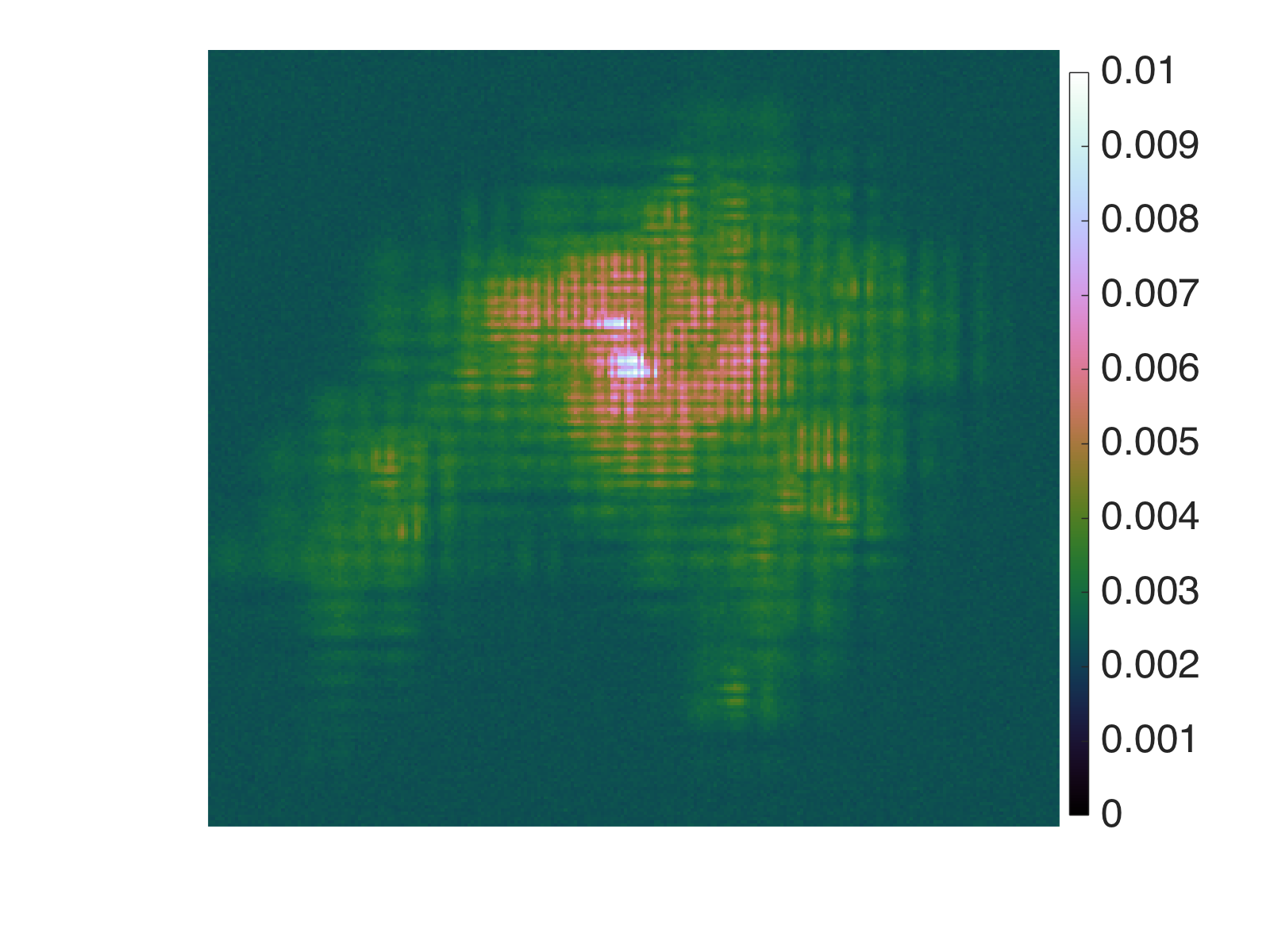} \put(-131,40){\rotatebox{90}{ M31}} &
		\includegraphics[trim={{.19\linewidth} {.07\linewidth} {.06\linewidth} {.017\linewidth}}, clip, width=0.245\linewidth, height = 0.215\linewidth]
		{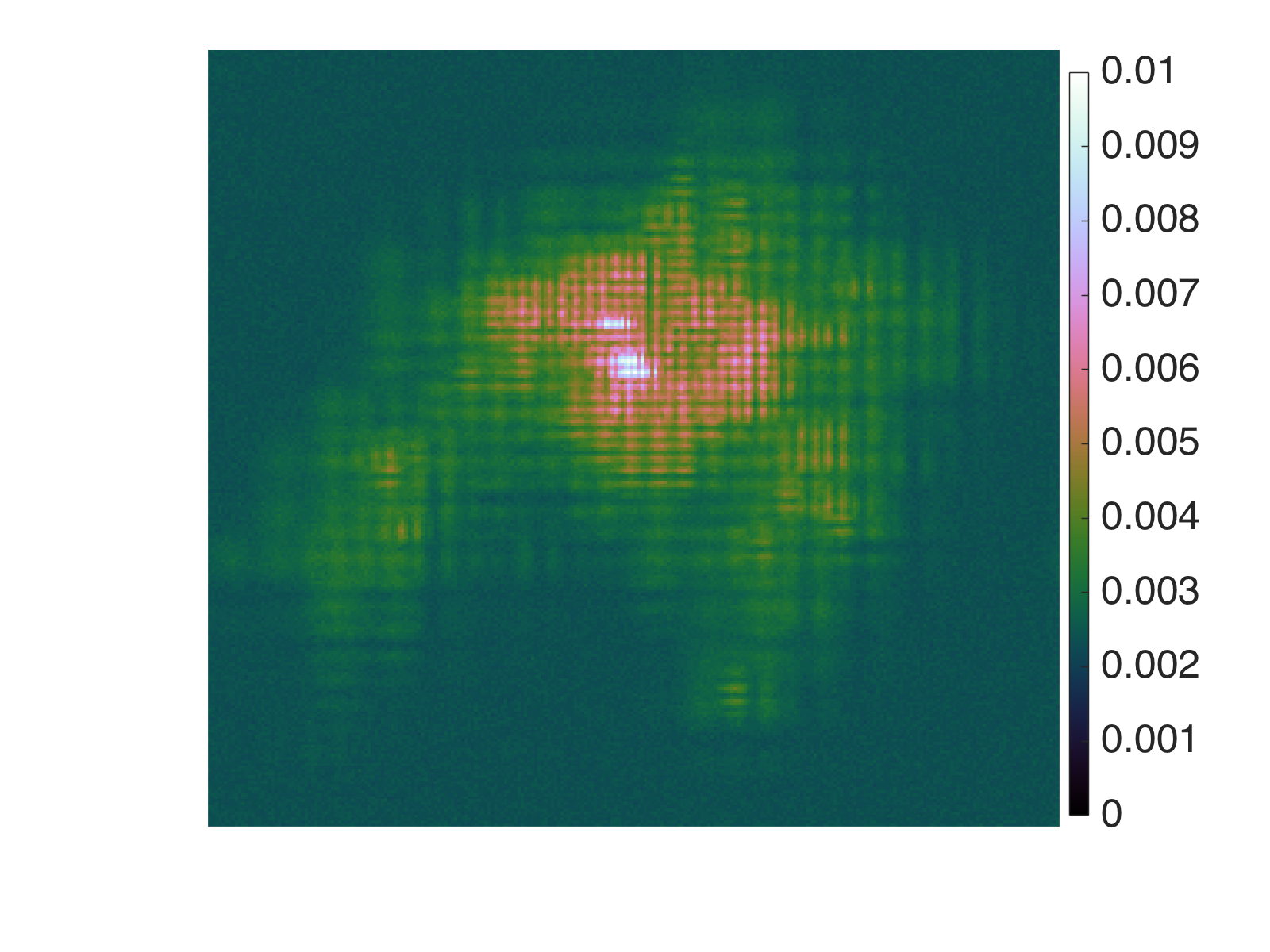} &
		\includegraphics[trim={{.19\linewidth} {.07\linewidth} {.06\linewidth} {.0155\linewidth}}, clip, width=0.245\linewidth, height = 0.215\linewidth]
		{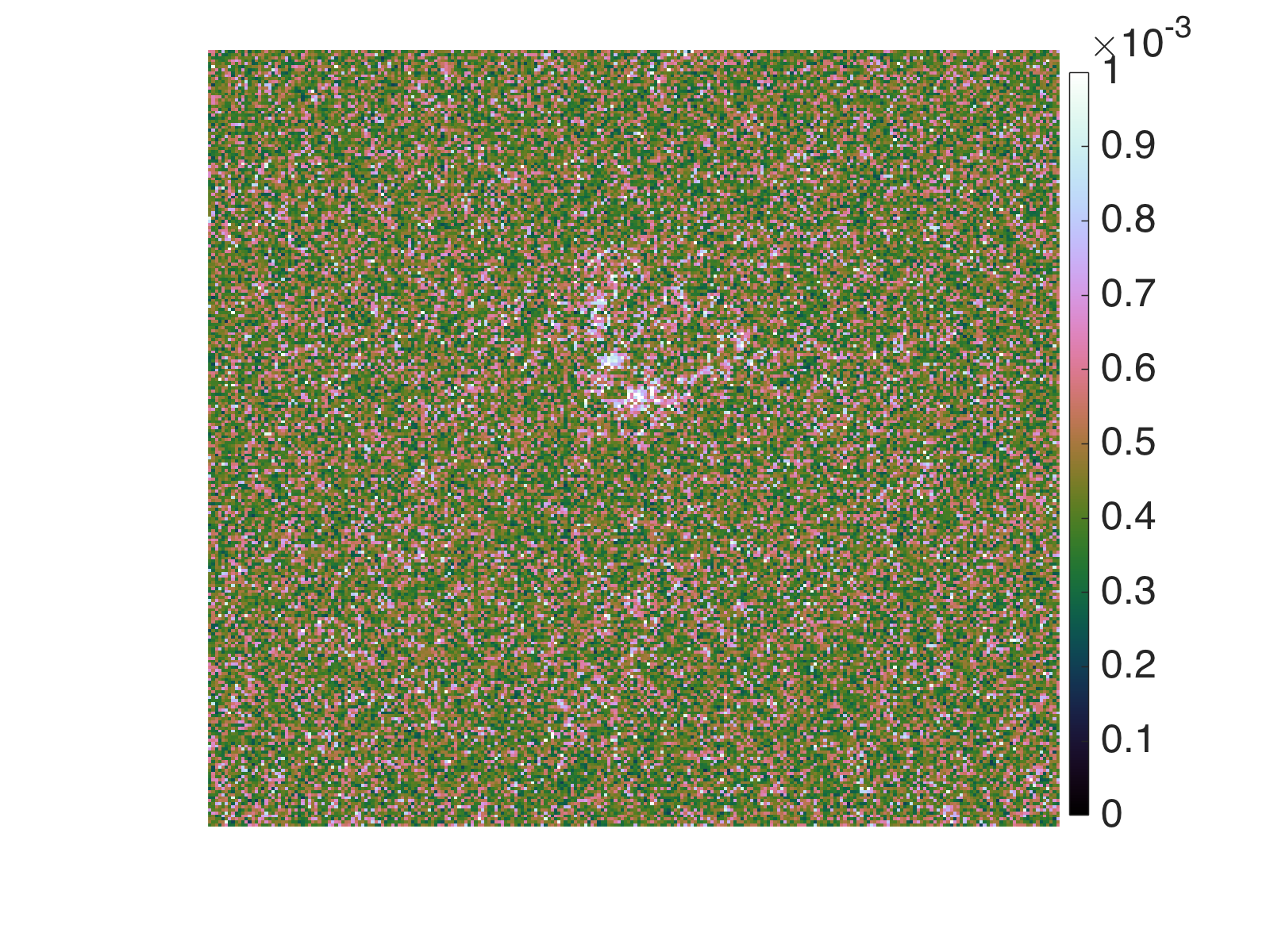} &
		\includegraphics[trim={{.19\linewidth} {.07\linewidth} {.06\linewidth} {.0155\linewidth}}, clip, width=0.245\linewidth, height = 0.215\linewidth]
		{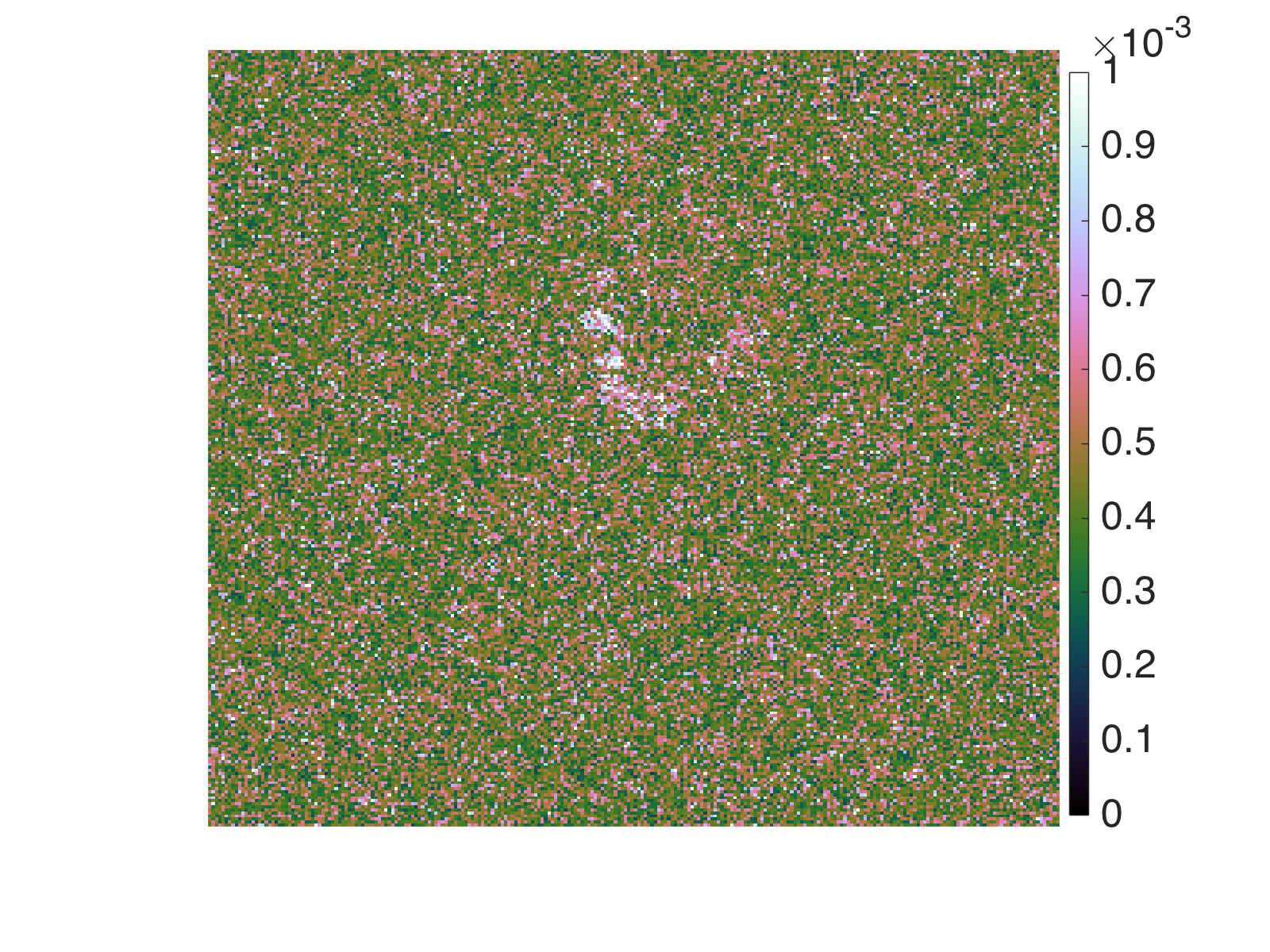} 
		\\
		\includegraphics[trim={{.35\linewidth} {.07\linewidth} {.015\linewidth} {.005\linewidth}}, clip, width=0.245\linewidth, height = 0.14\linewidth]
		{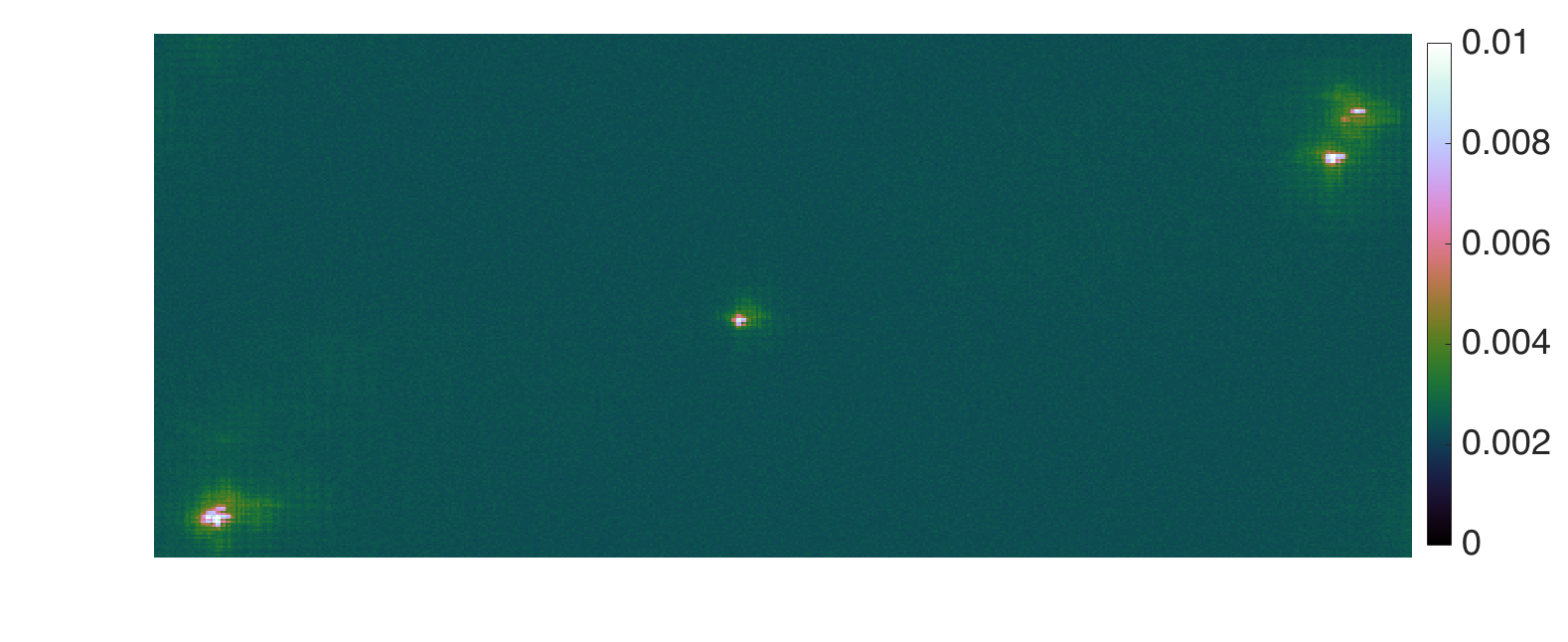}  \put(-131,11){\rotatebox{90}{ Cygnus A}} &
		\includegraphics[trim={{.35\linewidth} {.07\linewidth} {.015\linewidth} {.005\linewidth}}, clip, width=0.245\linewidth, height = 0.14\linewidth]
		{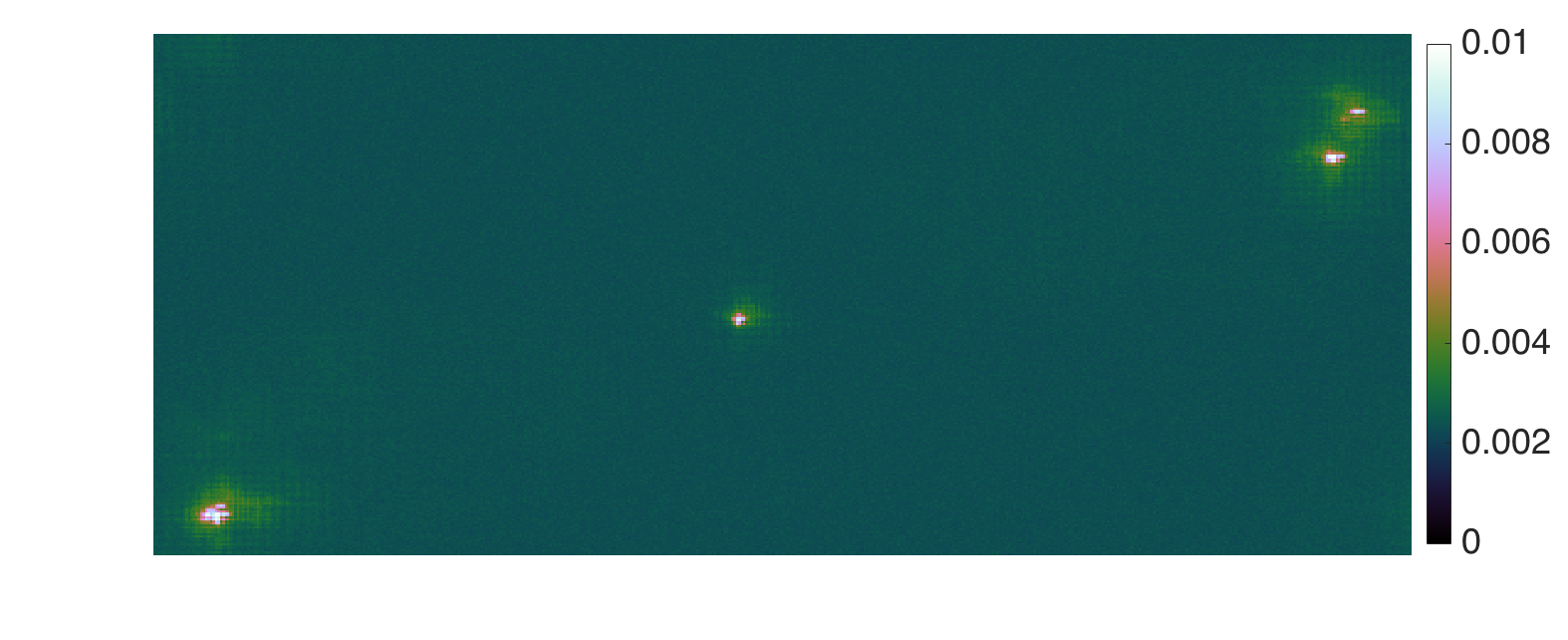} &
		\includegraphics[trim={{.35\linewidth} {.07\linewidth} {.015\linewidth} {.005\linewidth}}, clip, width=0.245\linewidth, height = 0.14\linewidth]
		{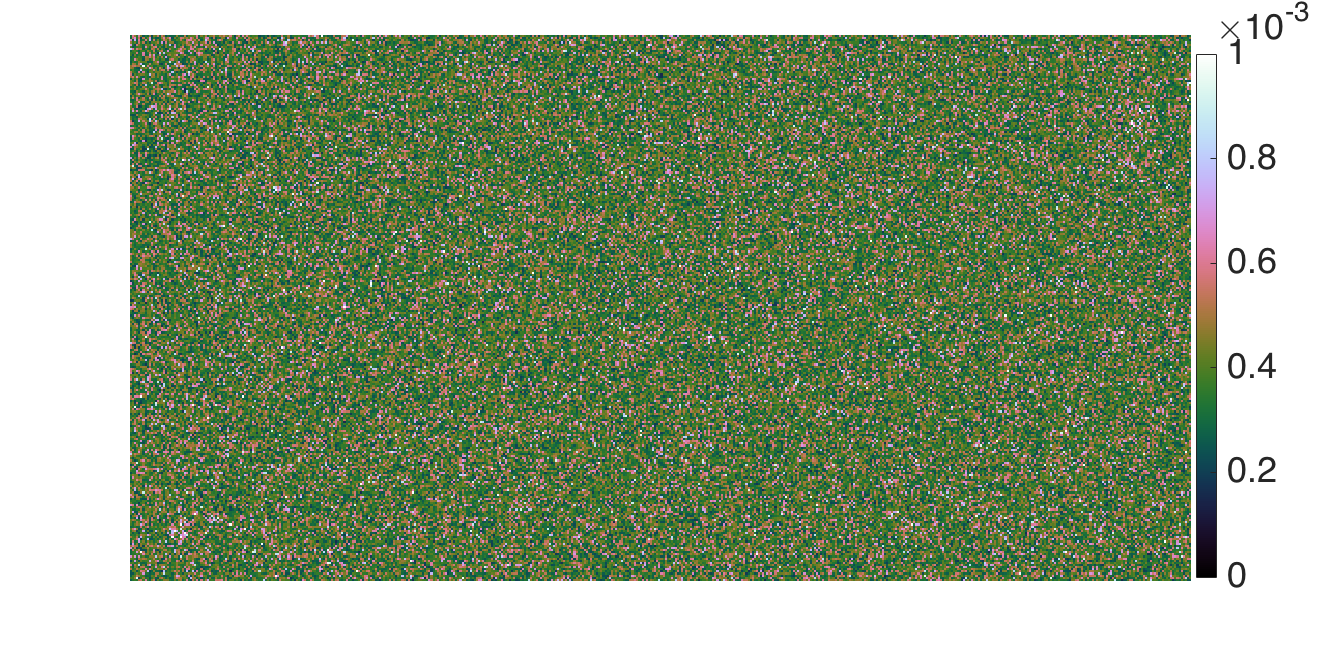} &
		\includegraphics[trim={{.35\linewidth} {.07\linewidth} {.015\linewidth} {.005\linewidth}}, clip, width=0.245\linewidth, height = 0.14\linewidth]
		{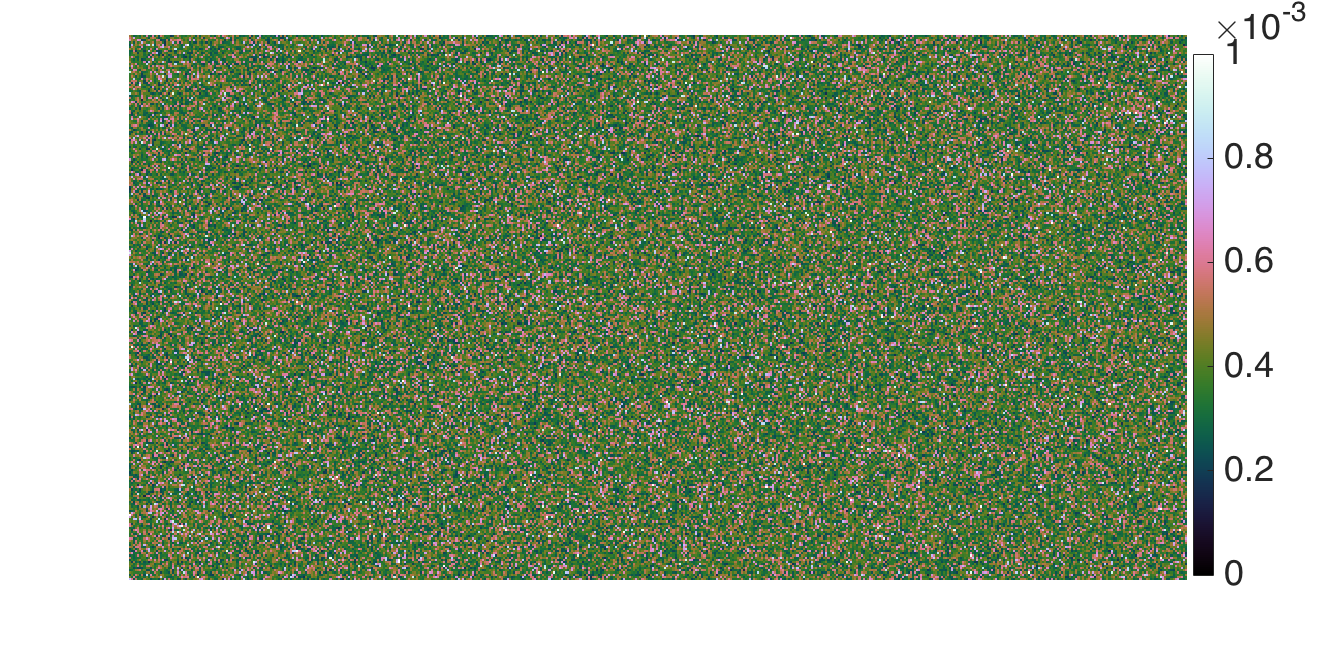} 
		\\	
		\includegraphics[trim={{.19\linewidth} {.07\linewidth} {.06\linewidth} {.017\linewidth}}, clip, width=0.245\linewidth, height = 0.215\linewidth]
		{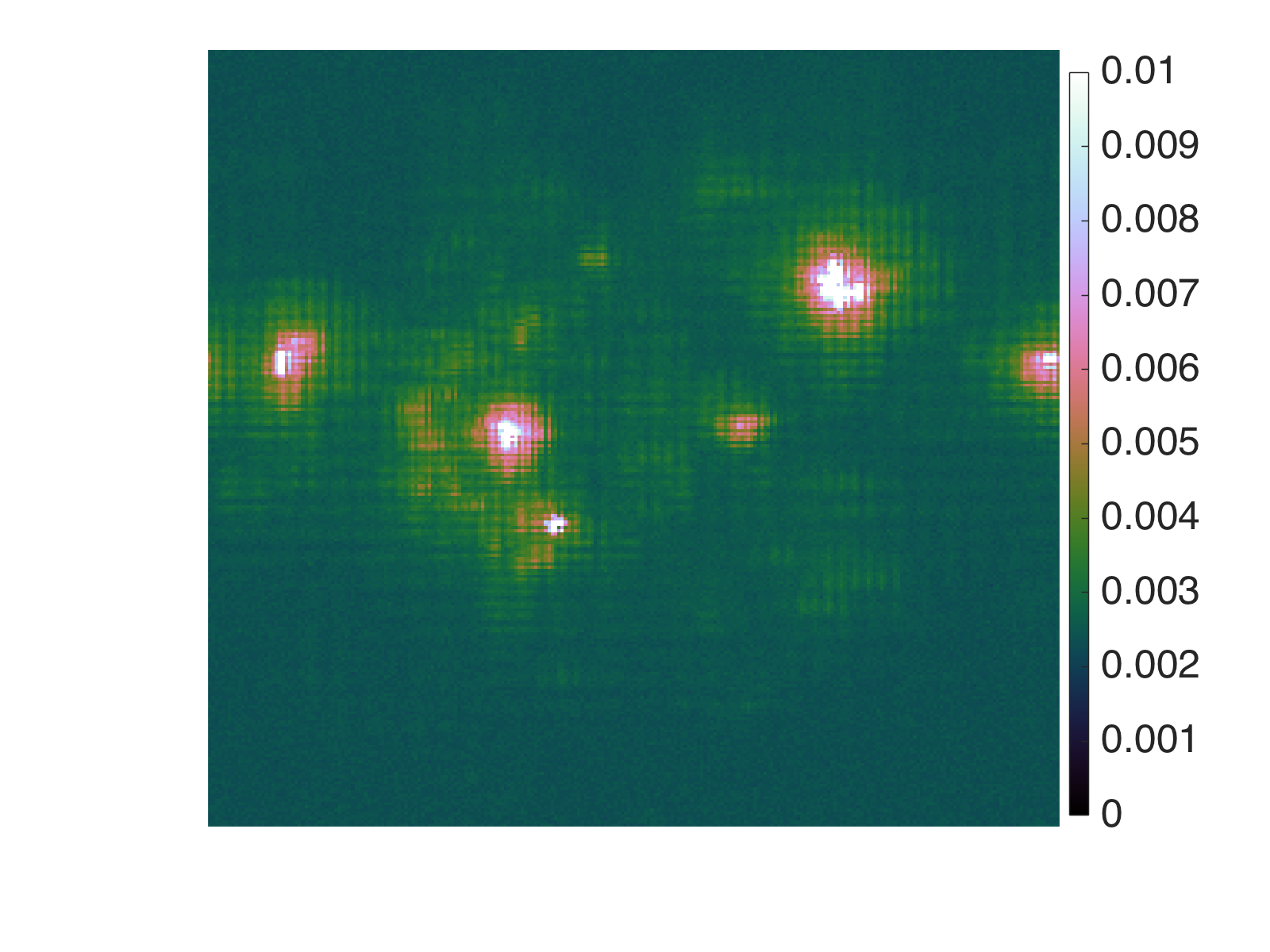} \put(-131,40){\rotatebox{90}{ W28}}  &
		\includegraphics[trim={{.19\linewidth} {.07\linewidth} {.06\linewidth} {.017\linewidth}}, clip, width=0.245\linewidth, height = 0.215\linewidth]
		{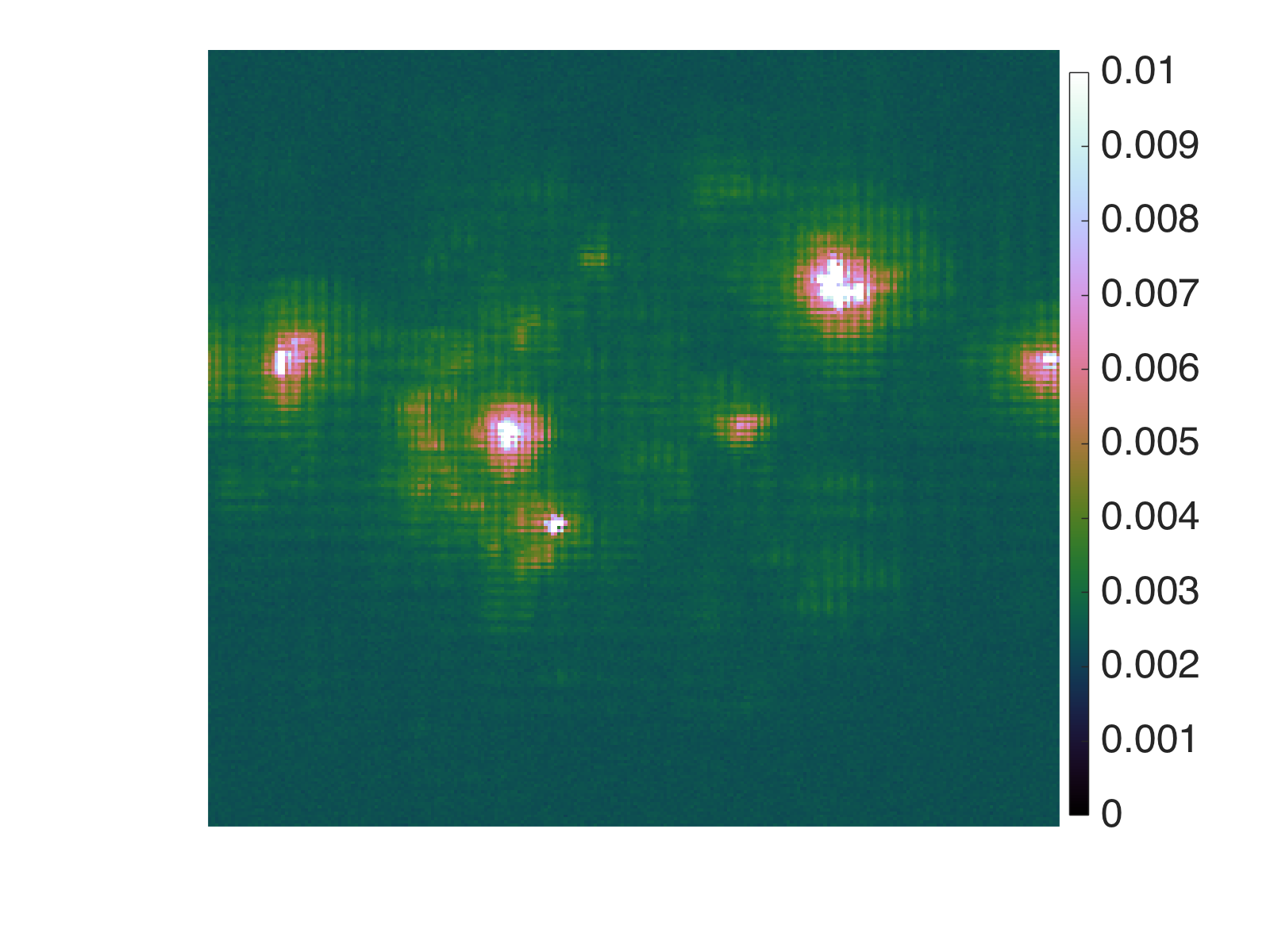} &
		\includegraphics[trim={{.19\linewidth} {.07\linewidth} {.06\linewidth} {.0155\linewidth}}, clip, width=0.245\linewidth, height = 0.215\linewidth]
		{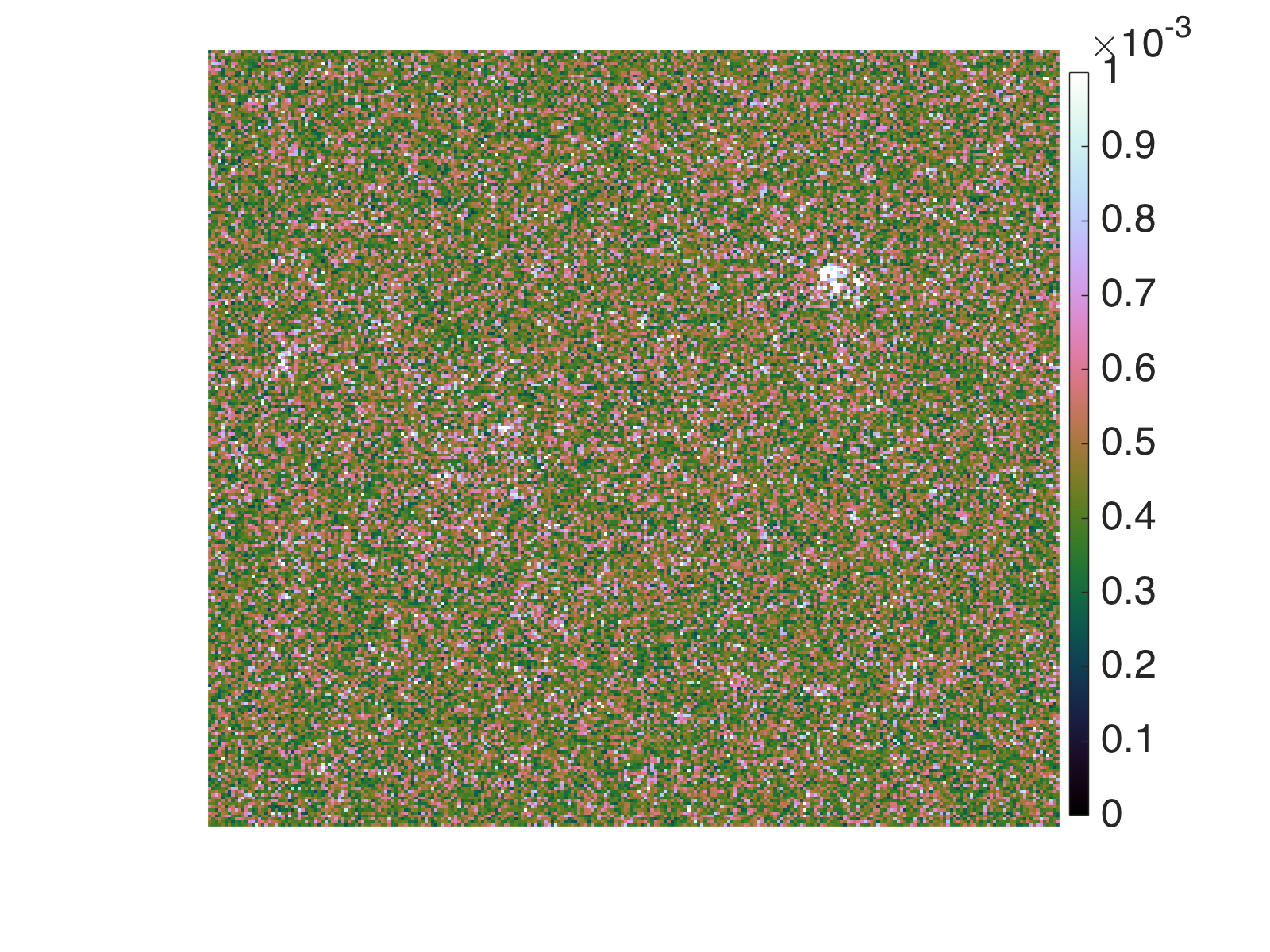} &
		\includegraphics[trim={{.19\linewidth} {.07\linewidth} {.06\linewidth} {.0155\linewidth}}, clip, width=0.245\linewidth, height = 0.215\linewidth]
		{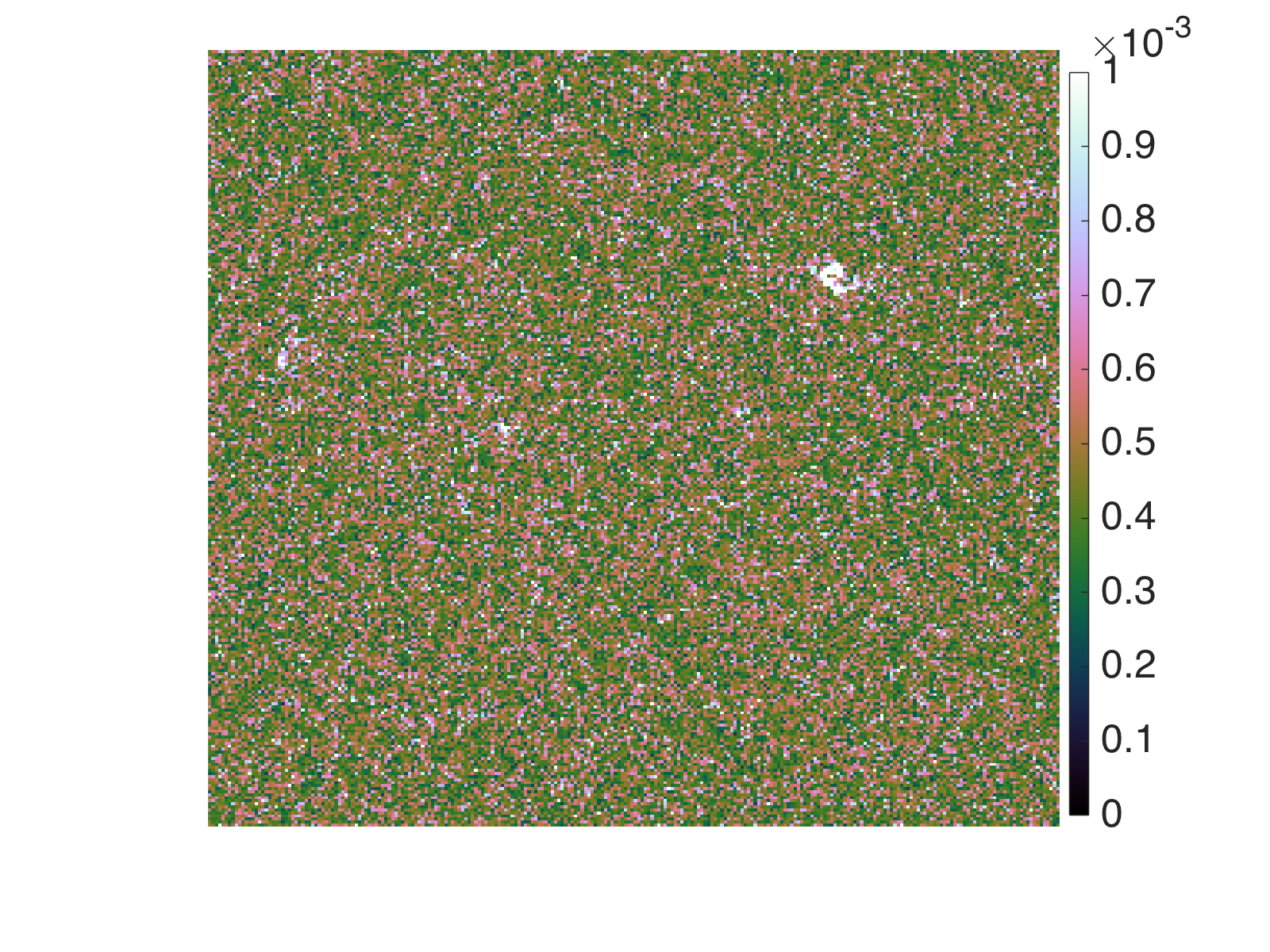}
		\\
		\includegraphics[trim={{.19\linewidth} {.07\linewidth} {.06\linewidth} {.017\linewidth}}, clip, width=0.245\linewidth, height = 0.215\linewidth]
		{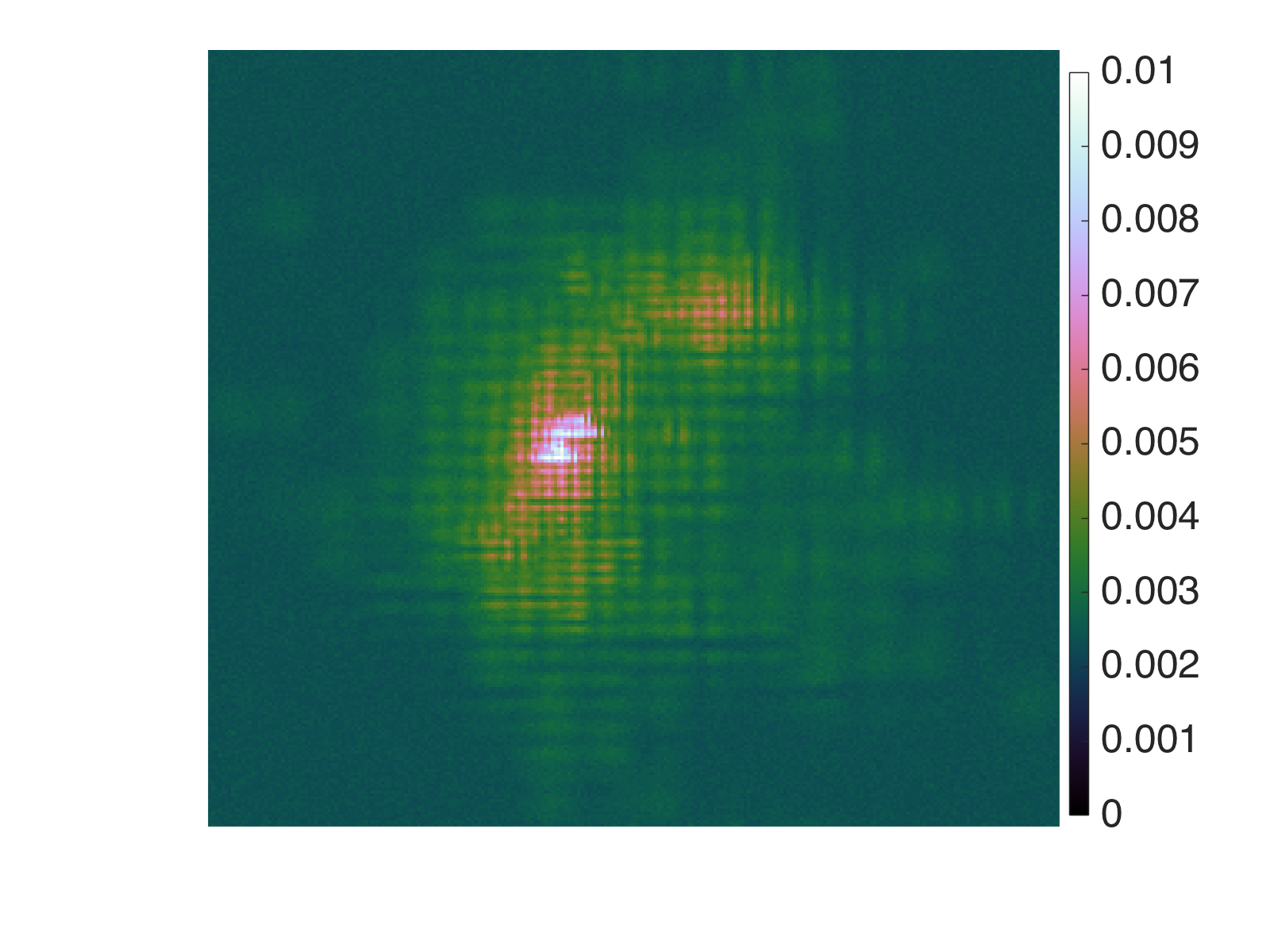} \put(-131,38){\rotatebox{90}{ 3C288}} &
		\includegraphics[trim={{.19\linewidth} {.07\linewidth} {.06\linewidth} {.017\linewidth}}, clip, width=0.245\linewidth, height = 0.215\linewidth]
		{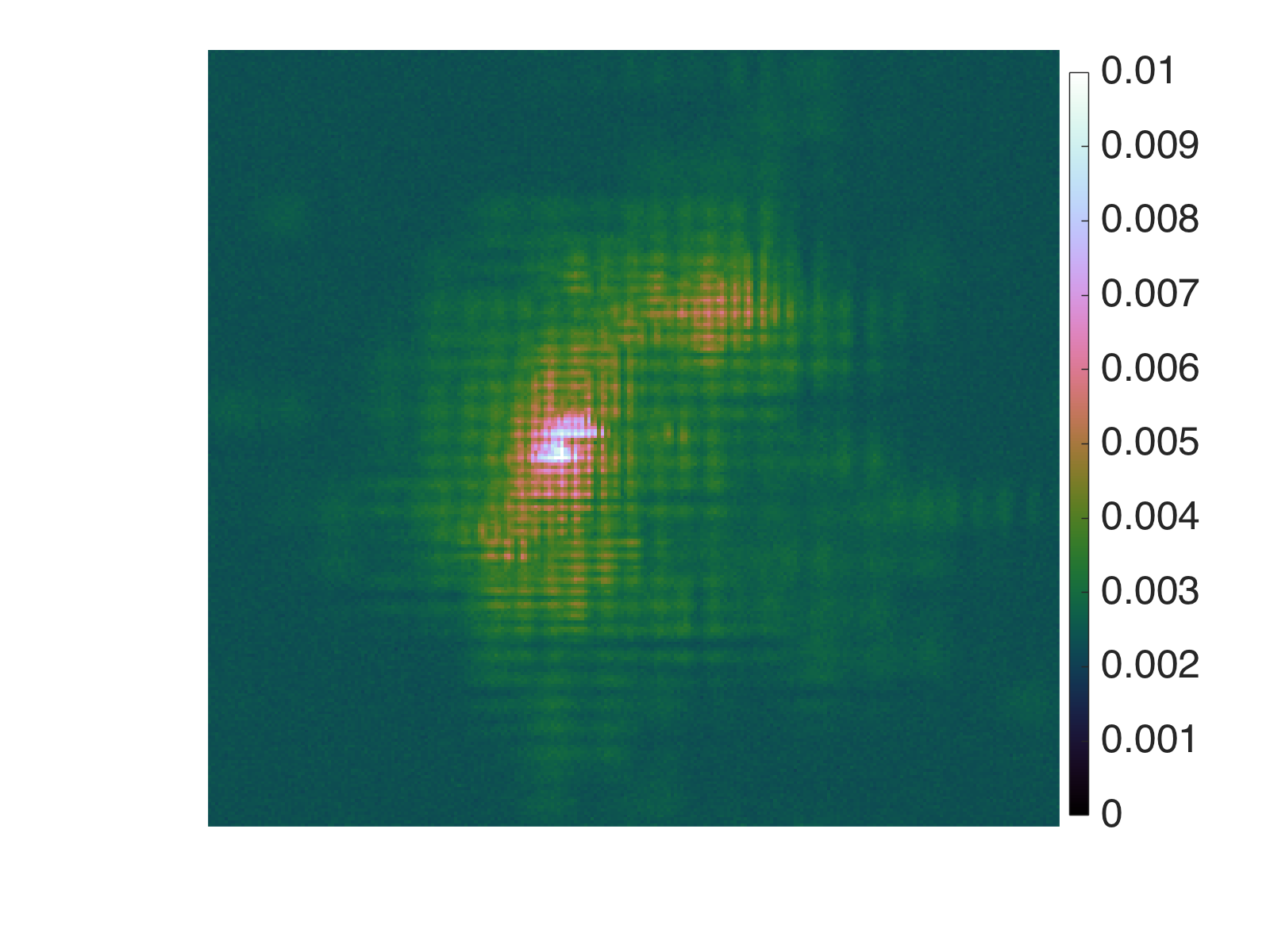} &
		\includegraphics[trim={{.19\linewidth} {.07\linewidth} {.06\linewidth} {.0155\linewidth}}, clip, width=0.245\linewidth, height = 0.215\linewidth]
		{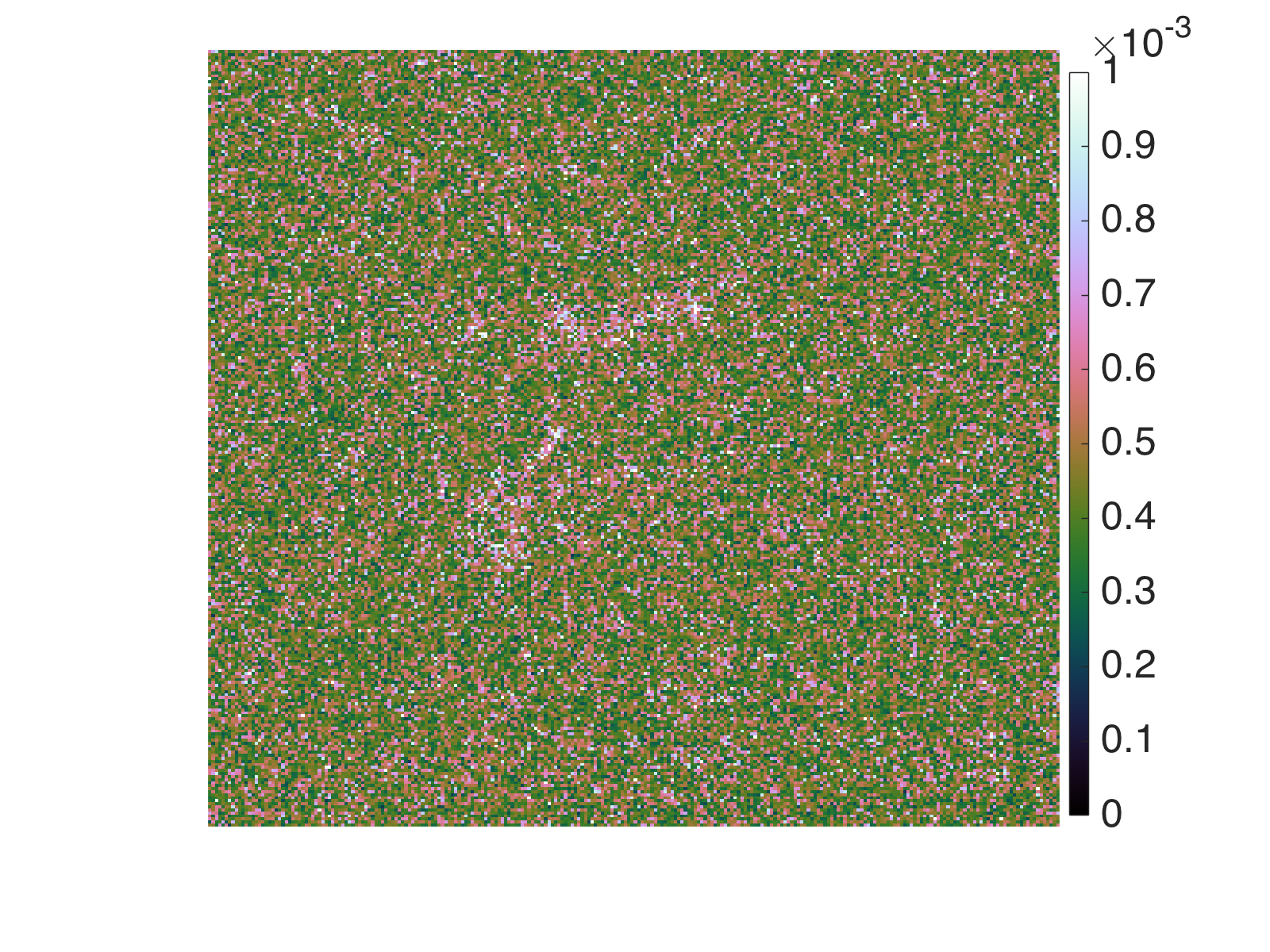} &
		\includegraphics[trim={{.19\linewidth} {.07\linewidth} {.06\linewidth} {.0155\linewidth}}, clip, width=0.245\linewidth, height = 0.215\linewidth]
		{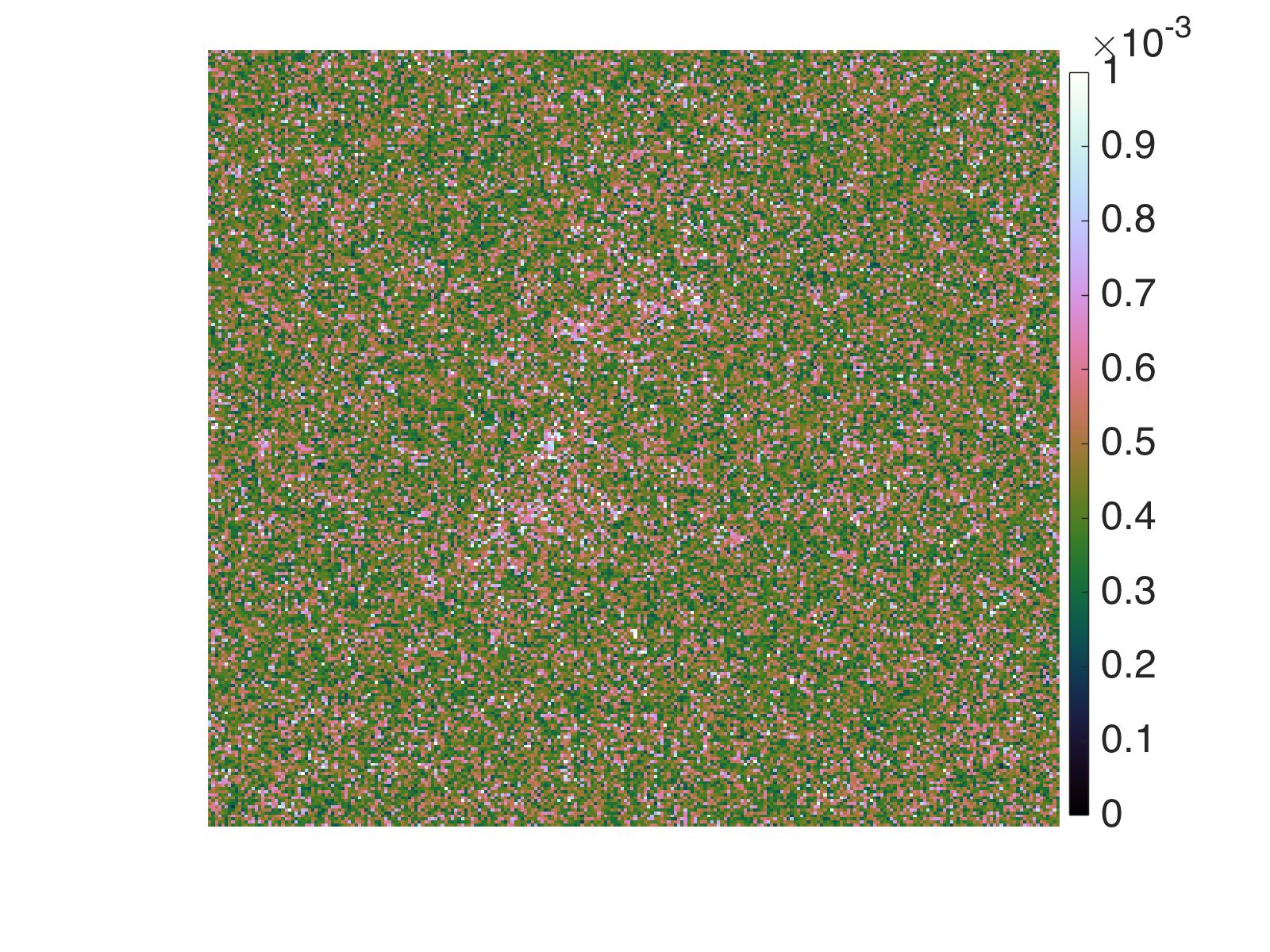}
		\\
		{\small (a) MYULA, analysis model} & {\small (b) MYULA, synthesis model} & {\small (c) Px-MALA, analysis model} & {\small (d) Px-MALA, synthesis model}
        \end{tabular}
	\caption{Length of pixel-wise credible intervals (95\% credible level). First to fourth rows are results for the images M31, Cygnus A, W28, and 3C288, respectively. 
	Columns (a) and (b) are results obtained with samples generated by MYULA using the analysis and synthesis models \eqref{eqn:ir-un-af} and \eqref{eqn:ir-un-sf}, respectively; columns (c) and (d) correspond to results obtained with Px-MALA. The results show that MYULA produces wider and smoother credible intervals, compared to those recovered by Px-MALA.  See further discussion in main text.
	}
	\label{fig-all-ci-pixel}
\end{figure*}
}
\addtolength{\tabcolsep}{\tabL}
%%%%

%--------
\subsection{Pixel-wise credible intervals}
%--------
Figure \ref{fig-all-ci-pixel} reports the length of the pixel-wise credible intervals \eqref{eqn:cr-local} for the M31, Cygnus A, W28, and 3C288 data, computed with MYULA and Px-MALA, and for the analysis and the synthesis models \eqref{eqn:baye-x} and \eqref{eqn:baye-a}. We observe that in this case MYULA delivers significantly better results than Px-MALA; the difference in the estimates illustrates clearly the bias-variance tradeoff related to the MH step in Px-MALA. Precisely, MYULA produces stable smooth estimates with low estimation variance, but which suffer from some estimation bias and overestimates uncertainties as a result. If necessary, this bias can be reduced by decreasing the value of $\lambda$. Conversely, the estimates obtained with Px-MALA are unstable and suffer from high estimation variance; however, they do not exhibit a noticeable bias as this is corrected by the MH step. Note that the amount of bias and variance observed are not universal properties of the MYULA and Px-MALA chains. They depend on the quantities that are estimated, and this is why they are visible in the marginal quantiles but not on the posterior means reported in Figure \ref{fig-others}. 

Furthermore, by inspecting Figure \ref{fig-all-ci-pixel}  we observe that the pixels close to object boundaries have wider credible intervals than the pixels in homogenous regions. This is related to the fact that there is uncertainty about the high frequency components of the image because of the sampling profile (see Figure \ref{fig-mask}). Similarly, we observe regular oscillations related to frequencies that are not measured by the sampling profile. Finally, as expected, we note that the analysis and synthesis models produce similar results.

%--------
\subsection{HPD credibility regions}
\begin{figure*}
	\centering
	\begin{tabular}{cccc}
		\includegraphics[trim={{.02\linewidth} {.05\linewidth} {.1\linewidth} {.01\linewidth}}, clip, width=0.245\linewidth, height = 0.21\linewidth]
		{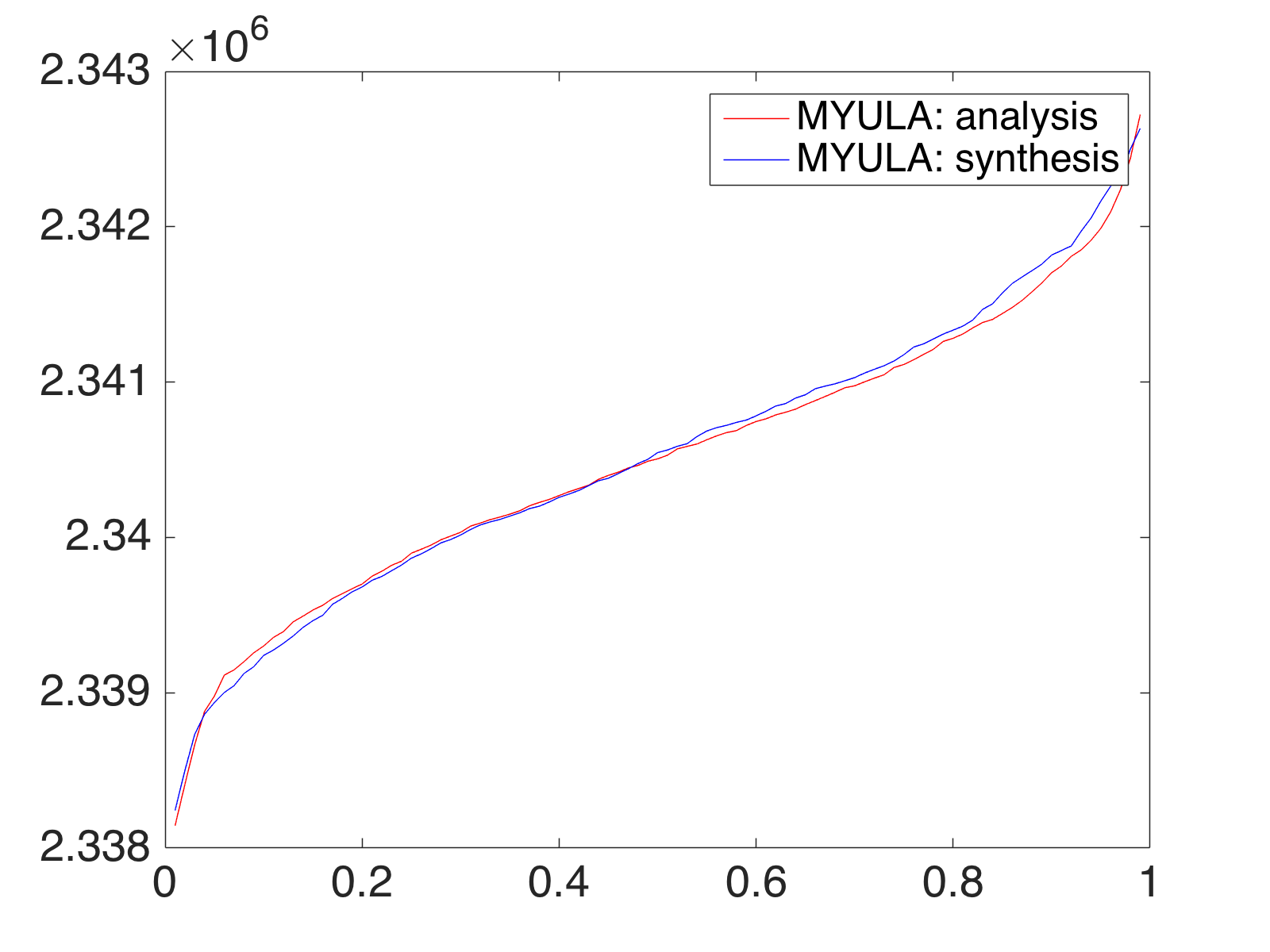}  \put(-60,-4){\tiny $1-\alpha$} &
		\includegraphics[trim={{.02\linewidth} {.05\linewidth} {.1\linewidth} {.01\linewidth}}, clip, width=0.245\linewidth, height = 0.21\linewidth]
		{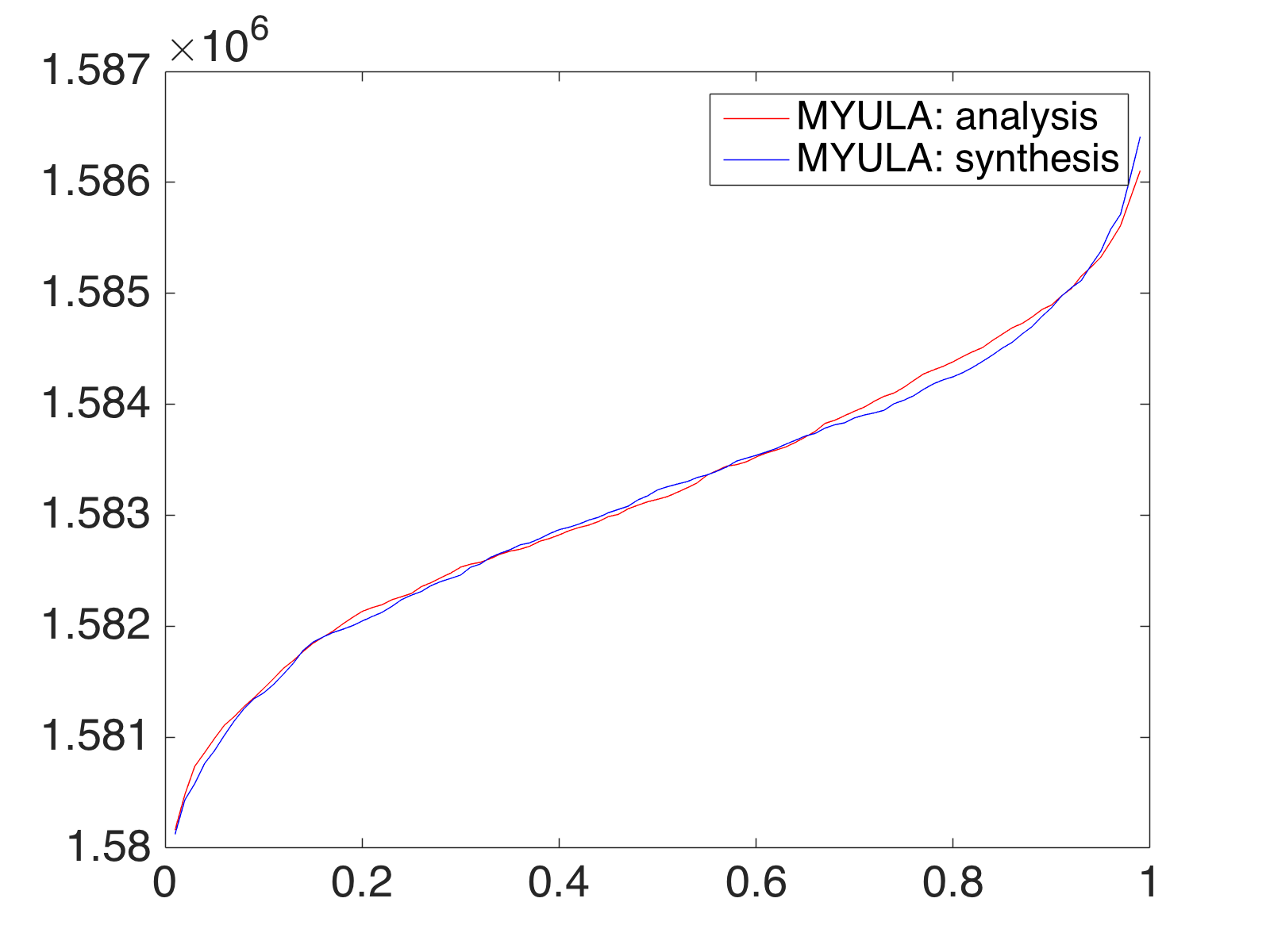}  \put(-60,-4){\tiny $1-\alpha$} &
		\includegraphics[trim={{.02\linewidth} {.05\linewidth} {.1\linewidth} {.01\linewidth}}, clip, width=0.245\linewidth, height = 0.21\linewidth]
		{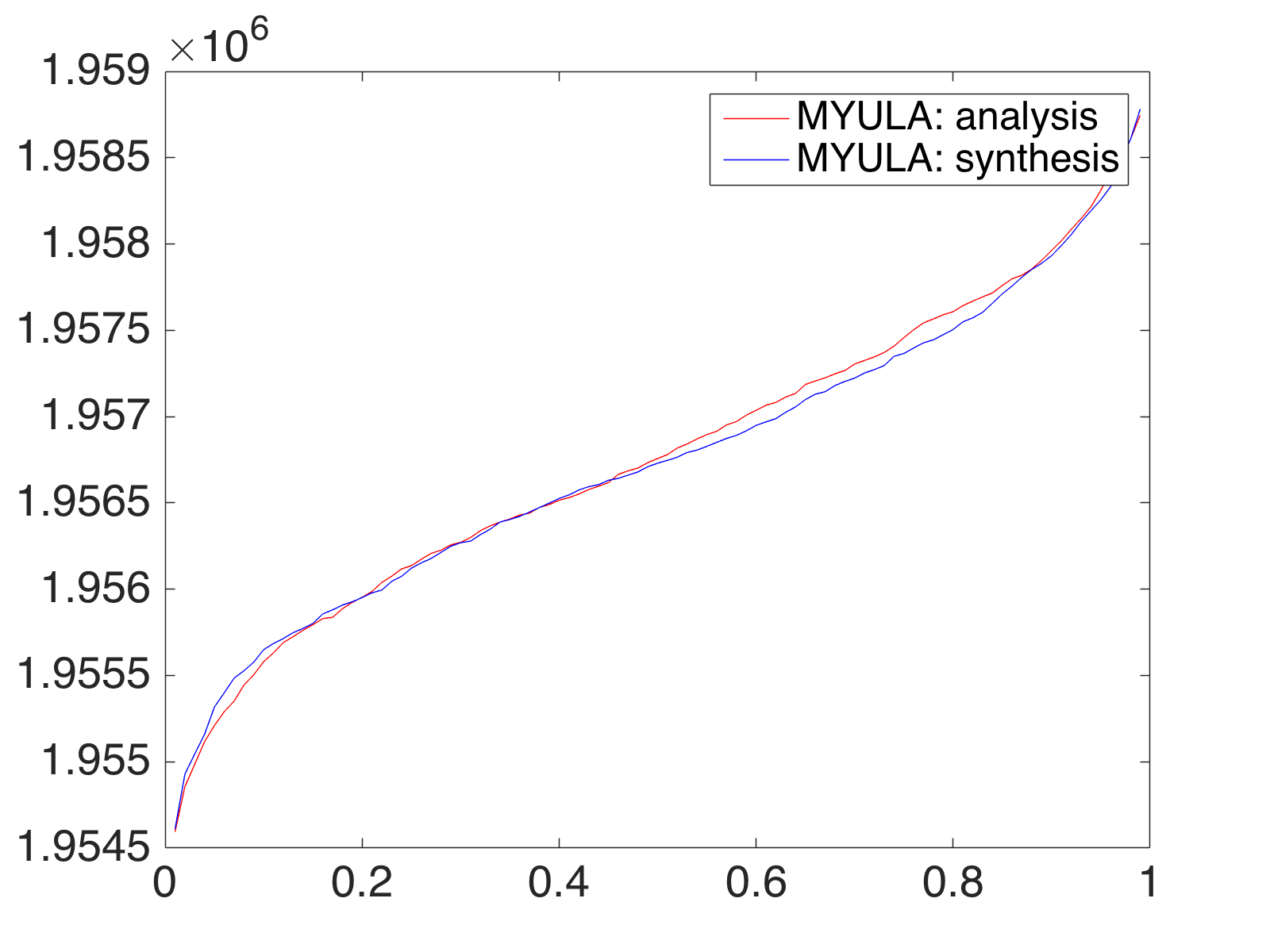}  \put(-60,-4){\tiny $1-\alpha$} &
		\includegraphics[trim={{.02\linewidth} {.05\linewidth} {.1\linewidth} {.01\linewidth}}, clip, width=0.245\linewidth, height = 0.21\linewidth]
		{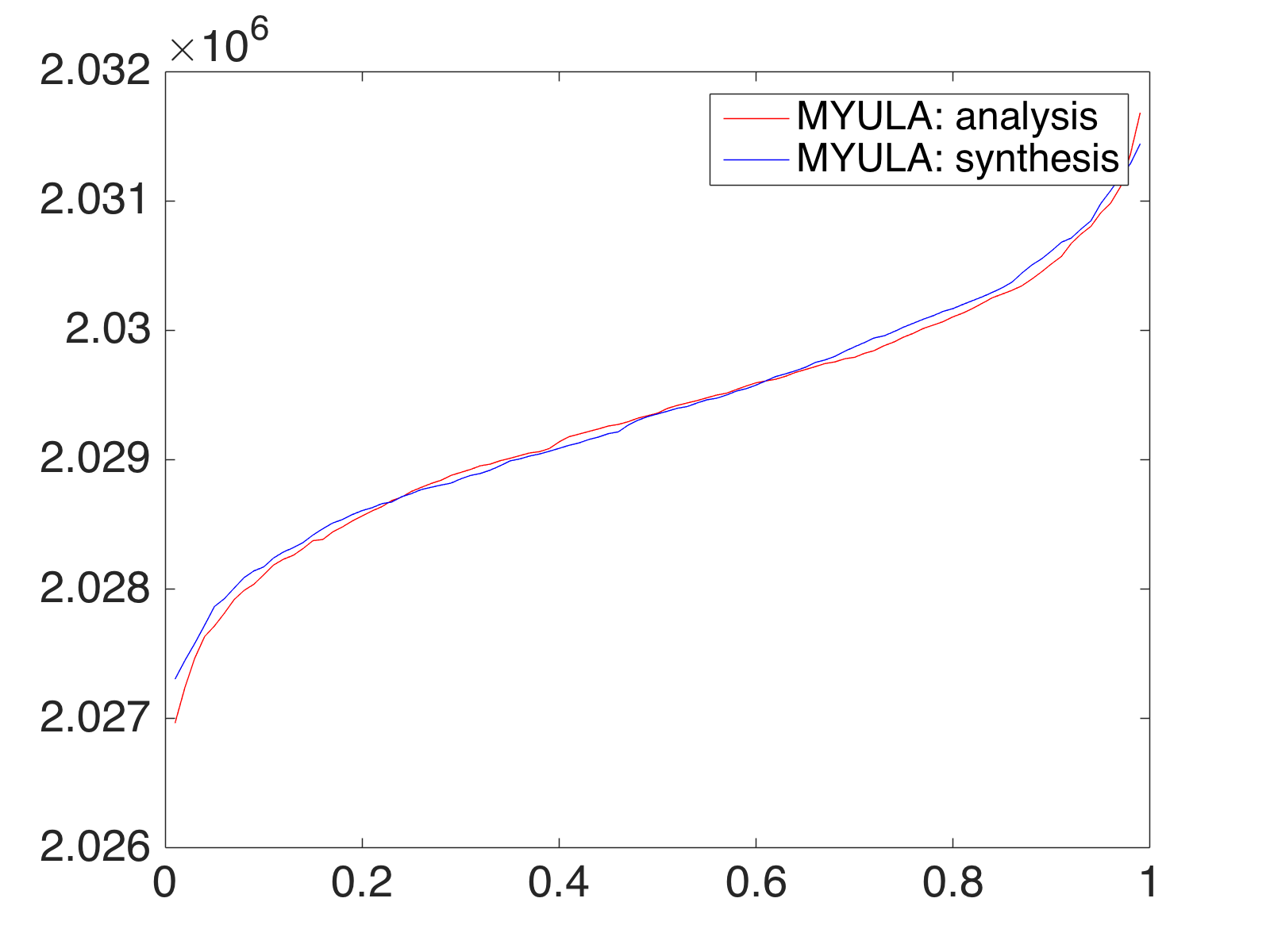}   \put(-60,-4){\tiny $1-\alpha$}
		\\
		\includegraphics[trim={{.02\linewidth} {.05\linewidth} {.1\linewidth} {.01\linewidth}}, clip, width=0.245\linewidth, height = 0.21\linewidth]
		{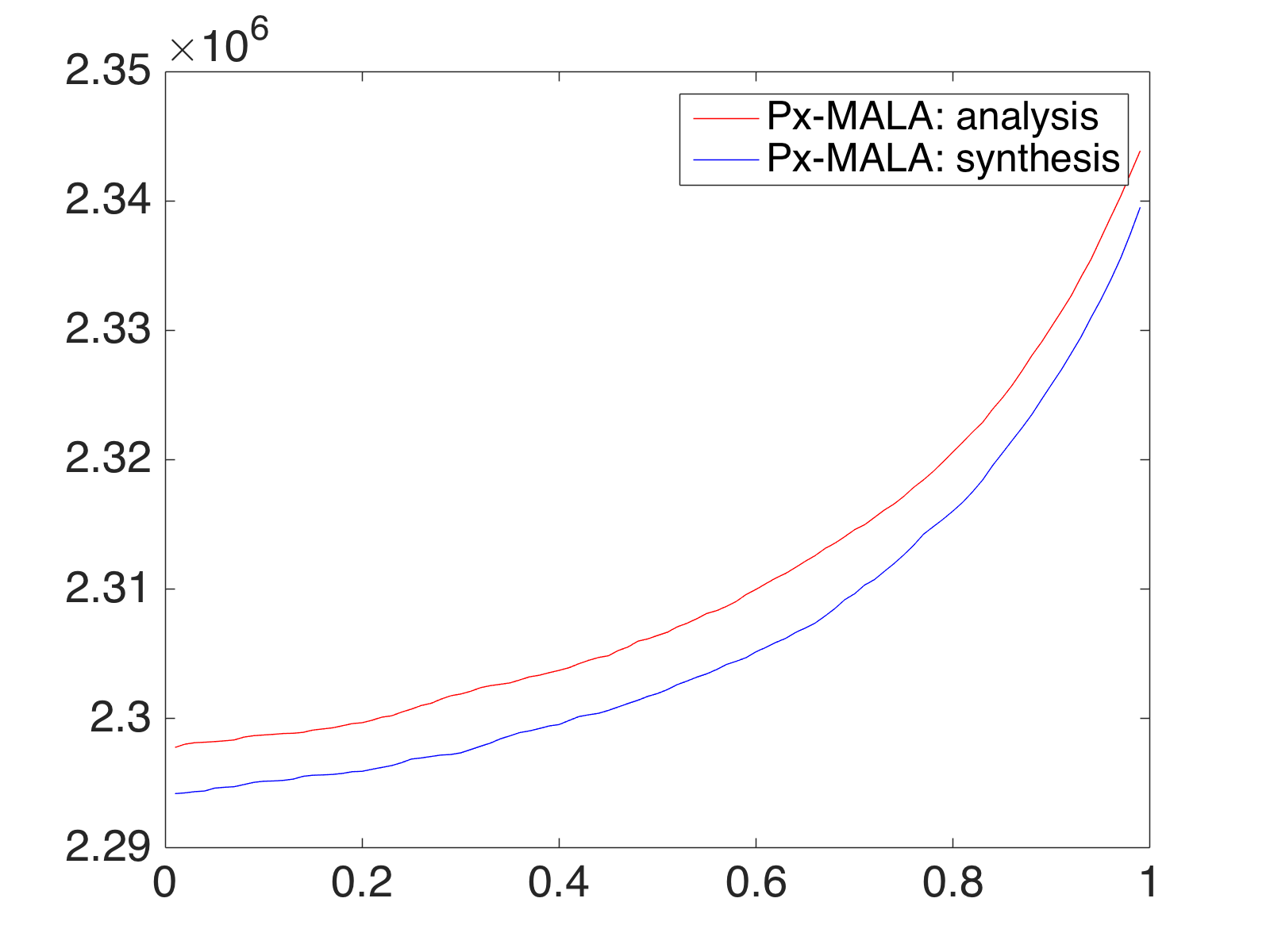}   \put(-60,-4){\tiny $1-\alpha$} &
		\includegraphics[trim={{.02\linewidth} {.05\linewidth} {.1\linewidth} {.01\linewidth}}, clip, width=0.245\linewidth, height = 0.21\linewidth]
		{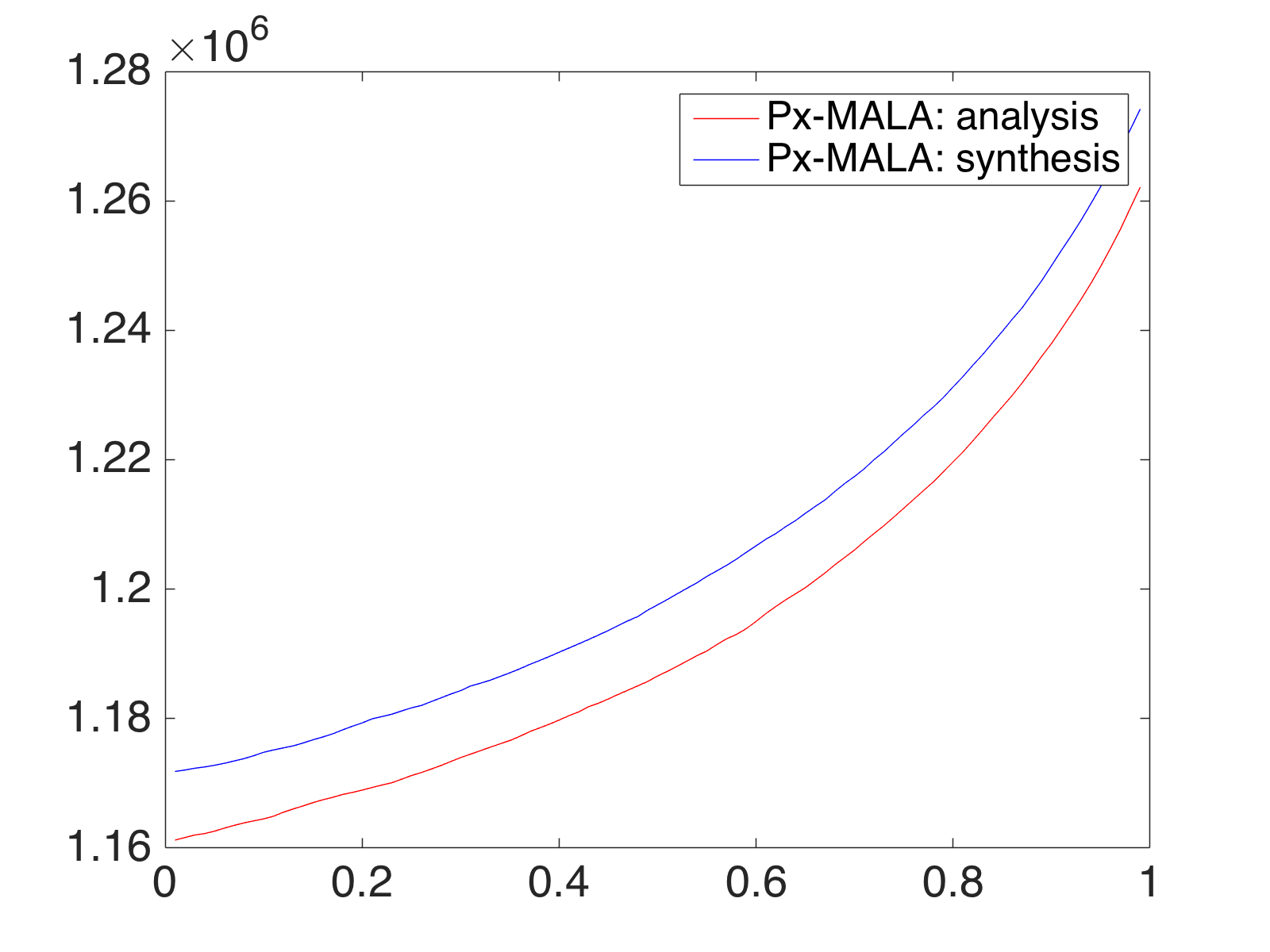}  \put(-60,-4){\tiny $1-\alpha$} &
		\includegraphics[trim={{.02\linewidth} {.05\linewidth} {.1\linewidth} {.01\linewidth}}, clip, width=0.245\linewidth, height = 0.21\linewidth]
		{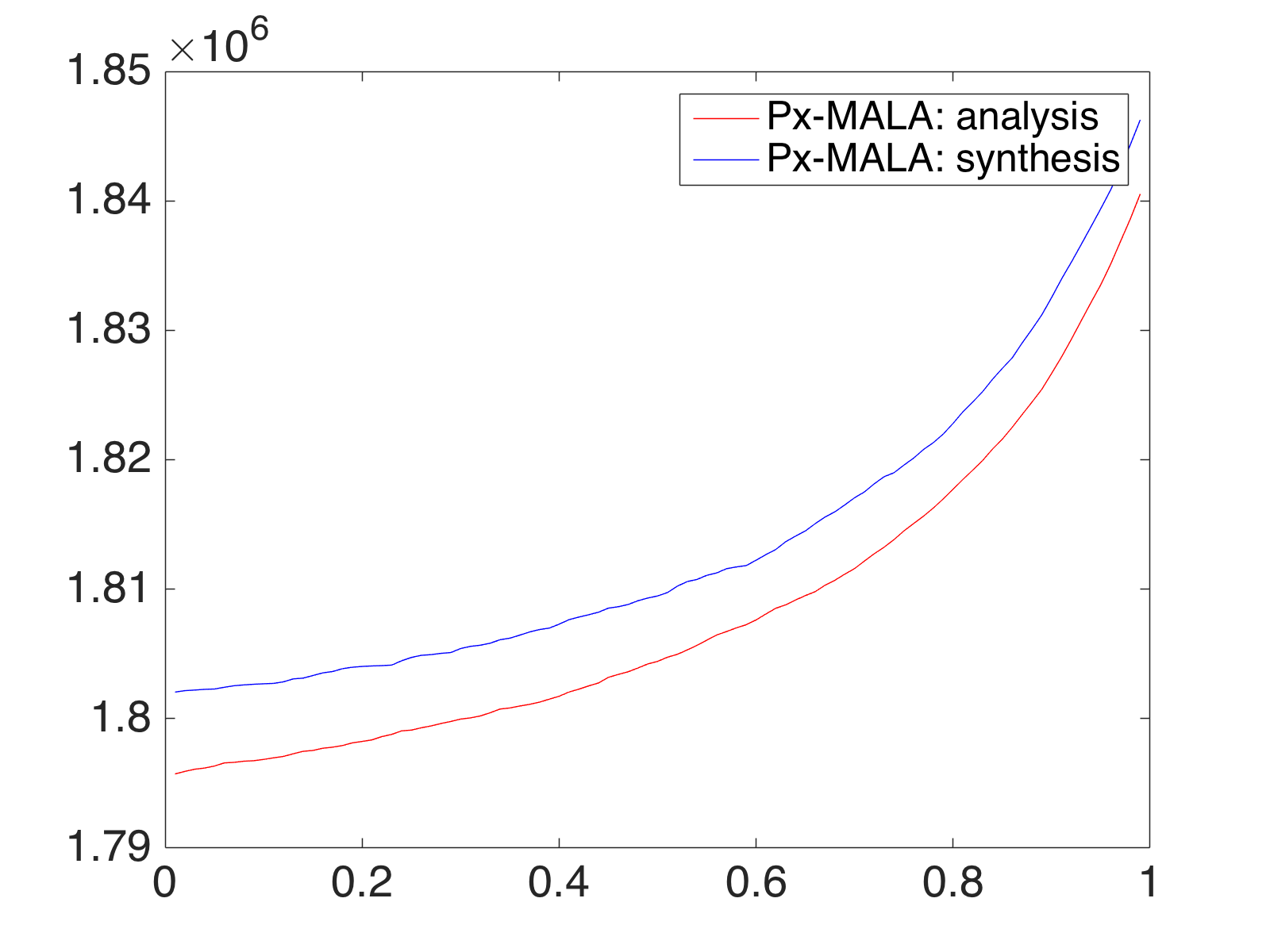}  \put(-60,-4){\tiny $1-\alpha$} &
		\includegraphics[trim={{.02\linewidth} {.05\linewidth} {.1\linewidth} {.01\linewidth}}, clip, width=0.245\linewidth, height = 0.21\linewidth]
		{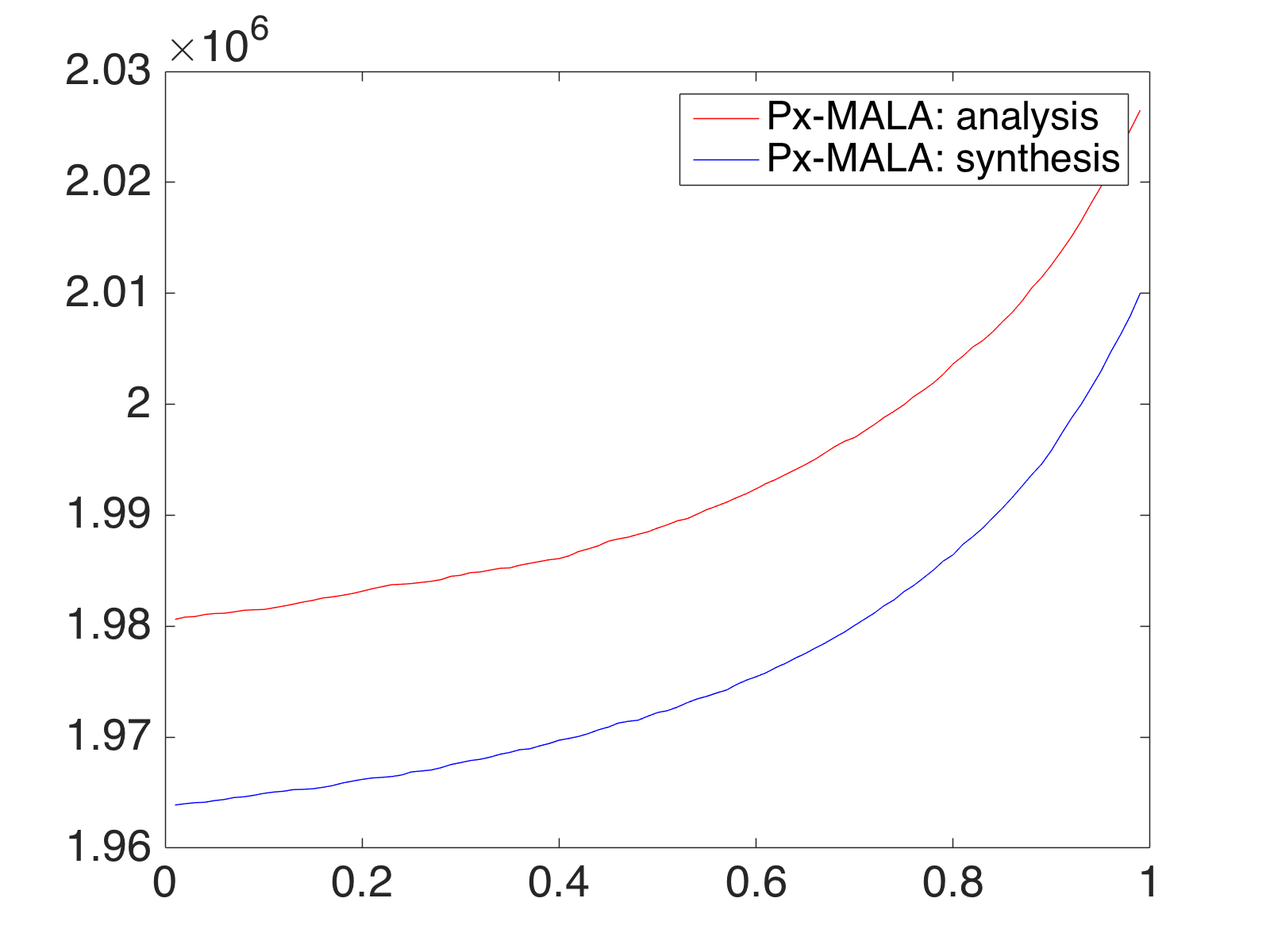}  \put(-60,-4){\tiny $1-\alpha$} \\

		 {\small  (a) M31 } & {\small (b) Cygnus A }  & {\small (c) W28}  & {\small (d) 3C288}  
        \end{tabular}
	\caption{HPD credible region isocontour levels $\gamma_{\alpha}$, 
	computed by MYULA (first row) and Px-MALA (second row), for test images (a) M31, (b) Cygnus A, (c) W28, and (d) 3C288, for the analysis 
	and synthesis models. Clearly, consistent results between Px-MALA and MYULA, and between the analysis and synthesis models, are obtained. Minor differences are discussed in the main text.
	}
	\label{fig-hpd-cr}
\end{figure*}

%--------
Figure \ref{fig-hpd-cr} shows the values of the HPD isocontour threshold ${\gamma}_{\alpha}$ \mbox{($\alpha \in [0.01, 0.99]$)}, 
defined in \eqref{eqn:cr-hpd}, computed with MYULA and Px-MALA using \eqref{eqn:hpd-r} for the synthesis and analysis models
 (red and blue colours are used to represent the results of the analysis and synthesis models, respectively). We observe that the MYULA and Px-MALA estimates are in agreement with each other. Similarly, the analysis and the synthesis models produce similar results. The minor differences in the estimates are again related to the bias-variance tradeoff of Px-MALA (MYULA produces estimates that are larger than Px-MALA but which are also more consistent, whereas Px-MALA estimates have less bias but are also less consistent because of a higher estimation variance). 
 
In the following section we use the HDP regions related to Figure \ref{fig-hpd-cr} to perform uncertainty quantification analyses and posterior 
checks for specific image structures.
 
 %%%%
\addtolength{\tabcolsep}{-\tabL}
\addtolength{\tabcolsep}{\tabL}
%%%%

%--------
\subsection{Hypothesis testing of image structure} \label{sec:exp-ht}
%--------
We now illustrate our methodology for testing structure in reconstructed images. We consider the five structures depicted in yellow in the first column of Figure \ref{fig-all-ci-hp}. All of these structures are physical (\emph{i.e.} present in the ground truth images), while for structure 2 in 3C288 is a reconstruction artefact.

Recall that the methodology proceeds as follows. First, we construct a surrogate test image ${\vect x}^{*, {\rm sgt}}$ by modifying a point estimator ({\it e.g.}, the sample mean or sample media image) by removing the structure of interest via segmentation-inpaiting ({\it e.g.}, by using \eqref{eqn:inpaint}, but results are generally not sensitive to the exact method used). Second, we check if ${\vect x}^{*, {\rm sgt}} \notin {C}_{\alpha}$ to determine whether there is strong evidence in favour of the structure considered. Conclusions are generally not highly sensitive to the exact value of $\alpha$; here we report results for $\alpha = 0.01$ related to a 99\% credible level.

The results of these experiments are summarised in Table~\ref{tab:hp-test} and Table~\ref{tab:hp-test-2}, which have been computed by using the posterior mean and the posterior median, respectively, to reconstruct $\hat{\vect x}^{*, {\rm sgt}}$. We observe that the same overall conclusions are largely obtained no matter which sampling method is used (MYULA or Px-MALA) or what model is applied (analysis model or synthesis model), indicating that the procedure is robust. Moreover, we observe that the three large physical structures are correctly classified and the reconstruction artefact is correctly highlighted as a structure for which there is lack of evidence. The structure in Cygnus A (see Figure \ref{fig-all-ci-hp}) is very small, containing only a few bright pixels that can easily be confused as noise, and it is typically highlighted as potentially non-physical. The only difference between Table \ref{tab:hp-test} and Table \ref{tab:hp-test-2} is the result of MYULA for the structure of Cygnus A, where the structure is correctly classified as physical when using the posterior median. This is due to the fact that the posterior median is closer to the boundary of ${C}_{\alpha}$ and has better sensitivity to small structures as a result. Therefore, we recommend using the median sample for testing. In summary, the proposed methodology, coupled with efficient MCMC sampling by MYULA, provides a powerful framework to perform detailed uncertainty analyses.

To conclude, we emphasise again that the standard methods for RI imaging, such as CLEAN-based methods, MEM
and CS-based methods, cannot provide error margins for their solutions, let alone support the detailed uncertainty quantification
 analyses presented in this article, which includes the calculation of local (pixel-wise) credible intervals, global HPD credible regions, and
tests for image structure.

%%%%
\begin{table*}
\begin{center}
\caption{Hypothesis test results for test structures shown in Figure~\ref{fig-all-ci-hp} for M31, Cygnus A, W28, and 3C288.
	Note that ${\gamma}_{\alpha}$ represents the isocontour defining the HPD credible region at credible level $(1-\alpha)$, where here $\alpha = 0.01$, 
	${\vect x}^{*, {\rm sgt}}$ represents the surrogate of point estimator ${\vect x}^{*}$ (sample mean), 
	and $(f+g)(\cdot)$ represents the objective function;  
	symbols with labels \ $\bar{}$ \ and \ $\hat{}$ \ are related to 
	the analysis model \eqref{eqn:ir-un-af} and the synthesis model \eqref{eqn:ir-un-sf}, respectively. 
	Symbol \xmark \ indicates that the test area is artificial (and no strong statistical statement 
	can be made as to the area), while \cmark \ indicates that the test area is physical. All values are in units $10^6$.
	Clearly, MYULA and Px-MALA give convincing and consistent hypothesis test results. 
	 } \label{tab:hp-test}
  \vspace{-0.05in}
\begin{tabular}{ccccHHIIc}
\toprule  
 \multirow{2}{*}{Images} &  Test  & Ground & \multirow{2}{*}{Method} &  \multicolumn{1}{c}{\multirow{2}{*}{$({\bar f} + {\bar g})(\bar{\vect x}^{*,{\rm sgt}})$ } }
 	&  \multicolumn{1}{c}{Isocontour} &  \multicolumn{1}{c}{\multirow{2}{*}{$({\hat f} + {\hat g})(\bm{\mathsf{\Psi}}^\dagger \hat{\vect x}^{*,{\rm sgt}})$ } }
 	& \multicolumn{1}{c}{Isocontour} & Hypothesis \\
	& areas & truth &  & \multicolumn{1}{c}{}  & \multicolumn{1}{c}{$\bar{\gamma}_{0.01}$} 
	&   \multicolumn{1}{c}{} & \multicolumn{1}{c}{$\hat{\gamma}_{0.01}$} & test
\\ \toprule 
\multirow{2}{*}{M31 (Fig. \ref{fig-all-ci-hp} ) } & \multirow{2}{*}{1}  &  \multirow{2}{*}{ \cmark}  &
MYULA & $\bf 2.20$  & $2.34$ &  $\bf 2.20$  & $2.34$ & \cmark
\\ 
& & & Px-MALA &  $\bf 2.44$  & $2.34$ & $\bf 2.43$   & $2.34 $ & \cmark
\\ \midrule
\multirow{2}{*}{Cygnus A  (Fig. \ref{fig-all-ci-hp} ) } &  \multirow{2}{*}{1} &  \multirow{2}{*}{ \cmark}  &
MYULA & $ 1.09$   & $ \bf 1.59$ & $ 1.09$  & $\bf 1.59$ &  \xmark
\\ 
& & & Px-MALA & $ 1.17$   & $ \bf 1.26$ & $ 1.18$  & $ \bf 1.27$ &  \xmark
\\ \midrule
\multirow{2}{*}{W28  (Fig. \ref{fig-all-ci-hp} ) } &   \multirow{2}{*}{1}  &   \multirow{2}{*}{ \cmark}  &
MYULA& $\bf 3.43$  & $1.96$ & $\bf 3.43$  &  $1.96$   &  \cmark
\\ 
& & &  Px-MALA & $\bf 3.38$   & $1.84$ &  $ \bf 3.37$   & $1.85$   &  \cmark
\\ \midrule
\multirow{4}{*}{3C288 (Fig. \ref{fig-all-ci-hp} ) } &  \multirow{2}{*}{1}  &   \multirow{2}{*}{ \cmark}  &
MYULA & $\bf 3.02$  & $2.03$  &   $\bf 3.02$ & $2.03$  &  \cmark
 \\ 
 & & & Px-MALA & $\bf 3.27$  & $2.02$  &  $\bf 3.25$ & $2.01$  &  \cmark    \\  \cdashline{2-9}
  &  \multirow{2}{*}{2} &  \multirow{2}{*}{ \xmark}  &
  MYULA & $1.752$ & $\bf 2.032$  & $1.752$ & $\bf 2.031$ & \xmark
 \\ 
   & & &  Px-MALA & $1.971$  & $\bf 2.027$ & $1.954$ & $\bf 2.010$ & \xmark
\\ \bottomrule
\end{tabular}
\end{center}
\end{table*}
%%%%

%%%%
\begin{table*}
\begin{center}
\caption{Same as Table~\ref{tab:hp-test} but based on the sample median instead of the sample mean (the mean is considered for Table~\ref{tab:hp-test}).
	This table shows that hypothesis tests based on the median, when using MYULA to generate samples, are able to detect very small structure, 
	such as the test region of Cygnus A. 
	 } \label{tab:hp-test-2}
  \vspace{-0.05in}
\begin{tabular}{ccccHHIIc}
\toprule  
 \multirow{2}{*}{Images} &  Test  & Ground & \multirow{2}{*}{Method} &  \multicolumn{1}{c}{\multirow{2}{*}{$({\bar f} + {\bar g})(\bar{\vect x}^{*,{\rm sgt}})$ } }
 	&  \multicolumn{1}{c}{Isocontour} &  \multicolumn{1}{c}{\multirow{2}{*}{$({\hat f} + {\hat g})(\bm{\mathsf{\Psi}}^\dagger \hat{\vect x}^{*,{\rm sgt}})$ } }
 	& \multicolumn{1}{c}{Isocontour} & Hypothesis \\
	& areas & truth &  & \multicolumn{1}{c}{}  & \multicolumn{1}{c}{$\bar{\gamma}_{0.01}$} 
	&   \multicolumn{1}{c}{} & \multicolumn{1}{c}{$\hat{\gamma}_{0.01}$} & test
\\ \toprule 
\multirow{2}{*}{M31 (Fig. \ref{fig-all-ci-hp} ) } & \multirow{2}{*}{1}  &  \multirow{2}{*}{ \cmark}  &
MYULA & $\bf 2.47$  & $2.34$ &  $\bf 2.48$ & $2.34$ & \cmark
\\ 
& & & Px-MALA &  $\bf 2.46 $  & $2.34$ & $\bf 2.46 $  & $2.34 $ & \cmark
\\ \midrule
\multirow{2}{*}{Cygnus A  (Fig. \ref{fig-all-ci-hp} ) } &  \multirow{2}{*}{1} &  \multirow{2}{*}{ \cmark}  &
MYULA & $ \bf 1.597$  & $  1.586$ & $ \bf 1.595$  & $ 1.586$ &  \cmark
\\ 
& & & Px-MALA & $ 1.205 $   & $ \bf 1.262$ & $ 1.216 $   & $ \bf 1.274$ &  \xmark
\\ \midrule
\multirow{2}{*}{W28  (Fig. \ref{fig-all-ci-hp} ) } &   \multirow{2}{*}{1}  &   \multirow{2}{*}{ \cmark}  &
MYULA & $\bf 3.67$  & $1.96$ & $\bf 3.67$  &  $1.96$   &  \cmark
\\ 
& & &  Px-MALA & $\bf 3.41 $  & $1.84$ &  $ \bf 3.39$   & $1.85$   &  \cmark
\\ \midrule
\multirow{4}{*}{3C288 (Fig. \ref{fig-all-ci-hp} ) } &  \multirow{2}{*}{1}  &   \multirow{2}{*}{ \cmark}  &
MYULA & $\bf 3.30$  & $2.03$  &   $\bf 3.30$ & $2.03$  &  \cmark
 \\ 
 & & & Px-MALA & $\bf 3.29$  & $2.02$  &  $\bf 3.27$ & $2.01$  &  \cmark    \\  \cdashline{2-9}
  &  \multirow{2}{*}{2} &  \multirow{2}{*}{ \xmark}  &
  MYULA & $2.026$ & $\bf 2.032$  & $2.027$ & $\bf 2.031$ & \xmark
 \\ 
   & & & Px-MALA & $ 1.994 $ & $\bf 2.027$ & $1.977 $  & $\bf 2.010$ & \xmark
\\ \bottomrule
\end{tabular}
\end{center}
\end{table*}
%%%%

%%%%
\addtolength{\tabcolsep}{-\tabL}
{ \renewcommand{\arraystretch}{0.0}
\begin{figure}
	\centering
	\begin{tabular}{cc}
		\includegraphics[trim={{.15\linewidth} {.07\linewidth} {.02\linewidth} {.155\linewidth}}, clip, width=0.49\linewidth, height = 0.41\linewidth]
		{./figs/M31_MYULA_mean_sample_ana} \put(-120,40){\rotatebox{90}{ M31}}  
		\put(-100,34){\yellow{\framebox(12,12){ }}}  \put(-100,34){\yellow{\bf 1}} 
		 &		 
		\includegraphics[trim={{.15\linewidth} {.07\linewidth} {.02\linewidth} {.155\linewidth}}, clip, width=0.49\linewidth, height = 0.41\linewidth]
		{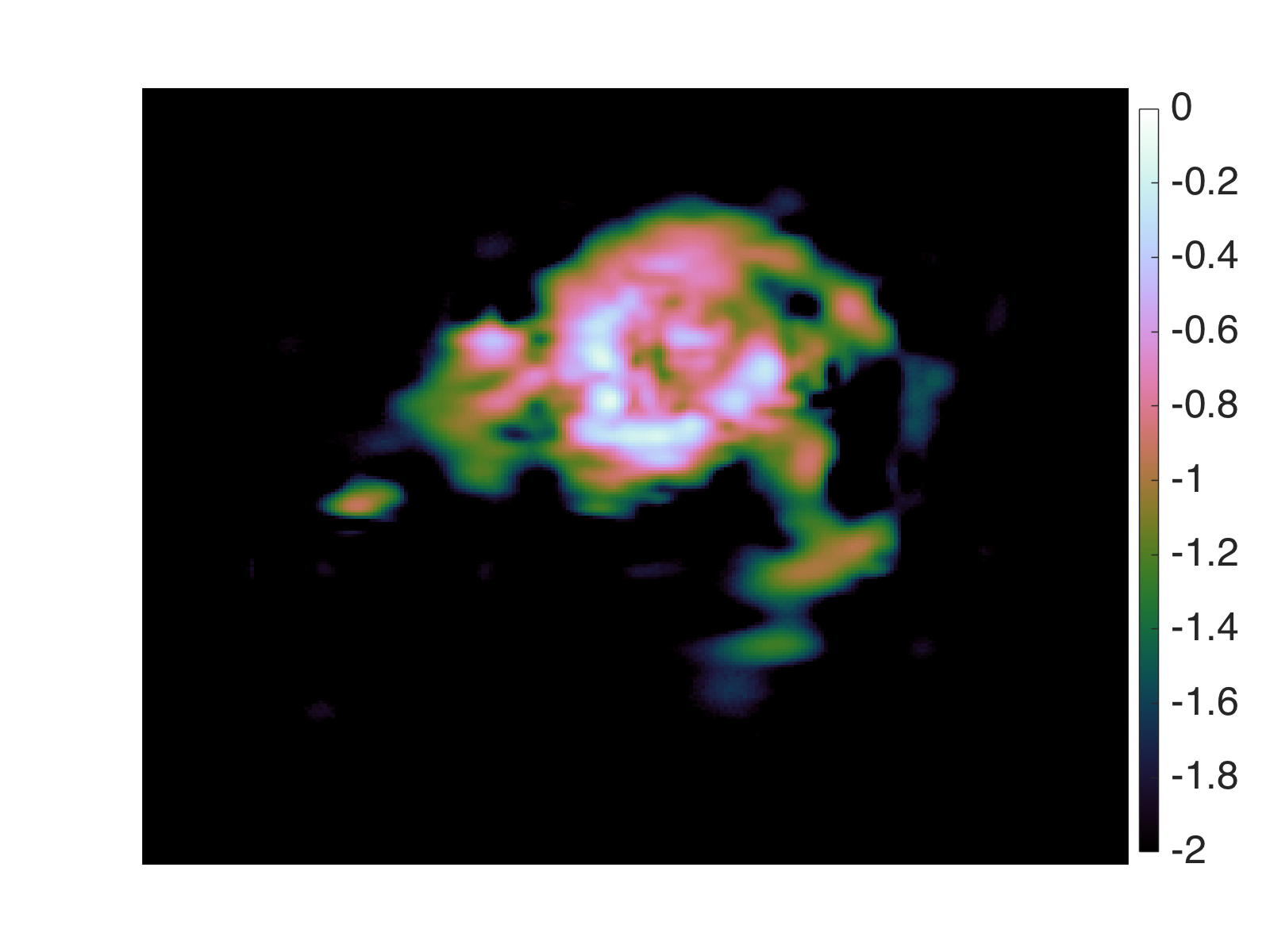}	 
		\\  
		\includegraphics[trim={{.15\linewidth} {.07\linewidth} {.02\linewidth} {.155\linewidth}}, clip, width=0.49\linewidth, height = 0.22\linewidth]
		{./figs/CYN_MYULA_mean_sample_ana}   \put(-120,11){\rotatebox{90}{ Cygnus A}}
		  \put(-69,21){\yellow{\framebox(10,10){ }}}  \put(-69,21){\yellow{\bf 1}} 
		&		
		\includegraphics[trim={{.15\linewidth} {.07\linewidth} {.02\linewidth} {.155\linewidth}}, clip, width=0.49\linewidth, height = 0.22\linewidth]
		{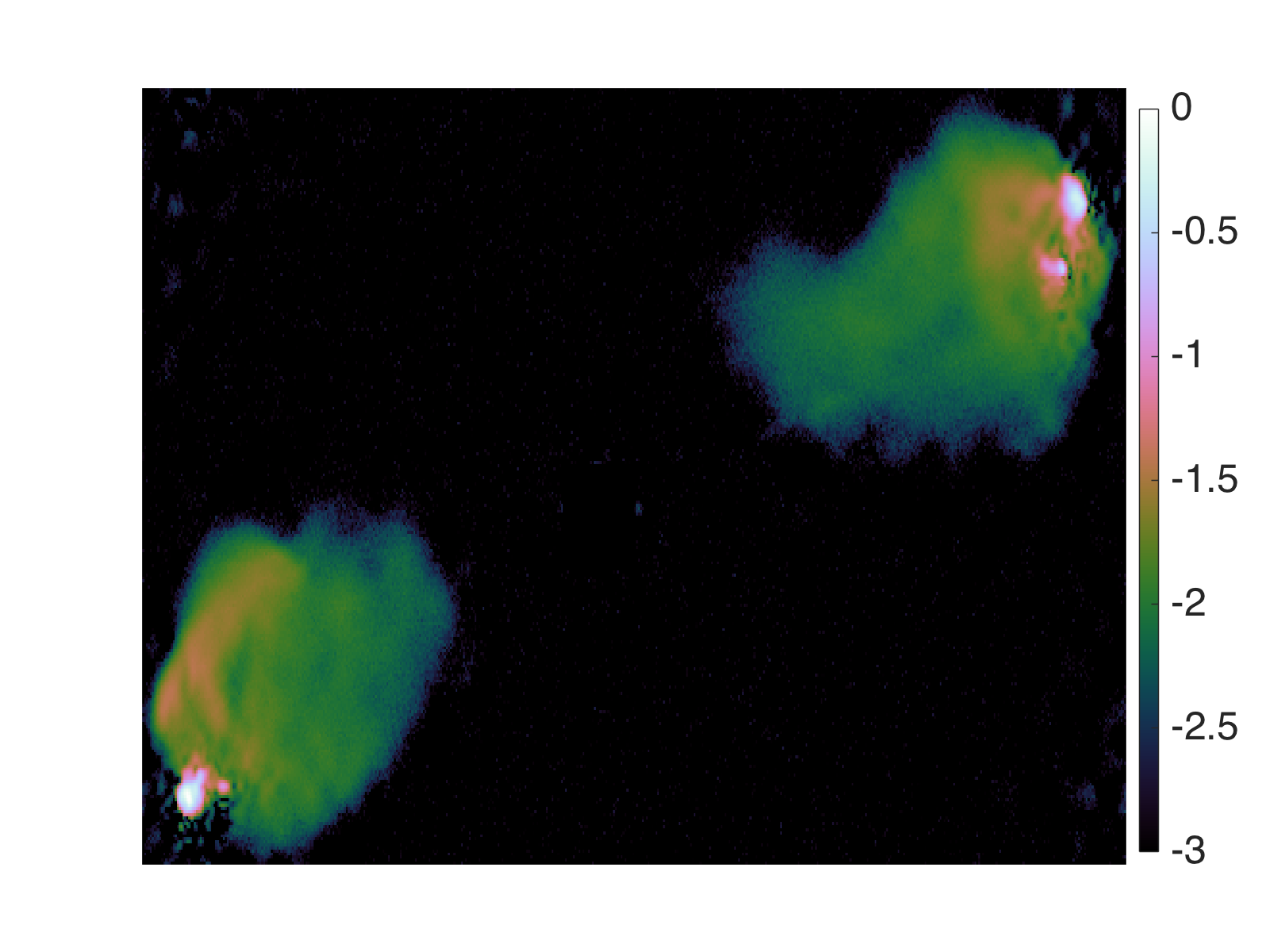} 	 
		 \\		 
		 \includegraphics[trim={{.15\linewidth} {.07\linewidth} {.02\linewidth} {.155\linewidth}}, clip, width=0.49\linewidth, height = 0.41\linewidth]
		{./figs/W28_MYULA_mean_sample_ana} \put(-120,40){\rotatebox{90}{ W28}}
		 \put(-109,57){\yellow{\framebox(10,10){ }}}  \put(-109,57){\yellow{\bf 1}}
		&
		\includegraphics[trim={{.15\linewidth} {.07\linewidth} {.02\linewidth} {.155\linewidth}}, clip, width=0.49\linewidth, height = 0.41\linewidth]
		{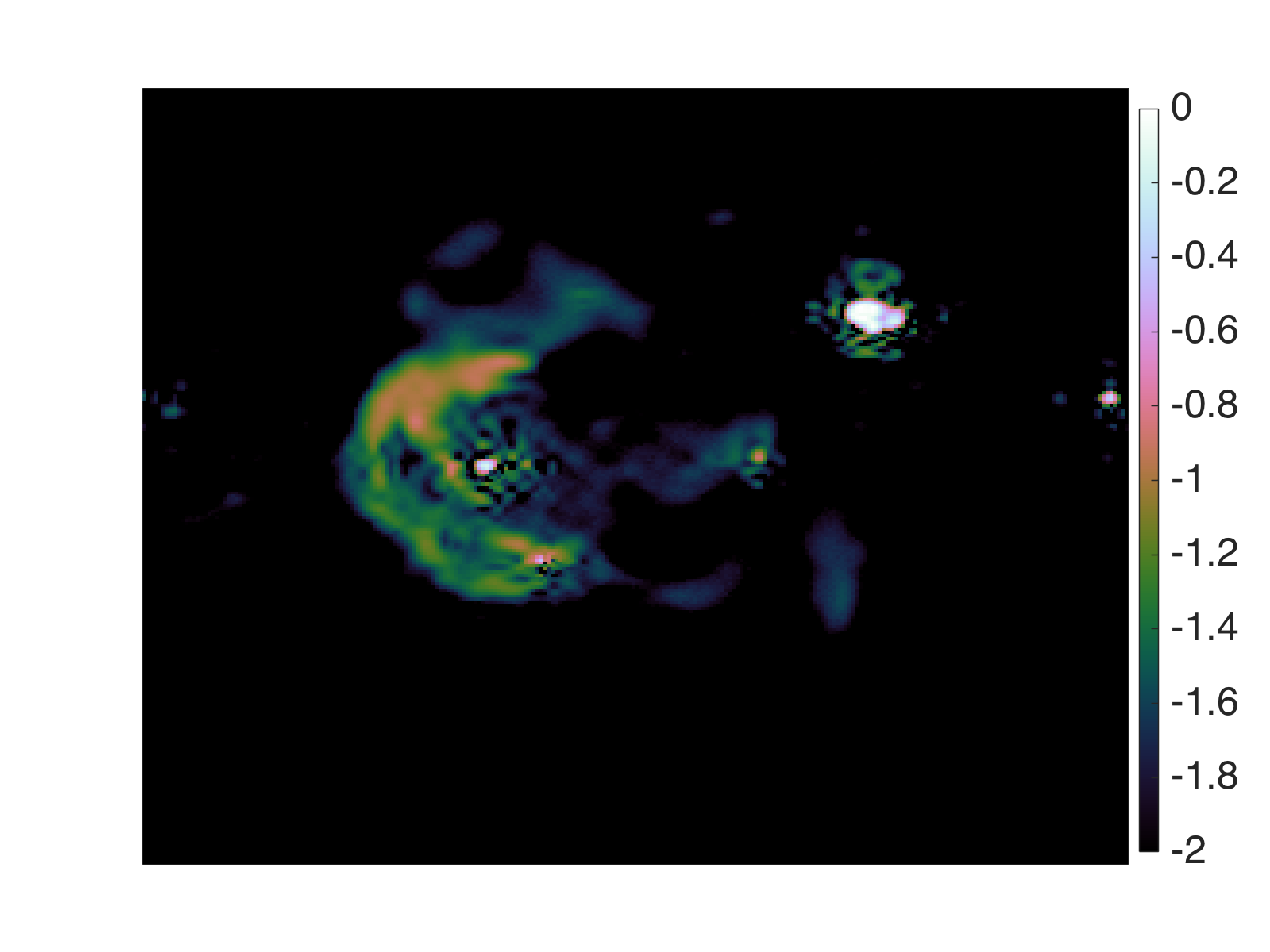}
		\\
		 \includegraphics[trim={{.15\linewidth} {.07\linewidth} {.02\linewidth} {.155\linewidth}}, clip, width=0.49\linewidth, height = 0.41\linewidth]
		{./figs/3C288_MYULA_mean_sample_ana} \put(-120,38){\rotatebox{90}{ 3C288}}
		 \put(-76,47){\yellow{\framebox(9,9){ }}}  \put(-76,47){\yellow{\bf 1}}
		 \put(-54,82){\yellow{\framebox(12,12){ }}}  \put(-54,82){\yellow{\bf 2}}
		&
		\includegraphics[trim={{.15\linewidth} {.07\linewidth} {.02\linewidth} {.155\linewidth}}, clip, width=0.49\linewidth, height = 0.41\linewidth]
		{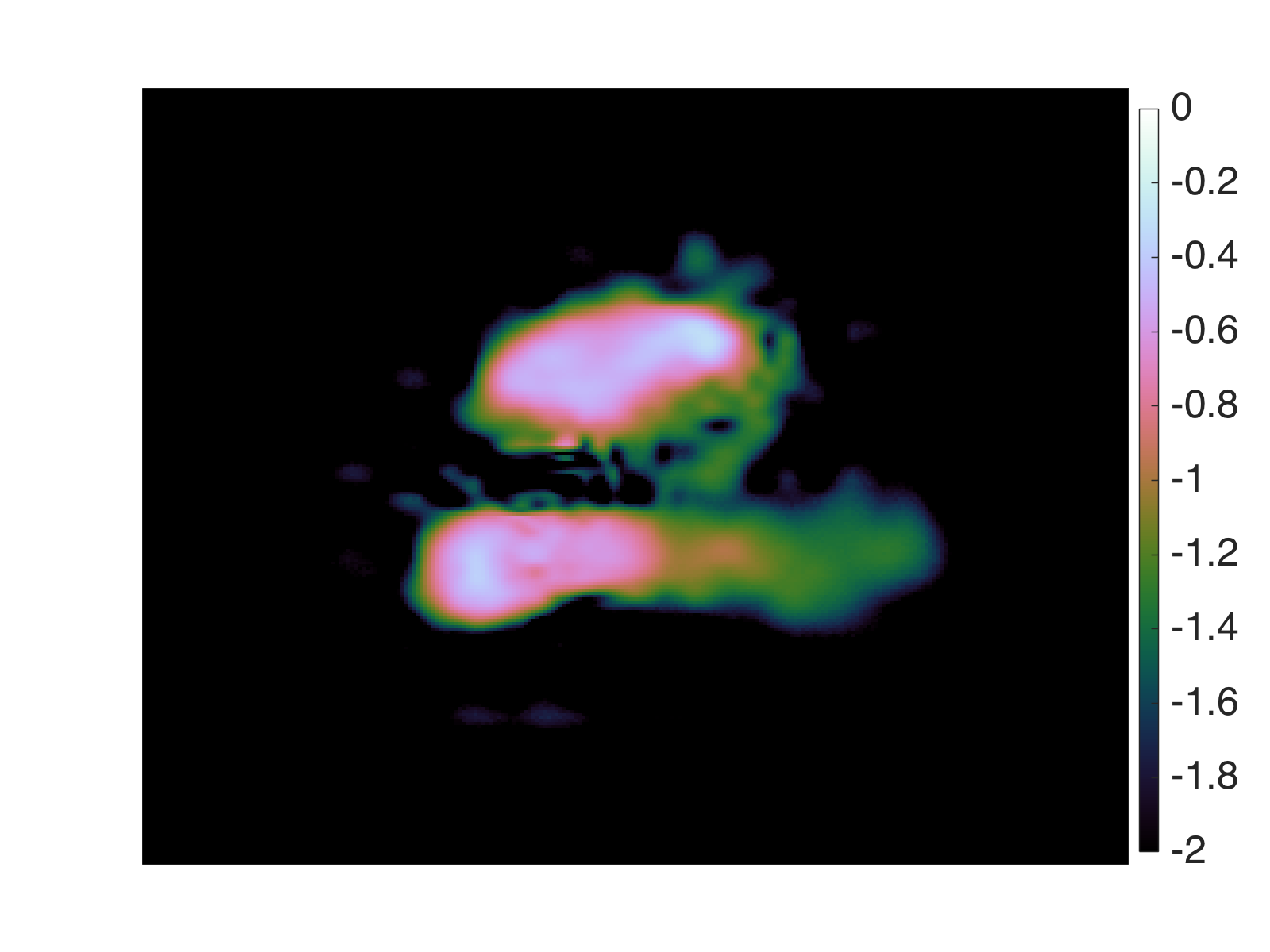}  \\
		{\small (a) MYULA point estimators } & {\small (b) inpainted surrogate }
        \end{tabular}
	\caption{Hypothesis testing for M31, Cygnus A, W28, and 3C288. 
  The five structures depicted in yellow are considered, all of which are physical ({\it i.e.} present in the ground truth images), except for structure 2 in 3C288, which is a reconstruction artefact.
	First column (a): point estimators obtained by MYULA for the analysis model \eqref{eqn:ir-un-af} (shown 
	in ${\tt log}_{10}$ scale). Second column (b): segmented-inpainted surrogate test images with information 
	in the yellow rectangular areas removed and replaced by inpainted background  (shown 
	in ${\tt log}_{10}$ scale). 
	Hypothesis testing is then performed to
	test whether the structure considered is physical by checking whether the surrogate test images shown in (b) fall outside of the HPD credible regions.  Results of these hypothesis tests are specified in Table~\ref{tab:hp-test} and Table~\ref{tab:hp-test-2}.
	Note that for the case shown in the last row the structures within areas 1 and 2 are tested independently.
	 }
	\label{fig-all-ci-hp}
\end{figure}
}
\addtolength{\tabcolsep}{\tabL}
%%%%

%-------------------------------------------------------------------
\section{Conclusions}\label{sec:con}
%-------------------------------------------------------------------
Uncertainty quantification is an important missing component in RI imaging that will only become increasingly important as the big-data era of radio interferometry emerges.  No existing RI imaging techniques that are used in practice (\textit{e.g.} CLEAN, MEM or CS approaches) provide uncertainty quantification.  Recent techniques that do provide some form of uncertainty information only support restrictive classes of priors (typically Gaussian or log-normal) and do not scale to big-data.  While sparsity-promoting priors have shown a great deal of promise for RI imaging \citep[\textit{e.g.}][]{PMdCOW16} and are receiving a great deal of attention, it has not previously been possible to quantify uncertainty information when adopting sparse priors.  Traditional MCMC sampling approaches that provide uncertainty information and scale to high dimensional settings, such as RI imaging, often exploit gradient information and cannot support non-differentiable sparse priors.  In the current article we solve precisely this problem.

We formulate the RI imaging problem in a Bayesian framework and consider two image models -- the analysis and synthesis models -- where sparse priors in a suitable signal representation (\textit{e.g.} wavelet basis) are adopted.  To perform Bayesian inference for models with sparse priors we consider two innovative MCMC sampling techniques, MYULA and Px-MALA, to sample the full, high-dimensional posterior image distribution.  These so-called proximal MCMC techniques exploit proximal calculus to handle non-differentiable prior distributions in high dimensional settings. 

Once the full posterior distribution is recovered, a single image is obtained from a point estimator and a variety of methods are presented to perform different types of uncertainty quantification.  Pixel-wise credible intervals are computed from the posterior distribution to provide, essentially, error bars for each individual pixel of the recovered image.  HPD credible regions are determined for the entire reconstruction, which are then used to perform hypothesis tests of image structure to determine whether the structure is physical or an artefact.  

We evaluated our methods on several test images that are representative in RI imaging.  Simple simulations of RI observations were performed and Px-MALA and MYULA were used to sample the full image posterior distribution, from which the uncertainty quantification techniques outlined above were applied.  Accurate point estimates of recovered images and meaningful uncertainty information were obtained.  While Px-MALA is guaranteed to converge to the target distribution, MYULA exhibits an asymptotic bias that can be made arbitrarily small.  MYULA, however, does not involve an MH accept-reject step which slows convergence considerably for Px-MALA.  

In summary, we develop proximal MCMC techniques to sample the full image posterior distribution for RI imaging for the sparse priors that have been shown in practice to be highly effective.  From the posterior distribution a point estimate of the image can be computed and uncertainty information regarding the accuracy of the reconstructed image can be quantified in a variety of ways.  These forms of uncertainty quantification provide rich information for analysing RI observations in a statistically robust manner.

In future work the techniques presented here will be extended to consider more complex models, for example with overcomplete dictionaries and for $\ell_p$ priors with $0\le p < 1$, which
can provide a stronger sparsity constraint than the $\ell_1$ prior.  
{Furthermore, we will investigate optimal techniques for setting the regularisation parameter in a hierarchical Bayesian framework, applying the strategies developed by \citet{MBF15}.  A more realistic measurement operator that better models real radio interferometry telescopes can be easily incorporated in our framework simply by replacing the measure operator $\bm{\mathsf{\Phi}}$ adopted.}

{We have so far considered the telescope calibration parameters to be estimated \textit{a priori} and then fixed.  Similarly to $\mu$, one can also consider hierarchical and empirical Bayesian approaches to fix or marginalise calibration parameters. In terms of uncertainty quantification, marginalisation has the advantage of integrating the uncertainty w.r.t.\ calibration parameters in the analyses, whereas methods that fix calibration parameters neglect this source of uncertainty. We emphasise at this point that  performing RI imaging and calibration jointly is a challenging problem because of the dimensionality involved, and this difficulty also extends to uncertainty quantification. Consequently, we leave this problem for future consideration.}

For massive data sizes, \textit{e.g.} big-data, like those anticipated from the SKA, it will be difficult if not impossible to apply any MCMC technique due to its inherent computational cost.
In the companion article \citep{CPMp217} we show how to scale the uncertainty quantification techniques presented in this article to big-data, exploiting recent developments in probability theory and again supporting the sparse priors that have been shown to be so effective in practice.

%-------------------------------------------------------------------
\section*{Acknowledgements}
%-------------------------------------------------------------------
This work is supported by the UK Engineering and Physical Sciences Research Council (EPSRC) by grant EP/M011089/1, and
Science and Technology Facilities Council (STFC) ST/M00113X/1. {We also thank the editor and the anonymous reviewer for their constructive comments, which have significantly improved this manuscript.}

\bibliographystyle{mnras}
\bibliography{refs_xhcai}

%=============================================================================
\label{lastpage}
 %---------------------------------------------------------------
\end{document}